\documentclass[prd,nofootinbib,preprint,superscriptaddress]{revtex4}
\pdfoutput=1
\usepackage[T1]{fontenc}
\usepackage{amsmath,amssymb}
\usepackage{epsfig}
\usepackage{graphicx}
\usepackage[usenames,dvipsnames]{color}
\usepackage{subfigure}
\usepackage{slashed}
\usepackage[colorlinks,citecolor=blue]{hyperref}
\usepackage{pdfpages}
\usepackage{color}

\begin{document}
\title{Low scale leptogenesis and dark matter in the presence of primordial black holes}

\author{Suruj Jyoti Das}
\email{suruj@iitg.ac.in}
\affiliation{Department of Physics, Indian Institute of Technology Guwahati, Assam 781039, India}

\author{Devabrat Mahanta}
\email{devab176121007@iitg.ac.in}
\affiliation{Department of Physics, Indian Institute of Technology Guwahati, Assam 781039, India}

\author{Debasish Borah}
\email{dborah@iitg.ac.in}
\affiliation{Department of Physics, Indian Institute of Technology Guwahati, Assam 781039, India}

\begin{abstract}
We study the possibility of low scale leptogenesis along with dark matter (DM) in the presence of primordial black holes (PBH). For a common setup to study both leptogenesis and DM we consider the minimal scotogenic model which also explains light neutrino mass at radiative level. While PBH in the mass range of $0.1-10^5$ g can, in principle, affect leptogenesis, the required initial PBH fraction usually leads to overproduction of scalar doublet DM whose thermal freeze-out occurs before PBH evaporation. PBH can lead to non-thermal source of leptogenesis as well as dilution of thermally generated lepton asymmetry via entropy injection, with the latter being dominant. The parameter space of scotogenic model which leads to overproduction of baryon or lepton asymmetry in standard cosmology can be made consistent in the presence of PBH with appropriate initial mass and energy fraction. On the other hand, for such PBH parameters, the scalar DM is constrained to be in light mass regime where its freeze-out occurs after PBH evaporation. We then discuss the possibility of fermion singlet DM with $N_2$ leptogenesis in the same model where due to singlet nature of DM, its connection with PBH parameters and hence leptogenesis becomes stronger compared to the previous case.

\end{abstract}

\maketitle

\section{Introduction}
The observed baryon asymmetry of the universe (BAU) has been a longstanding puzzle in particle physics and cosmology  \cite{Zyla:2020zbs,Aghanim:2018eyx}. While the universe is expected to start in a matter-antimatter symmetric manner, any primordial asymmetry set by initial conditions is expected to get diluted by exponential expansion phase during inflation. The observed BAU is often quoted in terms of baryon to photon ratio which, according to the latest data from Planck satellite  is given as \cite{Aghanim:2018eyx} 
\begin{equation}
\eta_B = \frac{n_{B}-n_{\bar{B}}}{n_{\gamma}} = 6.1 \times 10^{-10}.
\label{etaBobs}
\end{equation}
Such a ratio based on the cosmic microwave background (CMB) measurements agree with the big bang nucleosynthesis (BBN) estimates as well. Similar to the BAU, there has been another longstanding puzzle related to the presence of a mysterious, non-luminous form of matter, popularly known as dark matter (DM), giving rise to approximately $26\%$ of the present universe. In terms of density 
parameter $\Omega_{\rm DM}$ and $h = \text{Hubble Parameter}/(100 \;\text{km} ~\text{s}^{-1} 
\text{Mpc}^{-1})$, the present DM abundance is conventionally reported as \cite{Aghanim:2018eyx}
\begin{equation}
\Omega_{\text{DM}} h^2 = 0.120\pm 0.001
\label{dm_relic}
\end{equation}
at 68\% CL. Apart from cosmological evidences, presence of DM has also been suggested by several astrophysical evidences \cite{Zwicky:1933gu, Rubin:1970zza, Clowe:2006eq}. While none of the standard model (SM) particles satisfy the criteria of a particle DM candidate, SM also fails to satisfy the criteria (known as Sakharov's conditions \cite{Sakharov:1967dj}) in adequate amounts to dynamically generate the observed BAU. This has led to several beyond standard model (BSM) proposals offering possible solutions to these puzzles.

The simplest way to generate the BAU is to introduce  additional heavy particles which can decay (or annihilate) into SM particles in a way which satisfies all the conditions mentioned above, leading to successful baryogenesis \cite{Weinberg:1979bt, Kolb:1979qa}. A popular way to achieve baryogenesis through lepton sector physics is known as leptogenesis \cite{Fukugita:1986hr} where, instead of creating a baryon asymmetry directly from baryon number (B) violating processes, a lepton asymmetry is generated from lepton number (L) violating interactions. If this lepton asymmetry is generated before the electroweak phase transition (EWPT), then the $(B+L)$-violating EW sphaleron transitions~\cite{Kuzmin:1985mm} can convert it to the required baryon asymmetry\footnote{A review of leptogenesis can be found in \cite{Davidson:2008bu}.}. Another interesting aspect of leptogenesis is the fact that it can be implemented naturally in typical seesaw models explaining the origin of light neutrino masses \cite{Minkowski:1977sc, Mohapatra:1979ia, Yanagida:1979as, GellMann:1980vs, Glashow:1979nm, Schechter:1980gr}, another observed phenomena SM fails to address. On the other hand, among the BSM proposals for DM, the weakly interacting massive particle (WIMP) \cite{Kolb:1990vq} is perhaps the most widely studied one\footnote{A review of WIMP type scenarios can be found in \cite{Arcadi:2017kky}.}. In WIMP paradigm, a DM particle with mass and interaction strength with SM particles typically around the ballpark of electroweak scale physics can naturally give rise to the observed DM abundance after thermal freeze-out in the early universe. Due to sizeable DM-SM interactions in WIMP paradigm, one can have observable DM-nucleon scattering at direct search experiments like XENON1T \cite{Aprile:2018dbl} keeping the scenario predictive and verifiable.

One common aspect of both leptogenesis and WIMP DM is that they are high scale phenomena occurring in the radiation dominated universe. While a non-vanishing lepton asymmetry must be generated before the EWPT temperature $T \sim \mathcal{O}(100 \; \rm GeV)$ in order to get converted into BAU, a typical WIMP type DM with mass $M_{\rm DM}$ usually freezes out at temperature corresponding to $M_{\rm DM}/T \sim \mathcal{O}(20-30)$. Both these temperatures correspond to radiation dominated era of standard $\Lambda {\rm CDM}$ cosmology. However, there exists no experimental evidence to suggest that the universe was radiation dominated prior to the BBN era, typically around 1 s after the big bang, corresponding to temperature of order $T \sim \mathcal{O}(1 \; \rm MeV)$. This has led to several studies where a non-standard cosmological phase\footnote{A recent review of such non-standard cosmology can be found in \cite{Allahverdi:2020bys}.} leading to either faster or slower expansion compared to the standard radiation dominated phase is introduced to study its impact on either DM phenomenology  \cite{McDonald:1989jd, Kamionkowski:1990ni, Chung:1998rq, Moroi:1999zb, Giudice:2000ex, Allahverdi:2002nb, Allahverdi:2002pu, Acharya:2009zt, Davoudiasl:2015vba, Drees:2018dsj, Bernal:2018ins, Bernal:2018kcw, Arias:2019uol, Delos:2019dyh, Chanda:2019xyl, Bernal:2019mhf, Poulin:2019omz, Maldonado:2019qmp, Betancur:2018xtj, DEramo:2017gpl, DEramo:2017ecx, Biswas:2018iny, Visinelli:2015eka, Visinelli:2017qga, Catena:2004ba, Catena:2009tm, Meehan:2015cna, Dutta:2016htz, Dutta:2017fcn, Arcadi:2020aot, Bernal:2020bfj} or leptogenesis \cite{Abdallah:2012nm, Dutta:2018zkg, Chen:2019etb}. In one of our earlier works \cite{Mahanta:2019sfo}, we studied the impact of such non-standard cosmology on both DM and leptogenesis within the framework of a popular model known as the scotogenic model offering radiative origin of light neutrino masses and a stable DM candidate \cite{Ma:2006km}. For some more recent works studying the effect of non-standard cosmology on DM and leptogenesis, please see \cite{Konar:2020vuu, Chang:2021ose}.

In this work, we consider another non-standard scenario where the early universe can be dominated by primordial black holes (PBH)\footnote{For a recent review of PBH, refer to \cite{Carr:2020gox}.}. Initially proposed by Hawking \cite{Hawking:1974rv, Hawking:1974sw}, PBH can have several interesting consequences in cosmology \cite{Chapline:1975ojl, Carr:1976zz, Carr:2009jm}. While PBH can evaporate by emitting Hawking radiation, it can be stable on cosmological scales if sufficiently heavy, potentially giving rise to some or all of DM \cite{Carr:2020xqk}. Even if lighter PBH are not long lived enough to be DM, they can still play non-trivial roles in genesis of DM as well as baryogenesis. Since PBH evaporate to all particles, irrespective of their SM gauge interactions, it can lead to production of DM as well as other heavy particles responsible for creating baryon or lepton asymmetry. While the role of PBH evaporation on baryogenesis was first pointed out in \cite{Hawking:1974rv, Carr:1976zz} followed by some detailed study in \cite{Baumann:2007yr}, recently it has been taken up by several authors in different contexts \cite{Hook:2014mla, Fujita:2014hha, Hamada:2016jnq, Morrison:2018xla, Hooper:2020otu, Perez-Gonzalez:2020vnz, Datta:2020bht}. In \cite{DeLuca:2021oer}, the authors found that the rate of baryon number violation via sphaleron transitions in the standard model can be enhanced in the presence of PBH. On the other hand, the role of PBH evaporation on DM genesis has been studied in different DM scenarios, see \cite{Morrison:2018xla, Gondolo:2020uqv, Bernal:2020bjf} and references therein.

In this work, we consider a low scale leptogenesis scenario along with WIMP type DM in the presence of PBH. Unlike in earlier works where high scale thermal or non-thermal leptogenesis were studied in the presence of PBH \cite{Fujita:2014hha, Morrison:2018xla, Hooper:2020otu, Perez-Gonzalez:2020vnz}, here we adopt a radiative seesaw scenario where scale of leptogenesis can be as low as a few TeV even with hierarchical right handed neutrinos. We consider the scenario where PBH evaporate before EWPT while focusing on three specific cases where evaporation occurs before, during or after the scale of leptogenesis. We find the PBH parameter space which can give rise to the observed asymmetry in the model, assuming PBH domination in the early universe. Although the dependency of the generated asymmetry on the relevant parameters of the particle physics model is already well studied, we demonstrate the asymmetry in the PBH parameter space, and find some interesting deviation from the usual thermal leptogenesis scenario, in the presence of PBH. We also investigate the possible scenarios of DM genesis in the same setup and reach at some interesting conclusions. 

This paper is organised as follows. In section \ref{sec2} we briefly discuss the scotogenic model and thermal leptogenesis with standard cosmological history. In section \ref{sec:lepto}, we discuss leptogenesis in scotogenic model in the presence of PBH followed by discussion of scalar doublet DM in section \ref{sec:DM}. In section \ref{sec:N2lepto}, we discuss the possibility of fermion singlet DM with leptogenesis due to heavier right handed neutrino decay. We finally conclude in section \ref{sec:conclude}.

\section{Standard thermal leptogenesis at TeV scale}
\label{sec2}
As mentioned before, we consider a specific particle physics model which allows the possibility of TeV scale leptogenesis, DM and non-zero neutrino mass. This is the minimal scotogenic model \cite{Ma:2006km} where the SM is extended by three gauge singlet right handed neutrinos  $N_i$ (with $i=1,2,3$), one additional scalar doublet $\eta$. An additional $Z_2$ symmetry is imposed under which these newly added particles are odd while all SM particles are even. The leptonic Yukawa Lagrangian relevant for light neutrino mass is
\begin{equation}\label{IRHYukawa}
{\cal L} \ \supset \ \frac{1}{2}(M_N)_{ij} N_iN_j + \left(Y_{ij} \, \bar{L}_i \tilde{\eta} N_j  + \text{h.c.} \right) \ . 
\end{equation}
Clearly, there is no coupling of neutrinos to the SM Higgs doublet $\Phi_1$ due to the unbroken $Z_2$ symmetry. However, light neutrino masses arise at radiative level with $Z_2$ odd particles taking part in the loop. 

The scalar potential of the model can be written as
\begin{align}
V(\Phi_1,\eta) & \ = \   \mu_1^2|\Phi_1|^2 +\mu_2^2|\eta|^2+\frac{\lambda_1}{2}|\Phi_1|^4+\frac{\lambda_2}{2}|\eta|^4+\lambda_3|\Phi_1|^2|\eta|^2 \nonumber \\
& \qquad +\lambda_4|\Phi_1^\dag \eta|^2 + \left[\frac{\lambda_5}{2}(\Phi_1^\dag \eta)^2 + \text{h.c.}\right] \, . \label {c}
\end{align}
where $\Phi_1$ is the SM Higgs doublet. Light neutrino masses which arise at one loop level can be evaluated as ~\cite{Ma:2006km, Merle:2015ica}
\begin{align}
(m_{\nu})_{ij} \ & = \ \sum_k \frac{Y_{ik}Y_{jk} M_{k}}{32 \pi^2} \left ( \frac{m^2_{H^0}}{m^2_{H^0}-M^2_k} \: \text{ln} \frac{m^2_{H^0}}{M^2_k}-\frac{m^2_{A^0}}{m^2_{A^0}-M^2_k}\: \text{ln} \frac{m^2_{A^0}}{M^2_k} \right) \nonumber \\ 
& \ \equiv  \ \sum_k \frac{Y_{ik}Y_{jk} M_{k}}{32 \pi^2} \left[L_k(m^2_{H^0})-L_k(m^2_{A^0})\right] \, ,
\label{numass1}
\end{align}
where 
$M_k$ is the mass eigenvalue of the mass eigenstate $N_k$ in the internal line and the indices $i, j = 1,2,3$ run over the three neutrino generations as well as three copies of $N_i$. Also, $A^0, H^0$ are the neutral pseudoscalar and scalar respectively contained in $\eta$. The function $L_k(m^2)$ is defined as 
\begin{align}
L_k(m^2) \ = \ \frac{m^2}{m^2-M^2_k} \: \text{ln} \frac{m^2}{M^2_k} \, .
\label{eq:Lk}
\end{align}
Using the physical scalar mass expressions \cite{Mahanta:2019sfo}, one can write $m^2_{H^0}-m^2_{A^0}=\lambda_5 v^2$. Thus, light neutrino mass is directly proportional to the parameter $\lambda_5$. In upcoming discussions, we will discuss the effects of $\lambda_5$ in details.

Similar to the vanilla leptogenesis scenario, here also non-zero CP asymmetry is generated by out-of-equilibrium decay of the lightest right handed neutrino $N_1$. Successful leptogenesis is possible at a scale as low as 10 TeV, even with hierarchical right handed neutrinos. Such low scale leptogenesis with hierarchical right handed neutrinos has been discussed by several authors \cite{Hambye:2009pw, Racker:2013lua, Clarke:2015hta, Hugle:2018qbw, Borah:2018rca, Mahanta:2019gfe, Mahanta:2019sfo, Sarma:2020msa} while quasi-degenerate right handed neutrino scenario was discussed in earlier works \cite{Kashiwase:2012xd, Kashiwase:2013uy}. For hierarchical right handed neutrinos, such a low scale leptogenesis is a significant improvement over the usual Davidson-Ibarra bound $M_1 > 10^9$ GeV for vanilla leptogenesis in type I seesaw framework \cite{Davidson:2002qv}. For details of the corresponding Davidson-Ibarra bound in this model, please refer to Appendix \ref{appen:2}.

The CP asymmetry parameter for $N_i \rightarrow l \eta$ is defined as
\begin{equation}
\epsilon_{i} =\frac{\sum_{\alpha}\Gamma(N_{i}\rightarrow l_{\alpha}\eta)-\Gamma(N_{i}\rightarrow\bar{l_{\alpha}}\bar{\eta})}{\sum_{\alpha}\Gamma(N_{i}\rightarrow l_{\alpha}\eta)+\Gamma(N_{i}\rightarrow\bar{l_{\alpha}}\bar{\eta})},
\label{epsilon1}
\end{equation} 
\\
which, for specific lepton flavour $\alpha$ leads to 
\begin{equation}
\epsilon_{i \alpha} = \frac{1}{8 \pi (Y^{\dagger}Y)_{ii}} \sum_{j\neq i} \bigg [ f \left( \frac{M^2_j}{M^2_i}, \frac{m^2_{\eta}}{M^2_i} \right) {\rm Im} [ Y^*_{\alpha i} Y_{\alpha j} (Y^{\dagger} Y)_{ij}] - \frac{M^2_i}{M^2_j-M^2_i} \left( 1-\frac{m^2_{\eta}}{M^2_i} \right)^2 {\rm Im}[Y^*_{\alpha i} Y_{\alpha j} H_{ij}] \bigg ]
\label{epsilonflav}
\end{equation}
where, the function $f(r_{ji},\eta_{i})$ is coming from the interference of the tree-level and one loop diagrams and has the form
\begin{equation}
f(r_{ji},\eta_{i})=\sqrt{r_{ji}}\left[1+\frac{(1-2\eta_{i}+r_{ji})}{(1-\eta_{i}^{2})^{2}}{\rm ln}(\frac{r_{ji}-\eta_{i}^{2}}{1-2\eta_{i}+r_{ji}})\right]
\end{equation}
with $r_{ji}=M_{j}^{2}/M_{i}^{2}$ and $\eta_{i}=m_{\eta}^{2}/M_{i}^{2}$. The self energy contribution $H_{ij}$ is given by 
\begin{equation}
H_{ij} = (Y^{\dagger} Y)_{ij} \frac{M_j}{M_i} + (Y^{\dagger} Y)^*_{ij}
\end{equation}
Now, the CP asymmetry parameter, neglecting the flavour effects (summing over final state flavours $\alpha$) is
\begin{equation}
\epsilon_{i}=\frac{1}{8\pi(Y^{\dagger}Y)_{ii}}\sum_{j\neq i}{\rm Im}[((Y^{\dagger}Y)_{ij})^{2}]\frac{1}{\sqrt{r_{ji}}}F(r_{ji},\eta_{i})
 \label{eq:14}
\end{equation}
\\
where the function $F(r_{ji},\eta)$ is defined as 
\begin{equation}
F(r_{ji},\eta_{i})=\sqrt{r_{ji}}\left[ f(r_{ji},\eta_{i})-\frac{\sqrt{r_{ji}}}{r_{ji}-1}(1-\eta_{i})^{2} \right].
\end{equation}

The Boltzmann equations for comoving number densities of $N_1$ and $B-L$ are given by~\cite{Buchmuller:2004nz}
\begin{eqnarray}
\frac{dn_{N_1}}{dz}& \ = \ &-D_1 (n_{N_1}-n_{N_1}^{\rm eq}) \, , \label{eq:bol1} \\
\frac{dn_{B-L}}{dz}& \ = \ &-\epsilon_1 D_1 (n_{N_1}-n_{N_1}^{\rm eq})-W^{\rm Total}n_{B-L} \, , \label{eq:bol2}
\end{eqnarray}
where $n_{N_1}^{\rm eq}=\frac{z^2}{2}K_2(z)$ is the equilibrium number density of $N_1$ (with $K_i(z)$ being the modified Bessel function of $i$-th kind). The quantity on the right hand side of the above equations
\begin{align}
D_1 \ \equiv \ \frac{\Gamma_{1}}{Hz} \frac{K_1(z)}{K_2(z)} \ = \ K_{N_1} z \frac{K_1(z)}{K_2(z)}
\label{D1}
\end{align}
 measures the total decay rate of $N_1$ with respect to the Hubble expansion rate, and similarly, $W^{\rm Total} \equiv \frac{\Gamma_{W}}{Hz}$ measures the total washout rate. The washout term is the sum of two contributions, i.e. $W^{\rm Total}=W_1+W_{\Delta L}$, where the washout due to the inverse decays $\ell \eta$, $\bar\ell \eta^* \rightarrow N_1$ is given by 
\begin{align}
W_1=W_{\rm ID} \ = \ \frac{1}{4}K_{N_1} z^3 K_1(z).
\end{align}
The other contribution to washout $ W_{\Delta L}$ originates from scatterings which violate lepton number by $\Delta L=1, 2$. In equation \eqref{D1}, $K_{N_1}$ is defined as
\begin{equation}
K_{N_1}=\dfrac{\Gamma_{1}}{H(z=1)}
\end{equation}
where $\Gamma_{1}$ is the $N_{1}$ decay width, $H$ is the Hubble parameter and $z=M_{1}/T$ with $T$ being the temperature of the thermal bath.

\begin{figure}
\includegraphics[scale=.29]{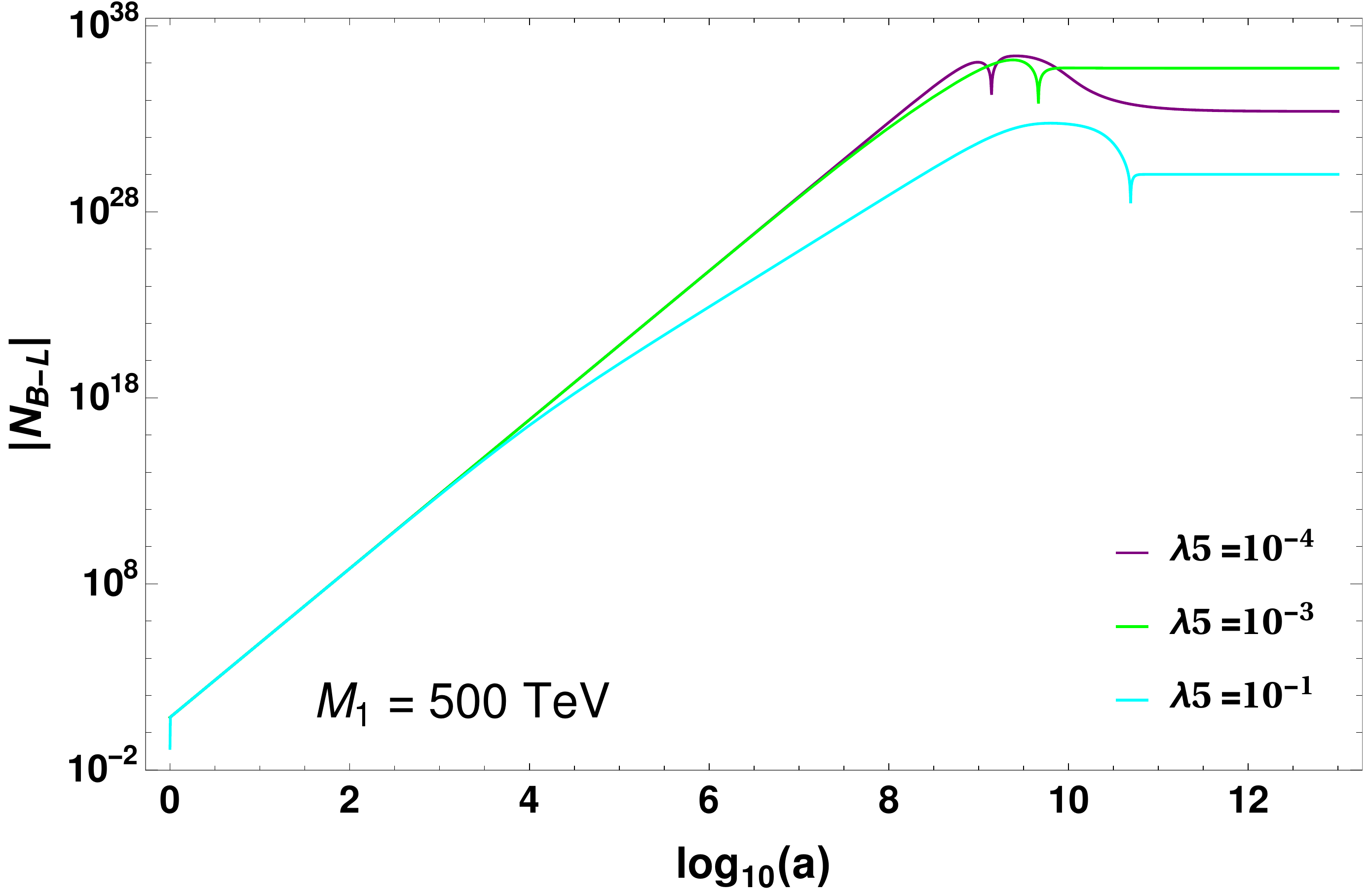}~~
\includegraphics[scale=.3]{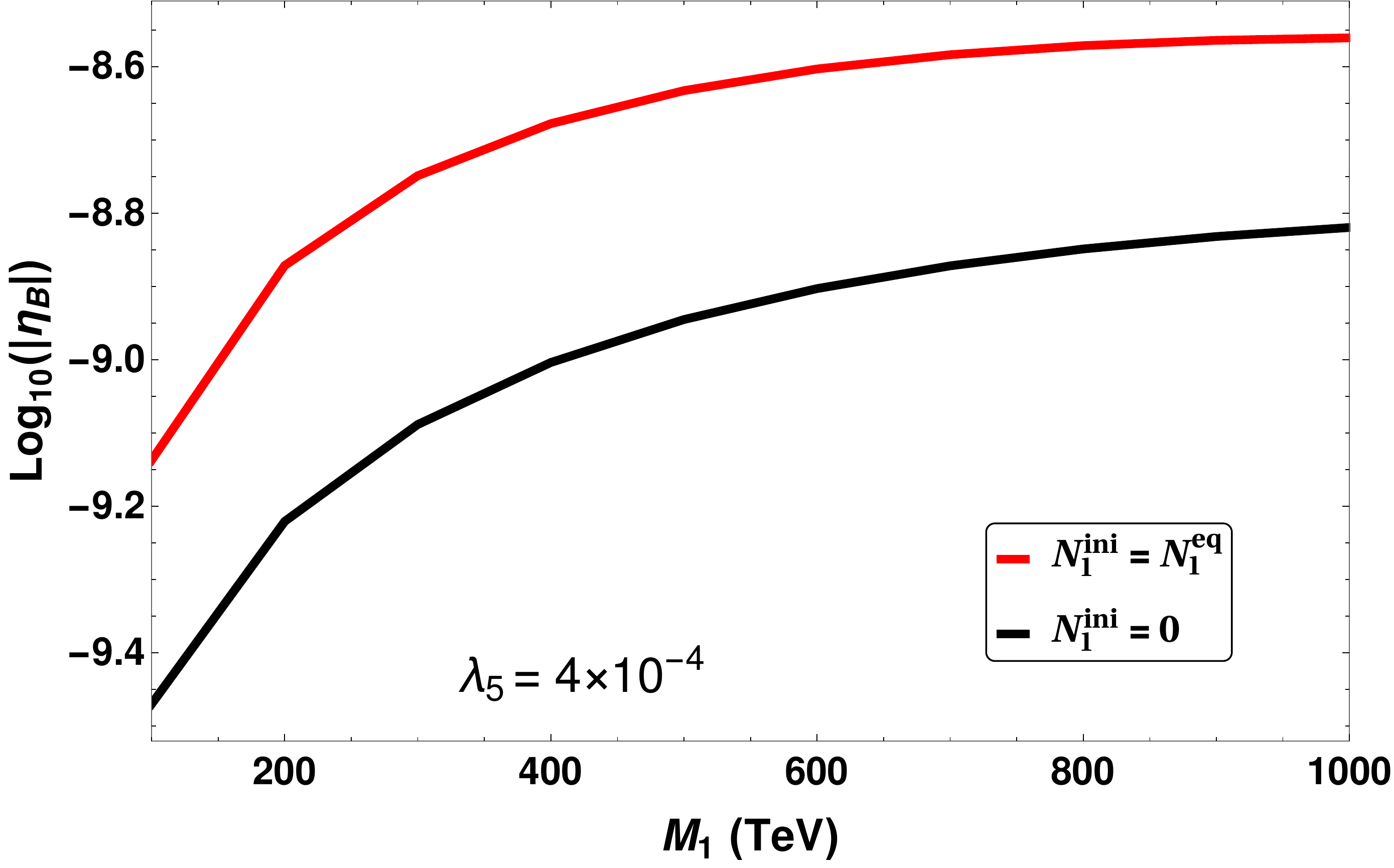}
\caption{Left panel: Evolution plot of the comoving number density of B-L for different $\lambda_{5}$ values in radiation dominated universe, taking $M_1=500$ TeV. The effect of the washout terms is clearly evident for $\lambda_5=10^{-4}$. Right panel: Baryon asymmetry as a function of leptogenesis scale for two different initial $N_1$ abundance and $\lambda_{5}=4\times10^{-4}$. The lightest neutrino mass has been fixed at $10^{-11}$ eV for both the plots. }
\label{fig:th}
\end{figure}

In the above analysis, the Hubble parameter is given by the usual expression in a radiation dominated universe of standard cosmology. While the details of leptogenesis in standard cosmology has already been worked out, we briefly comment on the differences arising due to initial number densities of $N_1$. In earlier works \cite{Hugle:2018qbw, Borah:2018rca, Mahanta:2019gfe, Mahanta:2019sfo}, the right handed neutrino was assumed to be in thermal equilibrium initially. While it is possible to produce TeV scale right handed neutrinos in equilibrium in the presence of additional interactions, it is not guaranteed in the minimal scotogenic model. This is due to the fact that $N_1$ couples to the SM bath only through Dirac Yukawa coupling, which can be small if $N_1$ is kept in the TeV regime. First, we consider $N_1$ to be initially in equilibrium, i.e., $n_{N_{1}^{\rm ini}}=n_{N_{1}^{\rm eq}}$ and next we consider the case of vanishing initial abundance of $N_1$ that is,  $n_{N_{1}^{\rm ini}} \approx 0$. As we will see, the latter case will be more realistic for us because of the small Yukawas and we continue with this assumption in subsequent analysis. Even if Yukawas are sizeable, at the epoch of PBH formation, the abundance of $N_1$ can be negligible as it is yet to enter thermal equilibrium with the bath particles. See Appendix \ref{appen:1} for justification behind this assumption of negligible initial abundance of $N_1$. While the initial abundance is vanishingly small, $N_1$ gets produced at later epochs from the inverse decays and scatterings. The same terms, later act as washout terms which tend to erase the asymmetry during the scale of leptogenesis. This can be seen in figure \ref{fig:th} (left), where we have shown the evolution of comoving number density of  $B-L$ with scale factor from a high temperature upto the sphaleron scale. In the right panel of figure \ref{fig:th}, we show the final baryon asymmetry as a function of leptogenesis scale, for both the cases mentioned above. It can be clearly seen that a vanishing initial abundance of $N_1$ decreases the asymmetry. We have considered a mass hierarchy: $M_2=10~M_1$ and $M_3=100~M_1$, which is also used in our subsequent analysis.

\section{Scotogenic leptogenesis in the presence of PBH}\label{sec:lepto}

The high density of the early universe provides the most suitable environment for the formation of PBH, with a wide range of masses. One of the most natural mechanisms  of PBH formation is from inhomogeneities large enough such that they collapse after re-entering the horizon. The origin of these inhomogeneities can be related to the quantum fluctuations during inflation. Since the large scale physics is tightly constrained from the CMB observations related to the amplitude of power spectra, an enhancement is needed at the smaller scales. These can be generated, for example, through some inflation models with running spectral index \cite{Drees:2011hb} or some inflection point inflation models which gives rise to an ultra slow-roll phase during inflation \cite{Bhaumik:2019tvl}. Some of the other possible mechanisms of forming PBH include collapse of loops formed by cosmic strings \cite{Hawking:1987bn} when they intersect with themselves and collisions of bubbles formed after a symmetry is broken \cite{Hawking:1982ga}. A comprehensive review of such mechanisms of PBH formation can be found in \cite{Carr:2020gox}. For our purpose, we consider PBH with a monochromatic mass function, i.e., all PBHs have the same mass. Although this is an ideal case, we do not expect significant changes to the qualitative behavior and conclusions reached in this work.

We consider PBH to be formed in the very early universe during the radiation dominated era, when the thermal plasma has a temperature, say $T_{\rm in}$. The initial mass of PBH is related to the mass enclosed in the particle horizon and is given by \cite{Bernal:2020bjf} 

\begin{equation}\label{eqn:Mini}
M_{\rm BH}(T_{\rm in})=\frac{4\pi}{3}\gamma\frac{\rho_{\rm Rad}(T_{\rm in})}{H^3(T_{\rm in}).}, 
\end{equation}

where $\gamma\approx0.2$ and $\rho_{\rm Rad}(T_{\rm in})$ is the initial radiation density. A lower bound on the initial PBH mass $M_{\rm in}$ can be immediately inferred from the upper bound on the inflationary scale \cite{Akrami:2018odb} 
\begin{equation}\label{eqn:infbnd}
H_I\leq 2.5 \times 10^{-5} M_{\rm Pl} \implies M_{\rm in}\gtrsim 0.1~g	
\end{equation}
with $M_{\rm Pl}$ being the Planck mass. On the other hand, the temperature of a black hole (BH) can be related to its mass as \cite{Hawking:1974sw}
\begin{equation}
T_{\rm BH}=\frac{1}{8\pi GM_{\rm BH}}\approx 1.06~\left(\frac{10^{13}\; {\rm g}}{M_{\rm BH}}\right)~{\rm GeV}.
\end{equation}
PBH may dominate the energy density of the universe depending on their initial abundance, characterised by the dimensionless parameter
\begin{equation}
\beta=\frac{\rho_{\rm BH}(T_{\rm in})}{\rho_{\rm Rad}(T_{\rm in})}.
\label{eqn:beta}
\end{equation}
Once the PBH form, they lose mass through Hawking evaporation at a rate given by \cite{MacGibbon:1991tj} 
\begin{equation}
\frac{dM_{\rm BH}}{da}  = -\dfrac{\epsilon(M_{\rm BH}) \kappa}{a H} \left( \frac{1\; {\rm g}}{M_{\rm BH}} \right)^{2}. \label{eq:MBH} \end{equation}  
Here $a$ is the scale factor and $\kappa=5.34\times10^{25} \; {\rm g \; s}^{-1}$ \cite{Lunardini:2019zob}. Note that all particles are emitted democratically from BH as evaporation is purely due to gravitational effects, and the evaporation function $\epsilon(M_{\rm BH})$\cite{Lunardini:2019zob} contains contribution from both SM and BSM particles. Assuming a radiation dominated universe, the temperature of the thermal plasma when the PBH have completely disappeared can be written as \cite{Bernal:2020bjf}
\begin{equation}
T_{\rm ev}=\left(\frac{9g_{*}(T_{\rm BH})}{10240}\right)^{\frac{1}{4}}\left(\frac{M_{\rm Pl}^{5}}{M_{\rm in}^{3}}\right)^{\frac{1}{2}}.\label{eq:Tev}
\end{equation}
Now, a PBH matter-like domination can be avoided if at all epochs, $\rho_{\rm BH} \ll \rho_{\rm Rad}$, which can be quantified in terms of $\beta$ as
\begin{equation}
\beta\ll\beta_c\simeq\frac{T_{\rm ev}}{T_{\rm in}}.\label{eqn:betac}
\end{equation}
As we will see, a very small $\beta$ or sub-dominant PBH energy density in the early universe does not change the predictions of thermal leptogenesis in scotogenic model significantly. Therefore, we will consider sizeable values of $\beta$ so that PBH can affect the results of thermal leptogenesis. On the other hand, such large values of $\beta$ often lead to over-production of dark matter as we will discuss in upcoming section. This actually forces us to have DM mass in a ballpark where PBH has no effect on its relic abundance. On the other hand, if PBH effect on DM relic is substantial, the corresponding values of $\beta$ will not have any significant effect on thermal leptogenesis. This complementary behaviour noticed in leptogenesis and DM with respect to fractional energy density of PBH is the main content of upcoming discussions.

For the discussion of leptogenesis in the presence of PBH, we closely follow the notations of \cite{Perez-Gonzalez:2020vnz}. Since we consider thermal leptogenesis in the presence of PBH, we solve the required Boltzmann equations for the PBH energy density, radiation energy density, right handed neutrinos and lepton asymmetry, which in terms of the scale factor $a$, can be written as
\begin{eqnarray}
\frac{d{\varrho_{\rm BH}}}{da}& \ = \ &\dfrac{1}{M_{\rm BH}}\dfrac{dM_{\rm BH}}{da}\varrho_{\rm BH} \, , \label{eq:bol1} \\
\frac{d{\varrho_{\rm Rad}}}{da}& \ = \ &-\dfrac{\epsilon_{\rm SM}(M_{\rm BH})}{\epsilon(M_{\rm BH})}\dfrac{1}{M_{\rm BH}} \dfrac{dM_{\rm BH}}{da}a\varrho_{\rm BH}  \, , \label{eq:bol2}\\
aH\frac{dn_{N_{1}}^T}{da} & \ = \ & - \left( n_{N_{1}}^T-n_{N_{1}}^{\rm eq} \right) \Gamma_{1}^{T}   \, ,\label{eq:bol5} \\
aH\frac{dn_{N_{1}}^{\rm BH}}{da} & \ = \ & - \left( n_{N_{1}}^{\rm BH} \right) \Gamma_{1}^{\rm BH} + n_{\rm BH}\Gamma_{{\rm BH}\rightarrow{N_1}}   \, ,\label{eq:bol6} \\
aH \frac{dn_{B-L}}{da} & \ = \ & \epsilon_{1} \left[ \left( n_{N_{1}}^T-n_{N_{1}}^{\rm eq}  \right)\Gamma_{1}^{T}+n_{N_{1}}^{\rm BH} \Gamma_{1}^{\rm BH} \right]-W n_{B-L} - \Delta W \frac{M_1}{T} H n_{B-L}. \label{eq:bol7}
\end{eqnarray}
Here, $\varrho_{\rm BH}=a^{3}\rho_{\rm BH}$ and $\varrho_{\rm Rad}=a^{4}\rho_{\rm Rad}$ are the comoving energy densities of PBH and radiation respectively.  In presence of the PBH, the Hubble parameter $H$ entering in the Boltzmann equations is given by\footnote{The contribution of $N_1$ in Hubble is found to be negligible.} $H=\sqrt{\dfrac{\varrho_{\rm BH}a^{-3}+\varrho_{\rm Rad}a^{-4}}{3M_{\rm Pl}^{2}}}$. In the above Boltzmann equations, $n_{N_{1}}^T$ and $n_{N_{1}}^{\rm BH}$ are $N_1$ densities generated from thermal bath and PBH (non-thermal source) respectively. $\Gamma_{{\rm BH}\rightarrow{N_1}}$ is the non-thermal production term for $N_{1}$ (originating from PBH evaporation) and can be written as \cite{Perez-Gonzalez:2020vnz} 
\begin{equation}
\Gamma_{{\rm BH}\rightarrow{N_1}}=\dfrac{27T_{\rm BH}}{32\pi^{2}}\left(-z_{\rm BH} {\rm Li}_{2}(-e^{-z_{\rm BH}})-{\rm Li}_{3}(-e^{-z_{\rm BH}}) \right),
\end{equation} 
where ${\rm Li}_{s}(z)$ are the poly-logarithm functions of order $s$ and $z_{\rm BH}=M_{1}/T_{\rm BH}$. The thermal average decay widths are defined as $\Gamma_{1}^{T}=\dfrac{K_{1}\left(M_{1}/T\right)}{K_{2}\left(M_{1}/T\right)} \Gamma_{1}$  and  $\Gamma_{1}^{\rm BH}=\dfrac{K_{1}\left(M_{1}/T_{\rm BH}\right)}{K_{2}\left(M_{1}/T_{BH}\right)} \Gamma_{1}$. 

\begin{figure}
\includegraphics[scale=0.29]{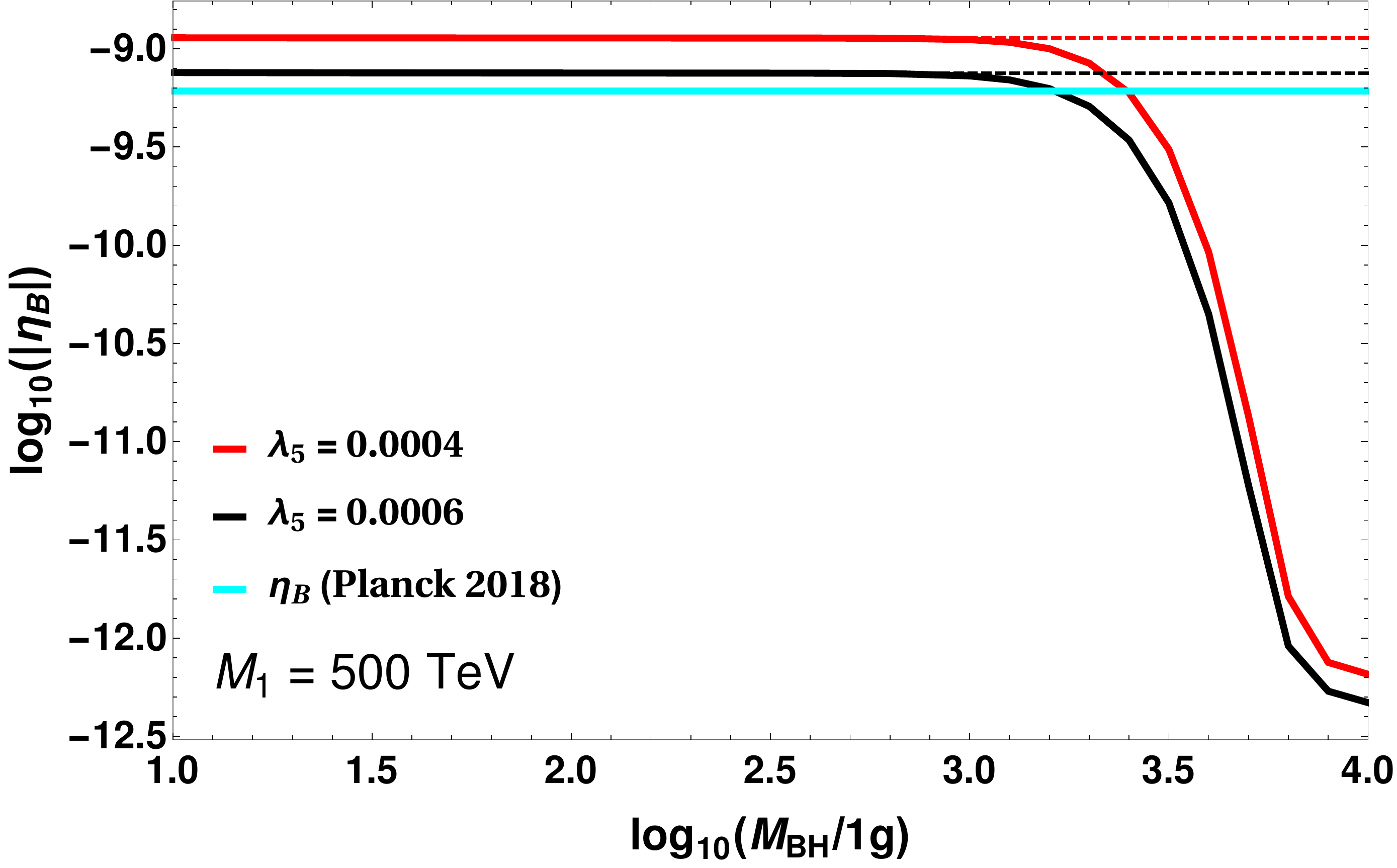}~~
\includegraphics[scale=.29]{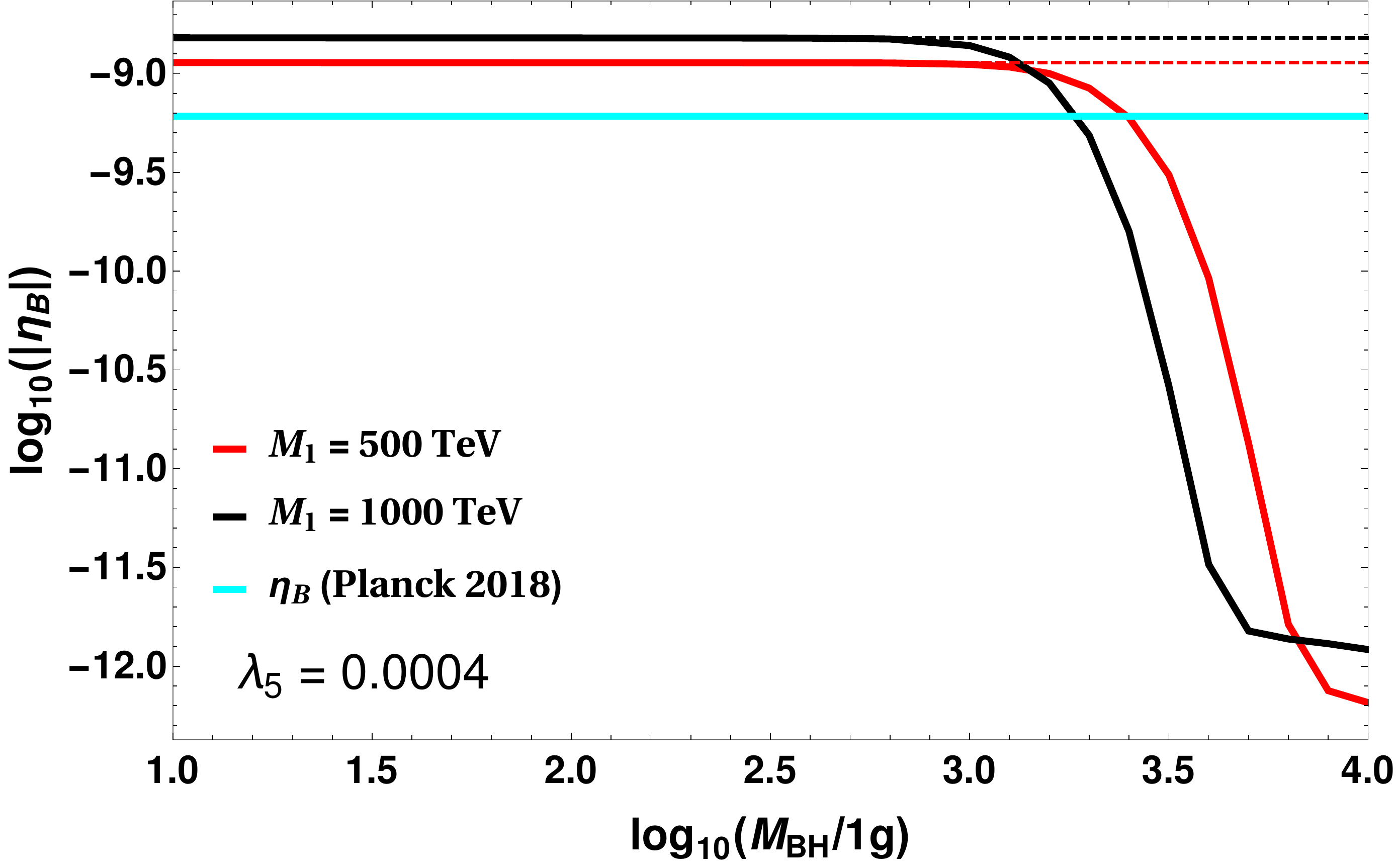}
\caption{Final baryon asymmetry as a function of the PBH mass, for two values of $\lambda_5$ keeping $M_1$ fixed at $500$ TeV (left panel), and for two values of $M_1$ keeping $\lambda_5$ fixed at $0.0004$ (right panel). Lightest neutrino mass is fixed at $m_\nu^1 = 10^{-11}$ eV for both. }
\label{fig:meta}
\end{figure}

In the asymmetry equation \eqref{eq:bol7}, $W=\dfrac{1}{4}\Gamma_{N_{1}}^{T}K_{2}(z)z^{2}$ is the washout term from the inverse decays. $\Delta W$ is the washout term because of $\Delta L =2$ scatterings of the type $l \eta \rightarrow \bar{l} \eta^*$ and can be written as \cite{Hugle:2018qbw}  
 \begin{equation}
 \Delta W=\frac{36\sqrt{5}M_{P}M_{1}\bar{m_{\xi}^{2}}}{2\sqrt{\pi}\sqrt{g_{*}}v^{4}z^{2}\lambda_{5}^{2}},\label{eqn:washout}
 \end{equation}
where $z=M_1/T$, v is the electroweak symmetry breaking scale and $\bar{m_{\xi}^{2}}$ is the effective mass parameter \cite{Hugle:2018qbw}. The $\Delta L =1$ scatterings are usually suppressed due to the presence of heavy right handed neutrinos in external legs.

\begin{figure}
\includegraphics[scale=0.29]{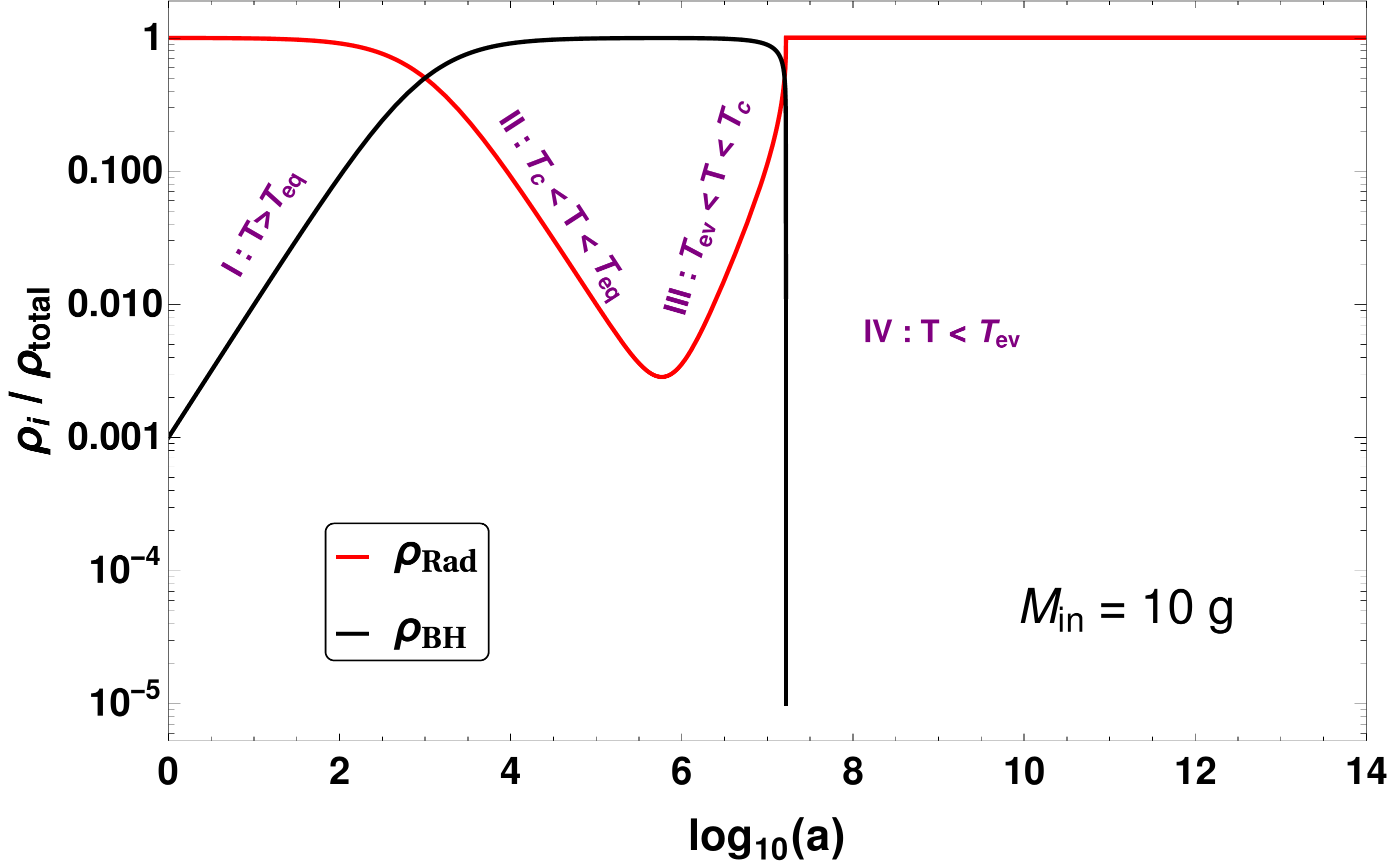}~~
\includegraphics[scale=0.29]{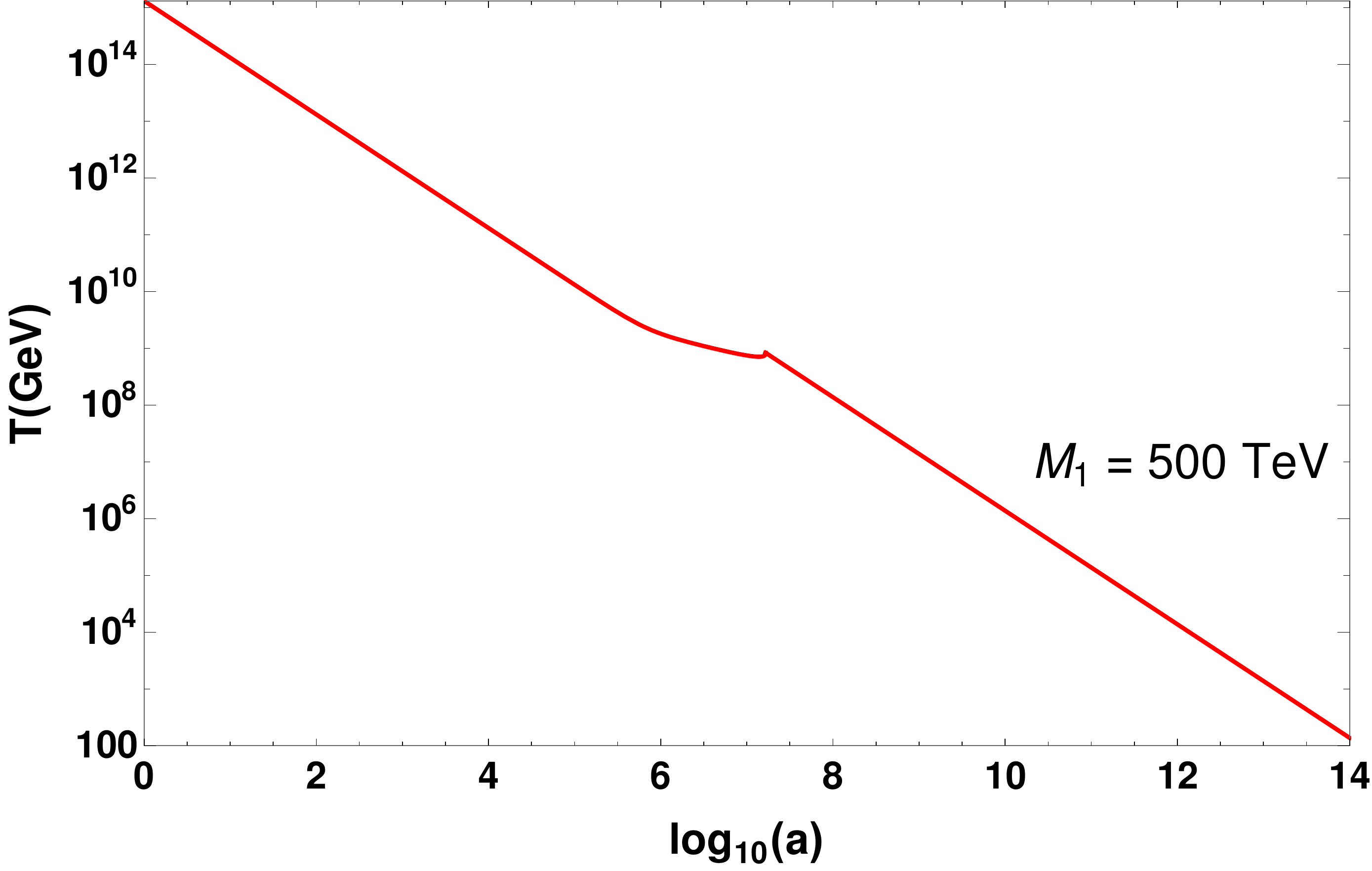}\\
\includegraphics[scale=0.29]{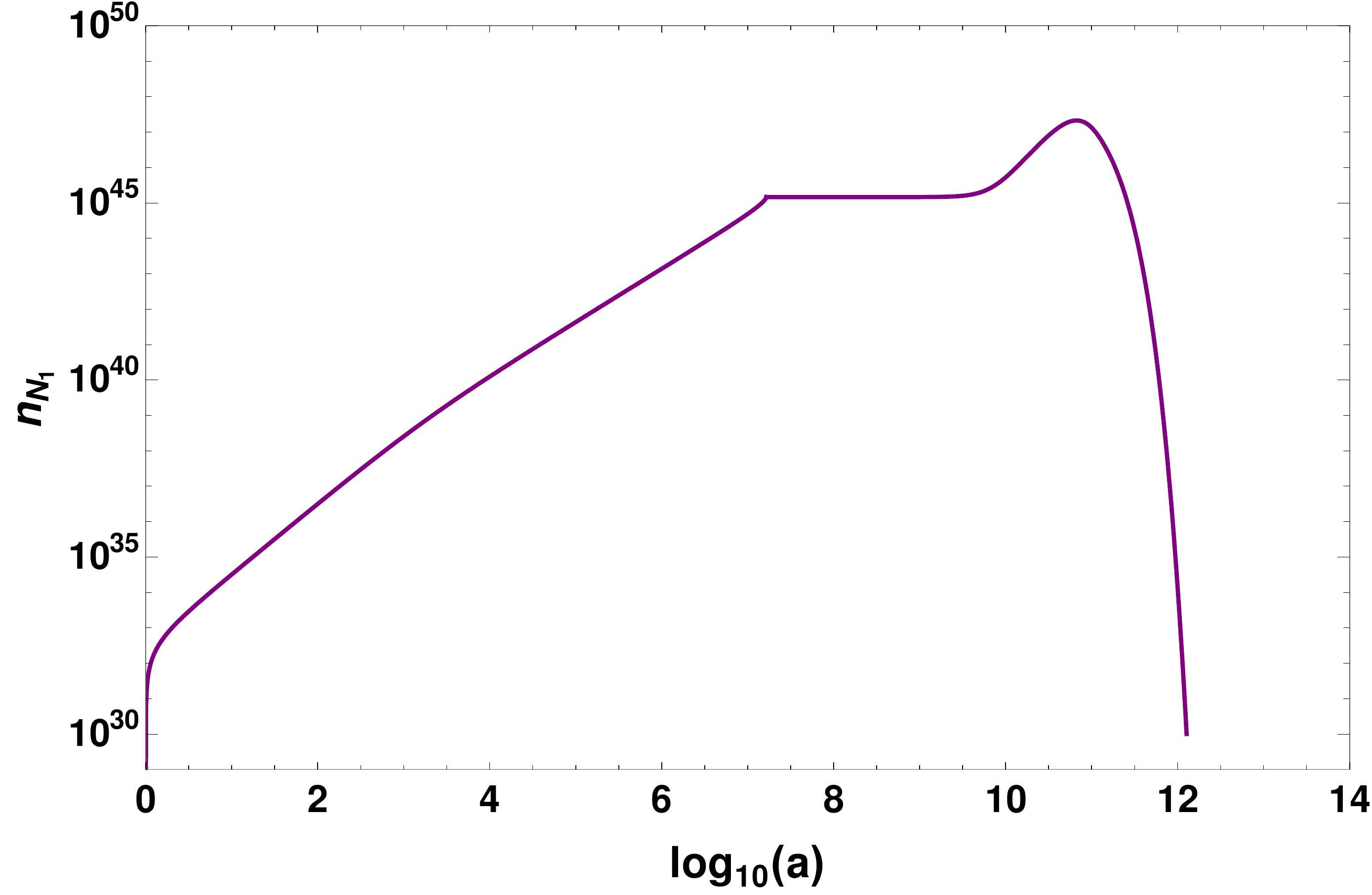}~~
\includegraphics[scale=0.29]{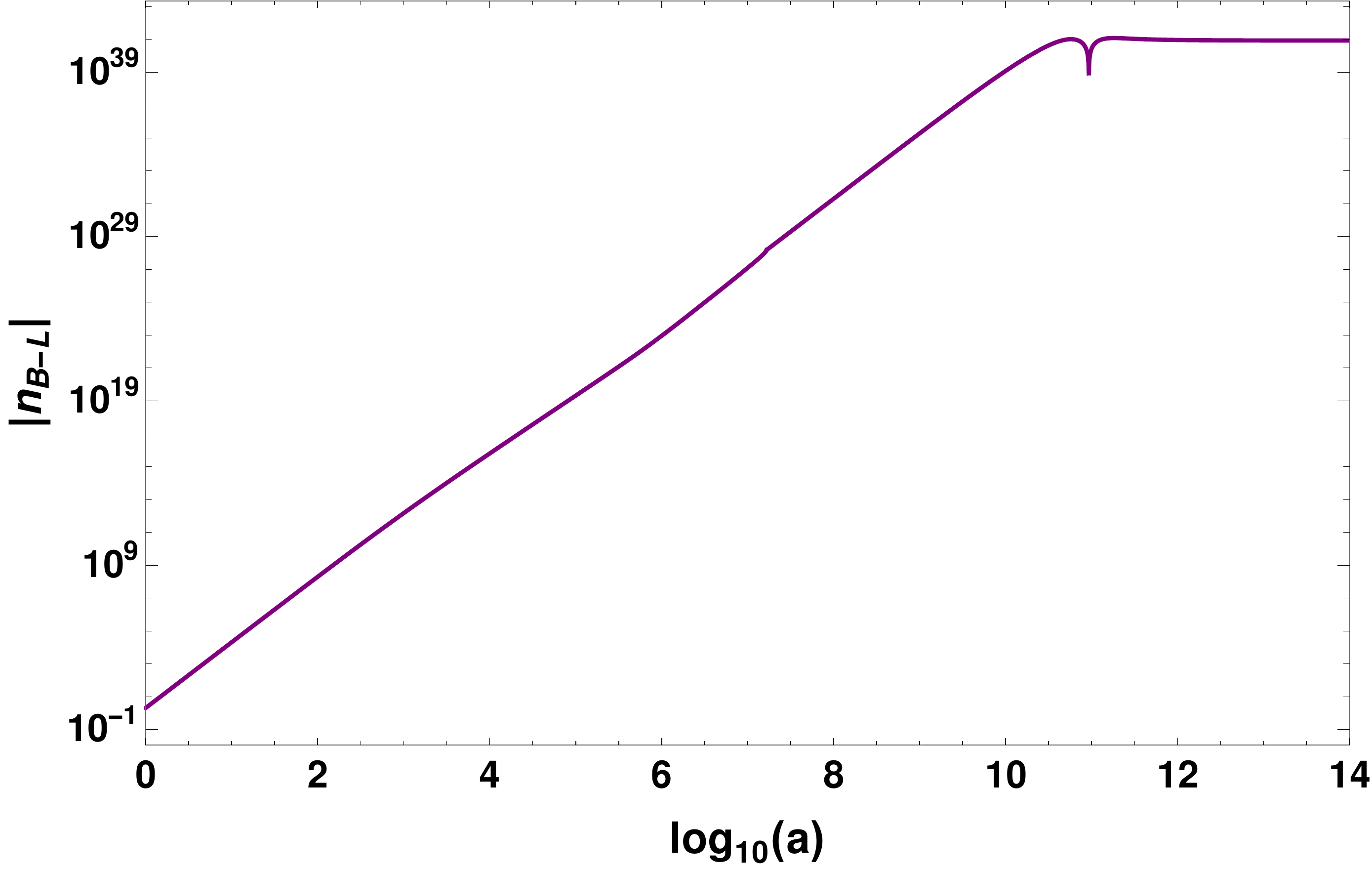}
\caption{Top panel: Evolution of the energy densities (left) and temperature of the thermal plasma (right), taking $M_{\rm in}=10$ g, $M_1=500$ TeV, $\lambda_5=4\times10^{-4}$, $m_\nu^1 = 10^{-11}$ eV. Bottom panel : Evolution of the comoving number densities of $N_1$ (left) and $B-L$ (right) for the same parameters. The different temperature regions shown in the top left panel are explained in Section \ref{sec:DM}. }
\label{fig:b4}
\end{figure}

Now, since the entropy of the universe is not conserved because of PBH evaporation, we need to follow the evolution of the temperature of the thermal plasma separately through the equation
\begin{equation}
\frac{dT}{da}  = -T\left(\frac{1}{a}+\frac{\epsilon_{SM}(M_{\rm BH})}{\epsilon(M_{\rm BH})}\frac{1}{M_{BH}}\frac{dM_{BH}}{da}\frac{a\varrho_{\rm BH}}{4\varrho_{\rm Rad}}\right). \label{eq:bol4}
\end{equation}
We solve the above coupled Boltzmann equations from the time of PBH formation upto the sphaleron scale, with a vanishing initial abundance of $N_1$. The initial PBH mass $M_{\rm in}$ and the initial temperature are related through equation \eqref{eqn:Mini}. Depending on the initial mass of PBH it can evaporate at different epochs. PBH with a  larger mass evaporate lately (equation \eqref{eq:Tev}). Now, PBH should fully evaporate before the BBN epoch, $T_{\rm BBN}\simeq 4~MeV$, such that they do not alter the successful BBN predictions. This translates into an upper bound on the initial PBH mass $M_{\rm in}$ as
\begin{equation}
T_{\rm ev}>T_{\rm BBN}\implies M_{\rm in}\lesssim 2\times 10^8 \; {\rm g}.\label{eqn:BBNbound}
\end{equation}
Also, from the leptogenesis point of view, we want PBH to evaporate before the sphaleron scale, $T_{\rm sph}\simeq 100~GeV$, which gives a tighter upper bound on the initial PBH mass,
\begin{equation}
M_{\rm in}\lesssim 2\times 10^5 \; {\rm g}.\label{eqn:sphbound}
\end{equation}
Note that PBH evaporation after sphaleron temperature will lead to overall dilution of baryon asymmetry only as the non-thermally produced $N_1$ from late PBH evaporation cannot contribute to baryon asymmetry.

\begin{figure}
\includegraphics[scale=0.29]{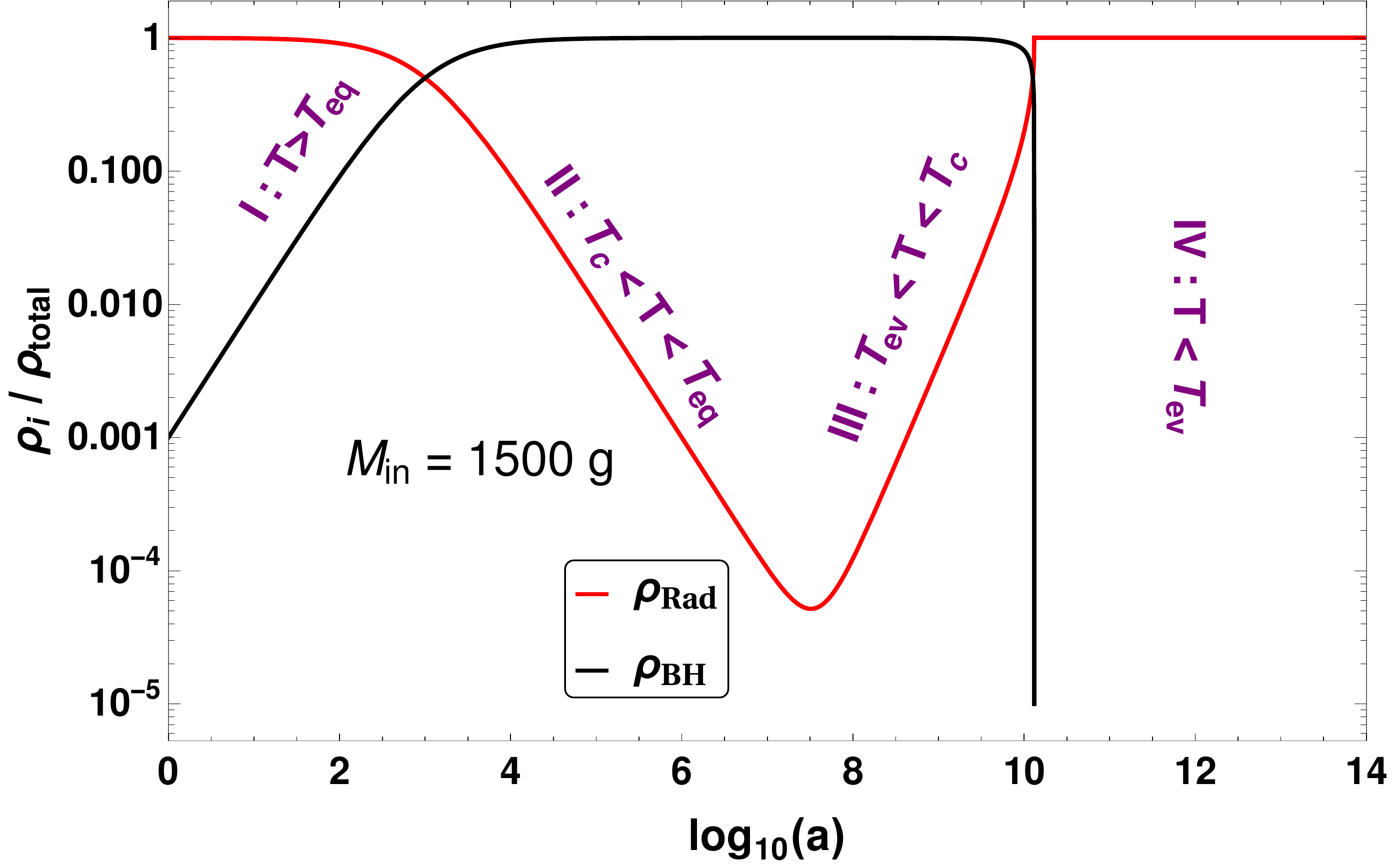}~~
\includegraphics[scale=.29]{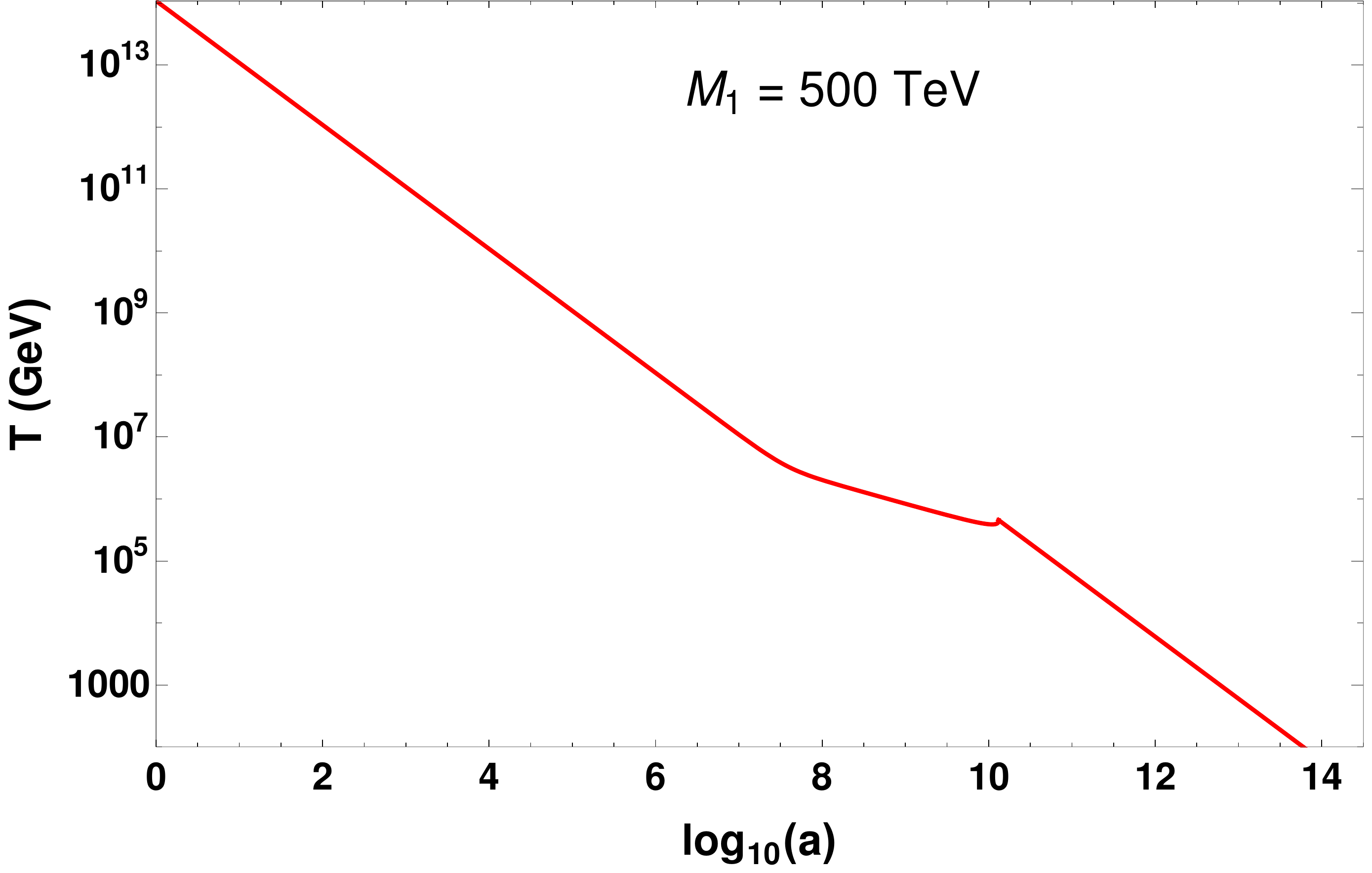}\\
\includegraphics[scale=.29]{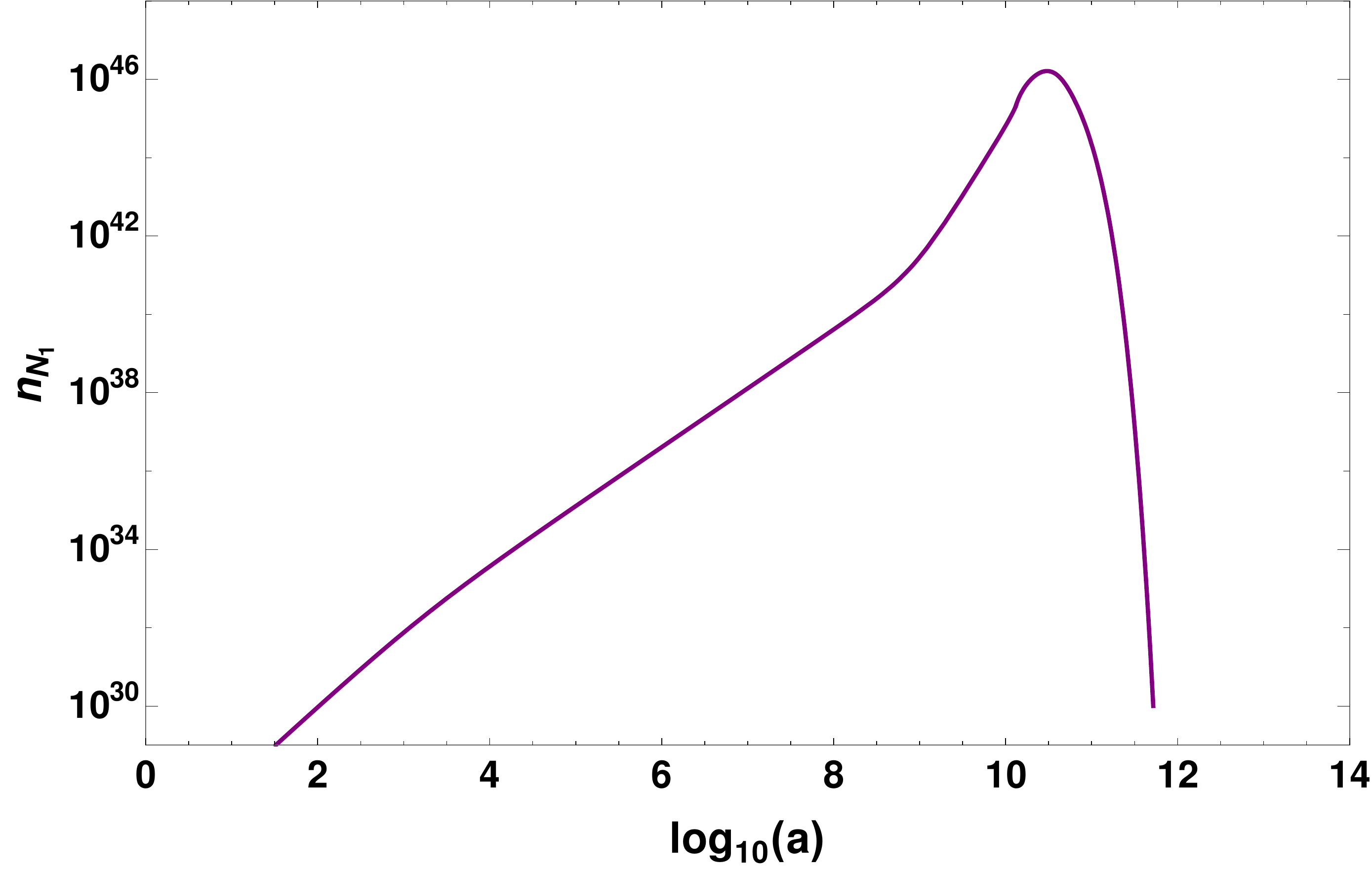}~~
\includegraphics[scale=.29]{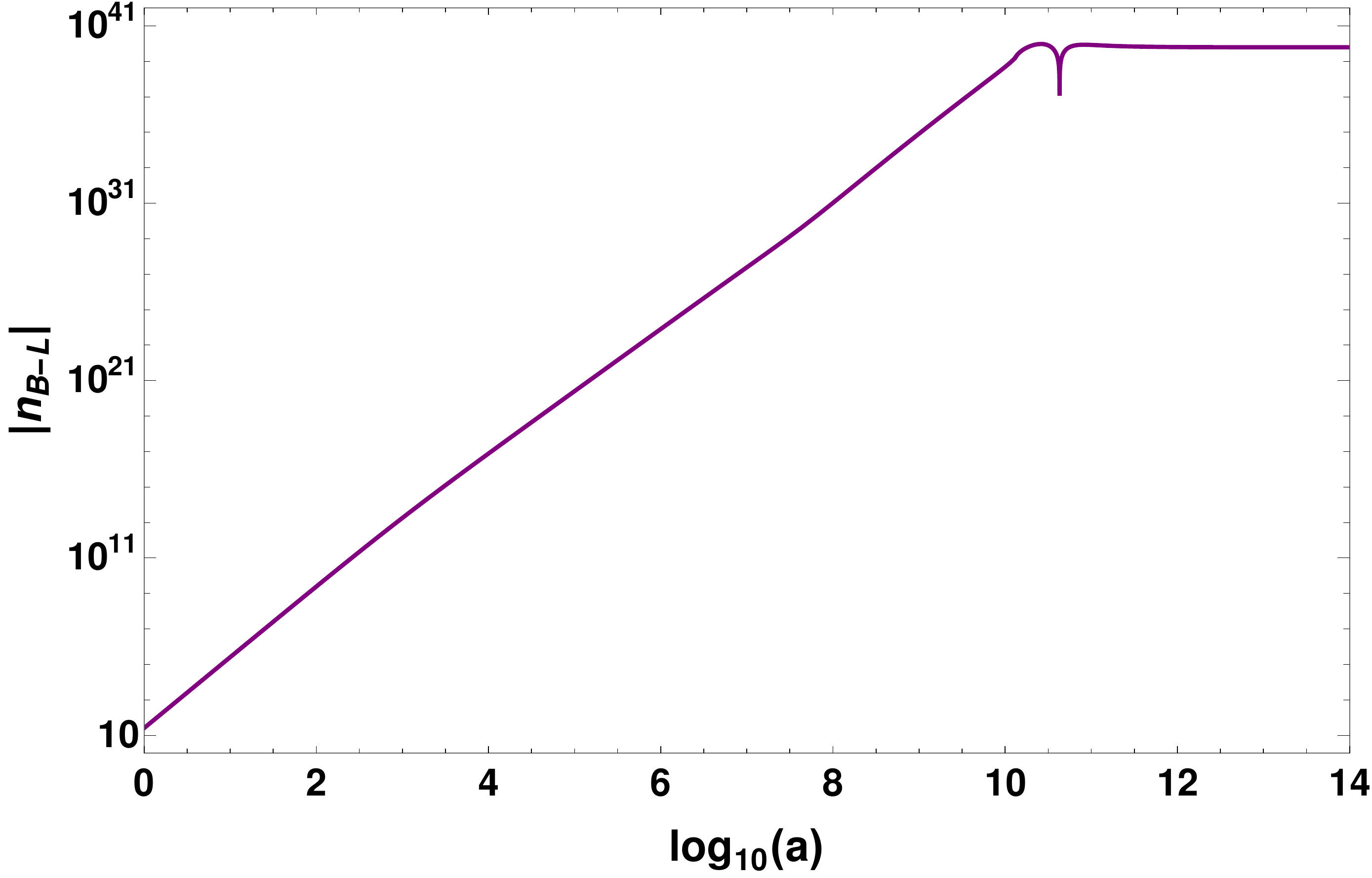}
\caption{Top panel: Evolution of the energy densities (left) and temperature of the thermal plasma (right), taking $M_{\rm in}=1500$ g, $M_1=500$ TeV, $\lambda_5=4\times10^{-4}$, $m_\nu^1 = 10^{-11}$ eV. Bottom panel: Evolution of the comoving number densities of $N_1$ (left) and $B-L$ (right) for the same parameters.}
\label{fig:dur}
\end{figure}

Now, depending on the PBH evaporation time relative to the scale of leptogenesis, we can have three different scenarios, also applicable in type I seesaw leptogenesis discussed in \cite{Perez-Gonzalez:2020vnz}. To be more specific, we can have scenarios where PBH evaporation takes place before, during and after the generation of lepton asymmetry. If we consider low scale leptogenesis at a scale of few hundreds of TeV, which is naturally possible in the scotogenic model, for most of the range of allowed initial PBH masses, they evaporate by the time thermal leptogenesis occurs. For instance, for leptogenesis scale around $300-800$ TeV, PBH with  mass less than around $600-1200$ g, evaporate before the scale of thermal leptogenesis. In this case, the right handed neutrinos produced by PBH evaporation only act as an initial condition for thermal leptogenesis. Thus, the asymmetry produced is almost the same as the thermal case. As PBH masses start increasing, they evaporate during or after the scale of leptogenesis. The evaporation injects entropy in the form of radiation and decreases the final baryon asymmetry. Although such evaporation can also generate additional sources of lepton asymmetry in terms of $N_1$, but entropy dilution effects dominate. This behaviour can be seen in figure \ref{fig:meta},  where we have shown the observed baryon asymmetry as a function of initial PBH mass, varying the relevant parameters in the scotogenic model. In the left panel plot, we keep the scale of leptogenesis fixed at 500 TeV, while  showing the variation in baryon asymmetry for two different values of $\lambda_5$. In the right panel plot, we keep $\lambda_5$ fixed at $0.0004$ and show the variation in baryon asymmetry for two different scales of leptogenesis. In figure \ref{fig:meta}, $\beta$ is kept fixed at $10^{-3}$. The effect of changing $\beta$ will be discussed later.

\begin{figure}
\includegraphics[scale=0.29]{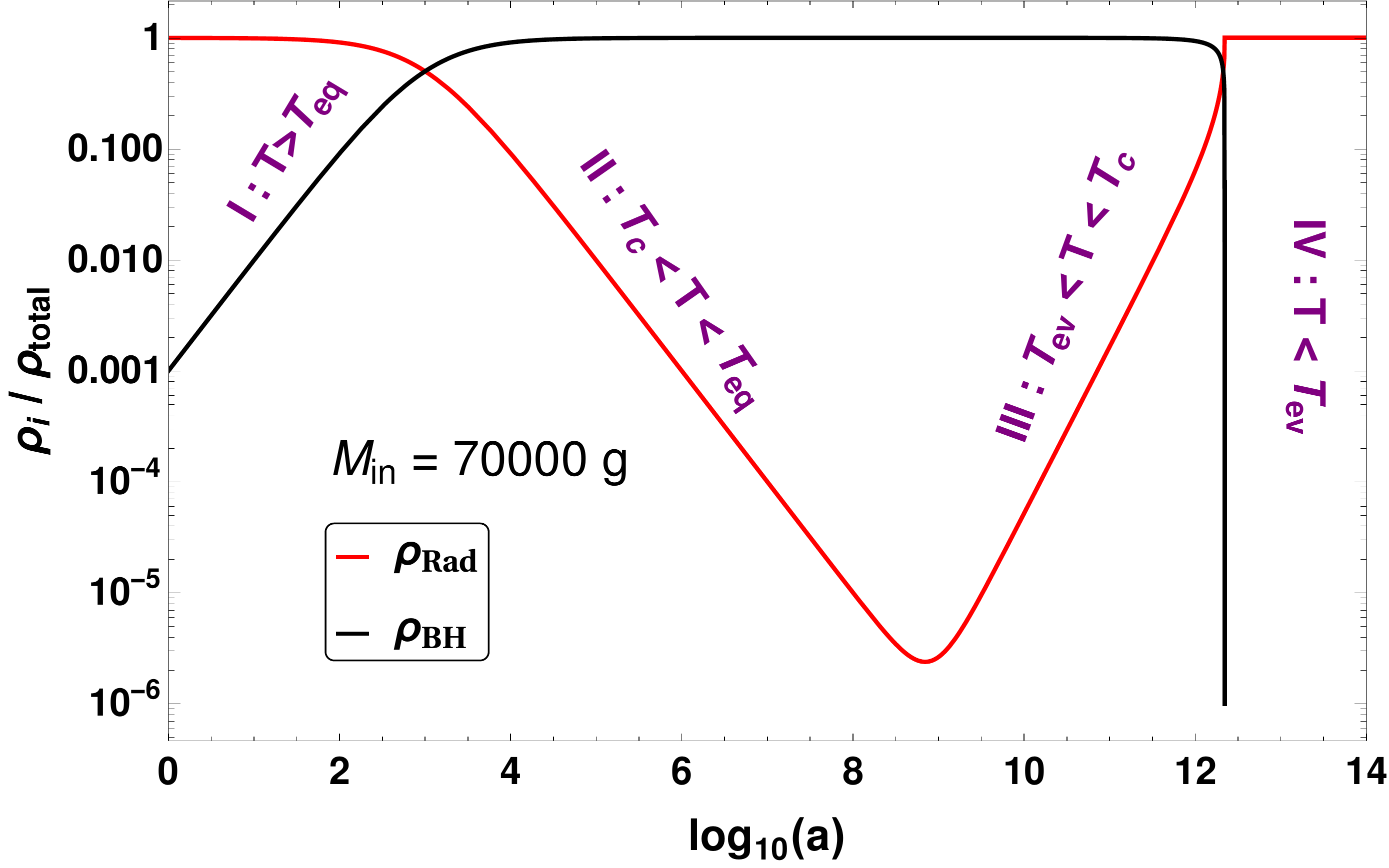}~~
\includegraphics[scale=.29]{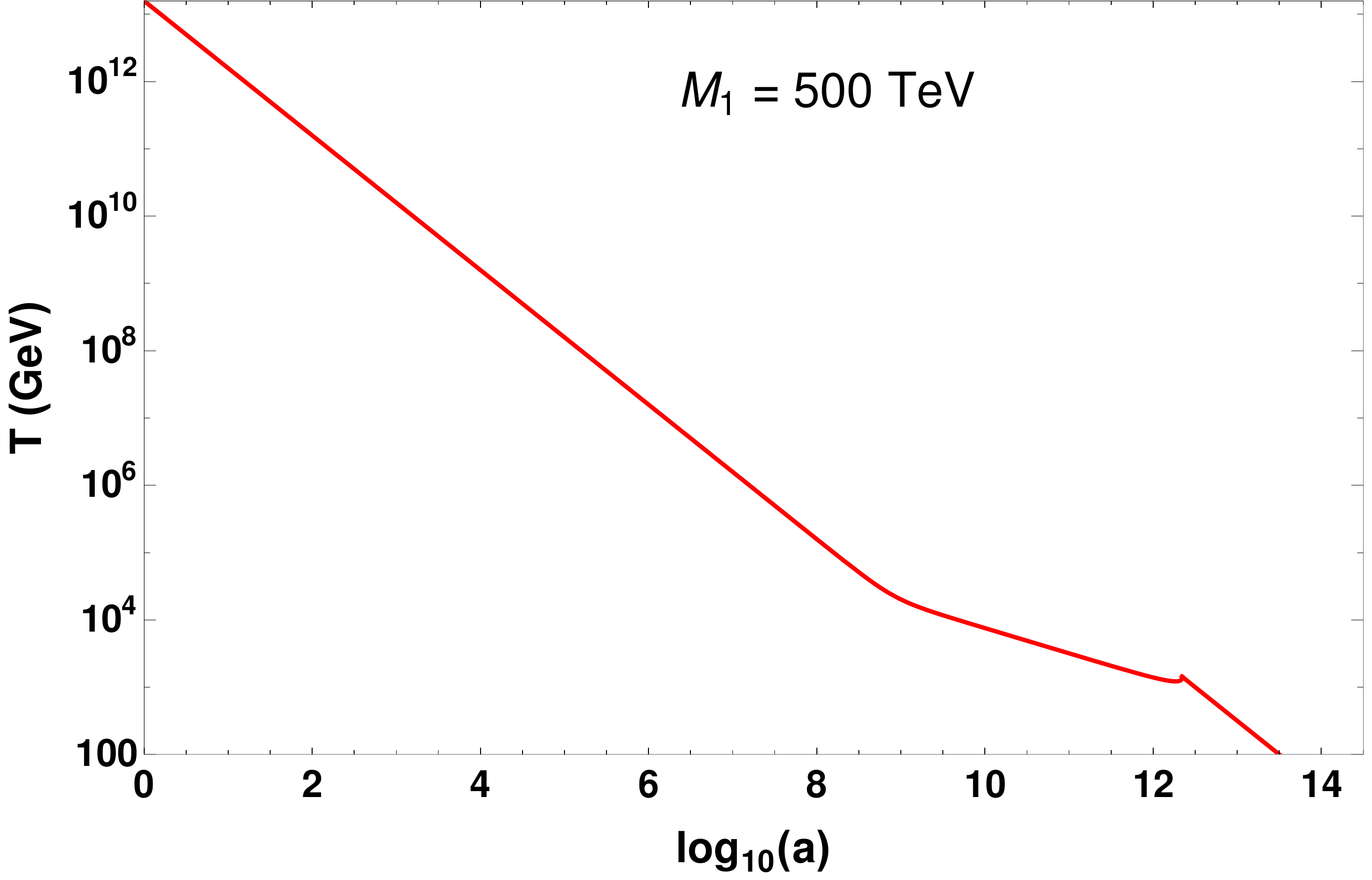}\\
\includegraphics[scale=.29]{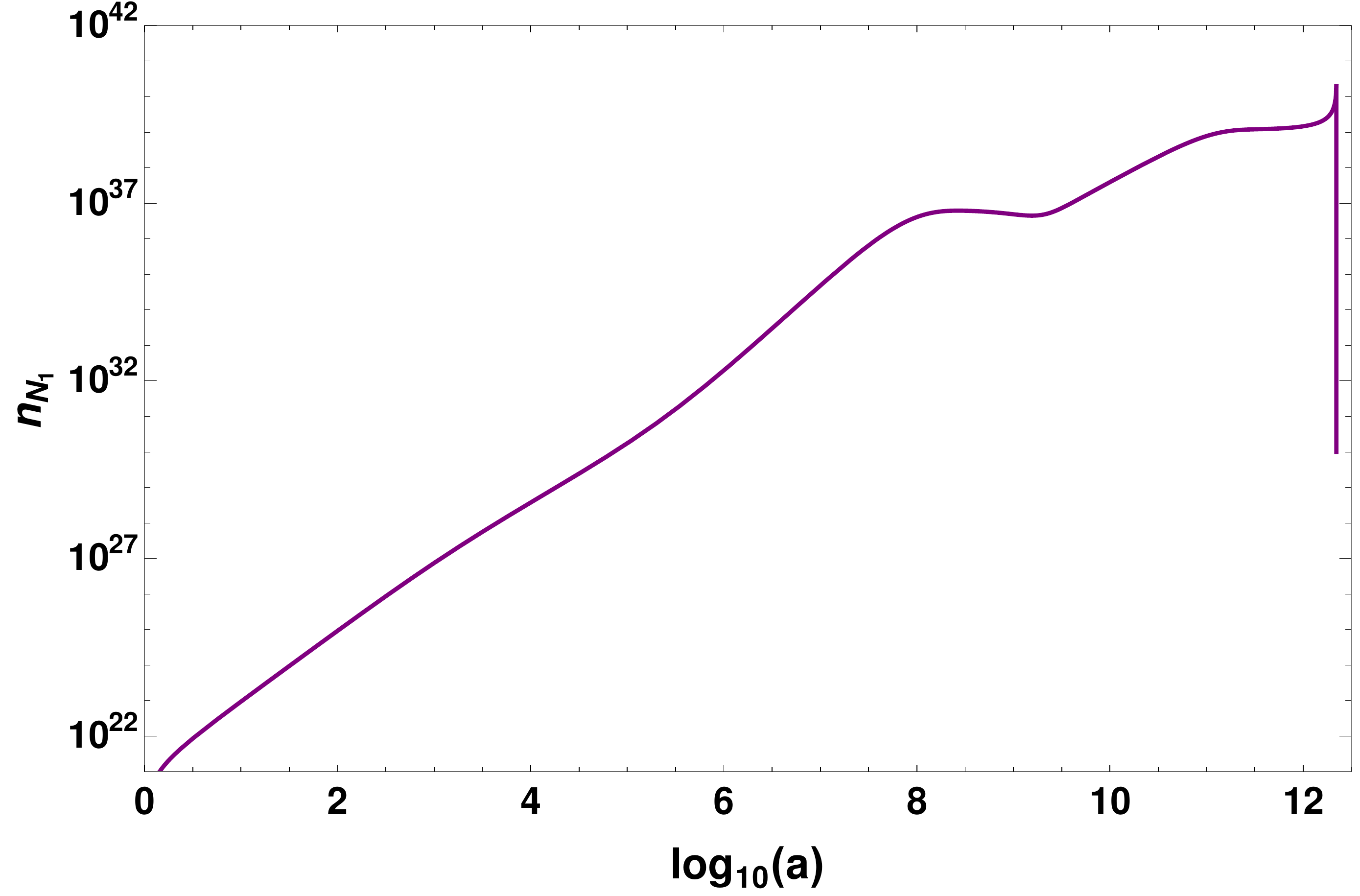}~~
\includegraphics[scale=.29]{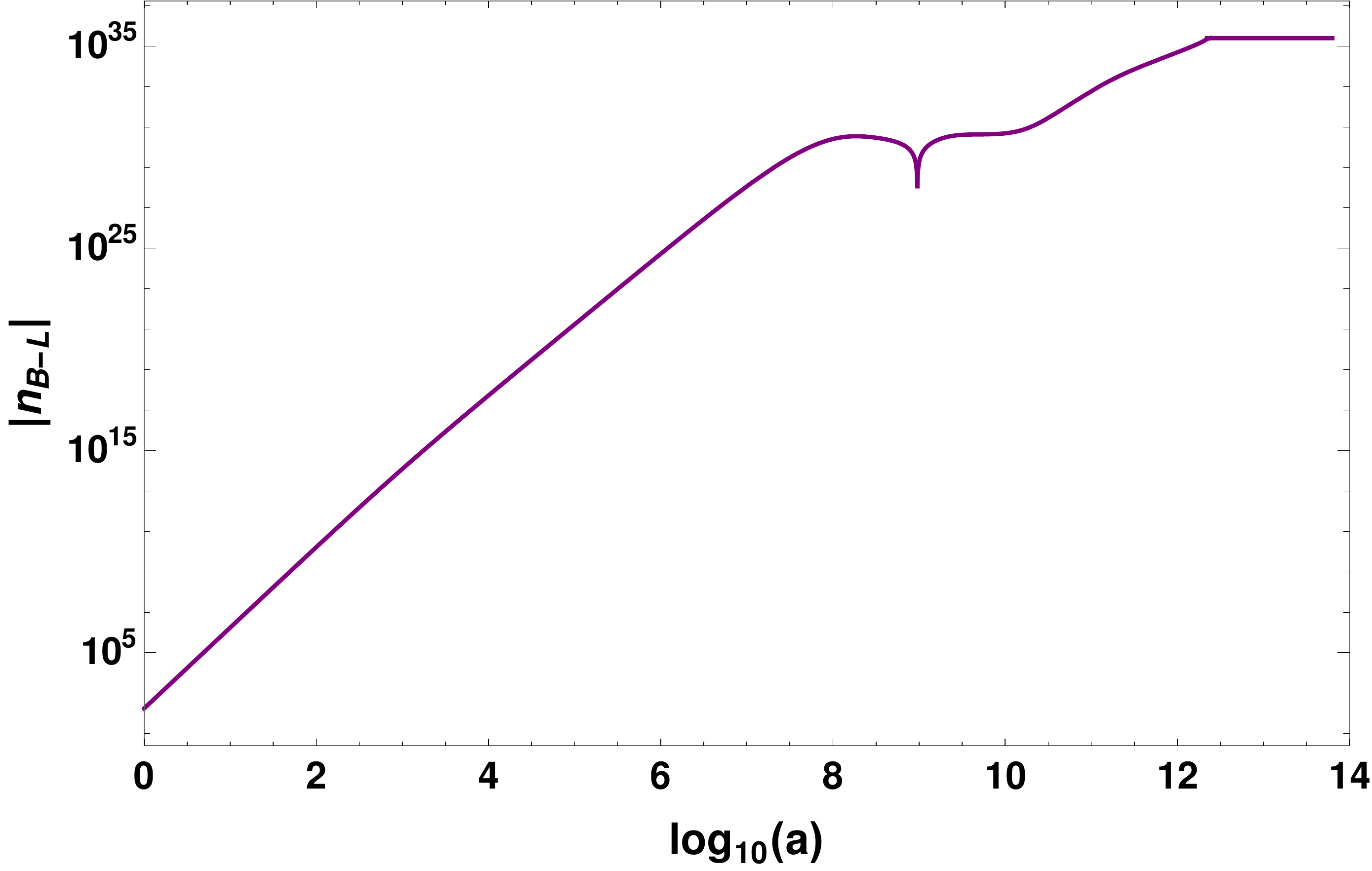}
\caption{Top panel : Evolution of the energy densities (left) and temperature of the thermal plasma (right), taking $M_{\rm in}=70000$ g, $M_1=500$ TeV, $\lambda_5=4\times10^{-4}$, $m_\nu^1 = 10^{-11}$ eV. Bottom panel : Evolution of the comoving number densities of $N_1$ (left) and $B-L$ (right) for the same parameters.}
\label{fig:after}
\end{figure}

In figures \ref{fig:b4}, \ref{fig:dur} and \ref{fig:after}, we show the evolution plots of the energy densities, temperature of the thermal plasma and the comoving number densities of $N_1$ and $B-L$, for the cases of PBH evaporation before, during and after the scale of leptogenesis respectively. The value of $\beta$ is chosen to be $10^{-3}$ such that a PBH domination is guaranteed. The PBH domination for a finite epoch can be clearly seen from upper left panel plots in these figures. In figure \ref{fig:b4}, the initial PBH mass is taken to be $10$ g, and hence it evaporates around ${\rm log} (a) \sim 7$, before the scale of thermal leptogenesis, which happens around ${\rm log} (a) \sim 11$. A kink in the temperature plot can be seen because of the entropy injection from PBH evaporation. The comoving number density of $N_1$ is increased initially from two contributions: the non-thermal PBH evaporation $n_{N_{1}}^{\rm BH}$ and from the inverse decays and scatterings $n_{N_{1}}^{T}$ of SM bath particles. $N_1$ can finally reach equilibrium, depending upon its Yukawa couplings, and its out of equilibrium decays produces a net asymmetry.  Figure \ref{fig:dur} corresponds to an initial PBH mass of $1500$ g, which evaporates during the scale of thermal leptogenesis, around ${\rm log} (a) \sim10$. In figure \ref{fig:after}, the initial PBH mass is taken to be $70000$ g such that it evaporates after thermal leptogenesis, around  ${\rm log} (a) \sim12$.

\begin{figure}
\includegraphics[scale=0.29]{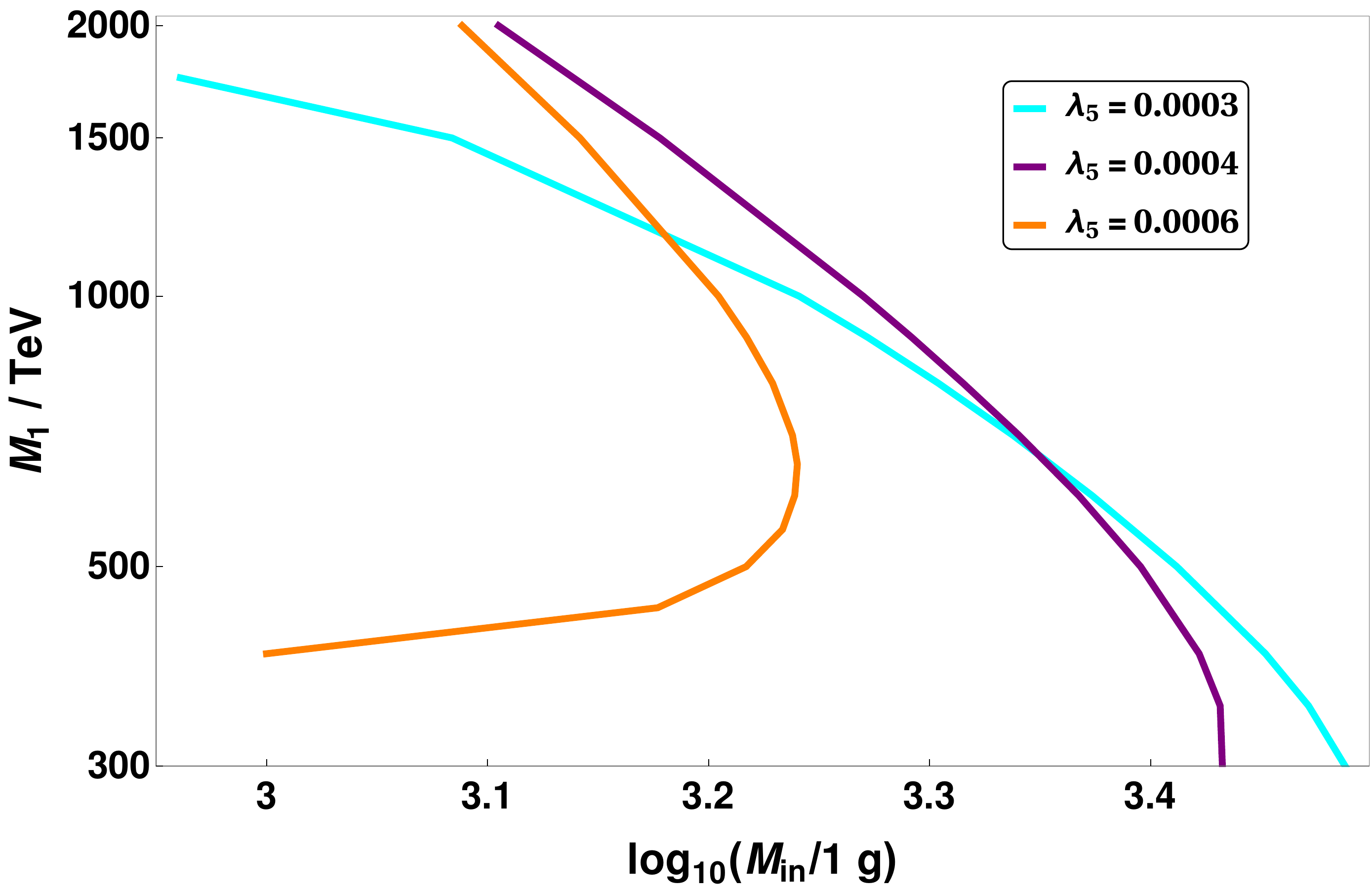}~~
\includegraphics[scale=0.29]{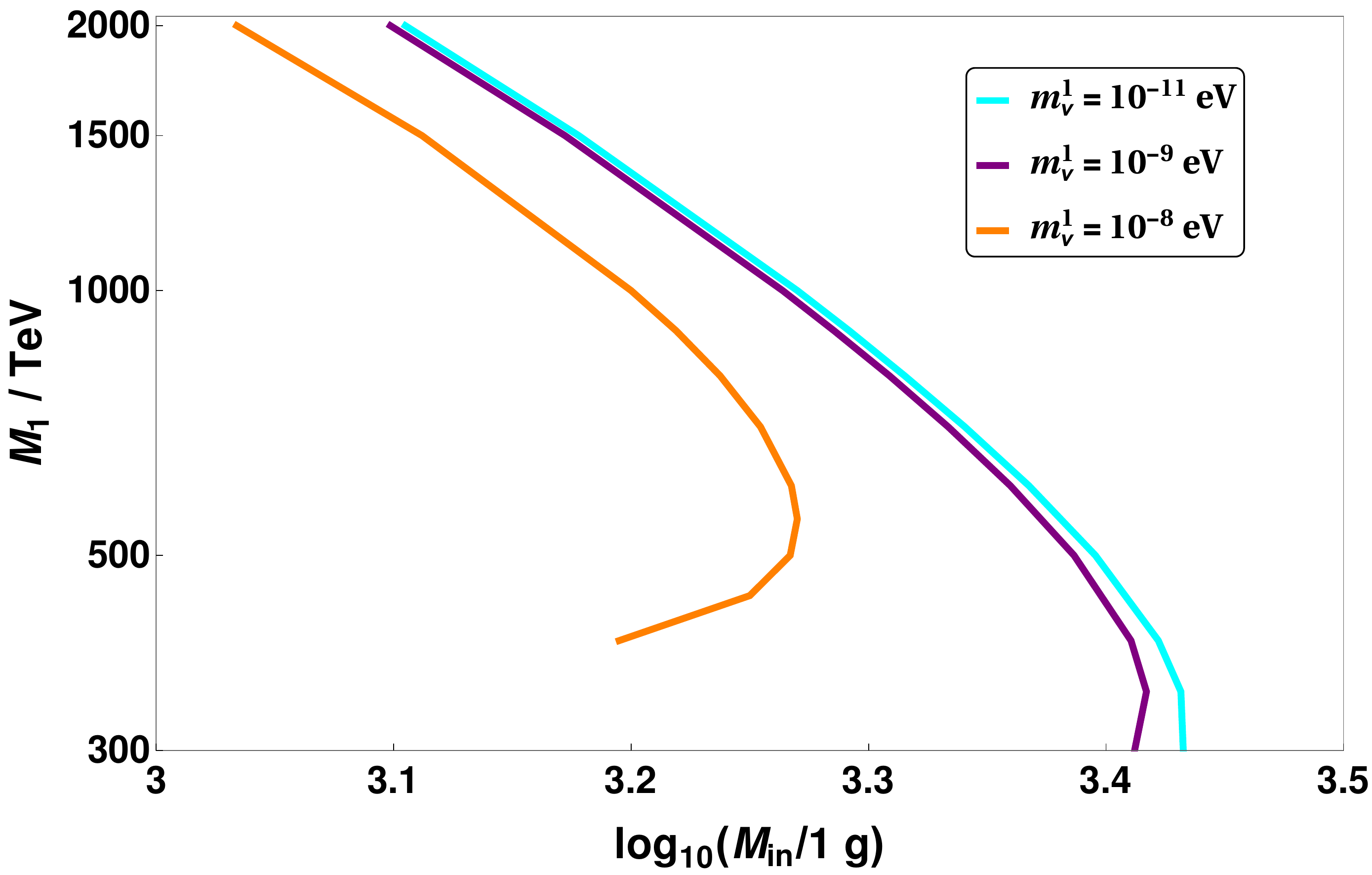}
\caption{Contours giving the observed baryon asymmetry of the Universe, in the $M_{\rm in}-M_1$ plane for different $\lambda_5$ values(left) and for different values of $m_\nu^1$ (right), taking a PBH dominated universe, $\beta=10^{-3}$. In the left panel, $m_\nu^1$ is fixed at $10^{-11}$ eV and $\lambda_5$ is taken to be $0.0004$ in the right panel.}
\label{fig:MBHvsM1}
\end{figure}

Finally, in figure \ref{fig:MBHvsM1} , we illustrate the effect of varying our model parameters $M_1, \lambda_5$ and the lightest neutrino mass $m_\nu^1$ (assuming normal ordering).  We show the contours giving the correct observed asymmetry in the $M_{\rm in}-M_{1}$ plane, for different values of $\lambda_5$ (left) and for different $m_\nu^1$ (right). While we focus on the possibility of low scale leptogenesis from a few hundreds of TeV to 2000 TeV or so, it is also possible to discuss high scale leptogenesis. Since high scale leptogenesis within type I seesaw has already been discussed in the context of PBH in earlier works including \cite{Perez-Gonzalez:2020vnz}, we outline this complementary window and study the effects of PBH domination in the early universe.
Now, in this regime, the decreasing behavior of $M_{in}$ with leptogenesis scale $M_1$ in order to give the correct asymmetry in figure \ref{fig:MBHvsM1} can be well understood from the right panel of figure \ref{fig:meta}. There, it can be noticed that for a higher leptogenesis scale, the intersection of $\eta_B$ with the observed asymmetry contour (cyan colour) occurs earlier. This is primarily because for a higher leptogenesis scale, the during case discussed above (PBH evaporation during the scale of leptogenesis) occurs earlier and hence for a lower PBH mass. Thus, the departure  from the thermal leptogenesis behavior starts occurring earlier. However, this trend of $M_{in}$ with $M_1$ is not always guaranteed, as can be seen in the orange contours at a lower value of $M_1$.   

The changing behavior with $\lambda_5$ in the left panel can be understood as follows : For a higher value of $\lambda_5$, the asymmetry produced is lower (see left of figure \ref{fig:th}). Hence, the dilution by PBH required in order to give the correct asymmetry should be minimal. This explains the lower value of PBH mass required for giving the correct asymmetry. The departure from this behavior seen in the figure can be attributed to the washout effects, which starts becoming important for a lower $\lambda_5$ and also for a higher value of $M_1$ (see equation(\eqref{eqn:washout})). Similar conclusion can be drawn from the right panel of figure \ref{fig:MBHvsM1}, since asymmetry for a higher neutrino mass $m_\nu^1$ is lower. 

Note that in figure \ref{fig:MBHvsM1} (left panel), for $\lambda_5=0.0006$, the asymmetry produced for $M_1\lesssim 400$ TeV is already below the observed value and hence additional dilution due to PBH evaporation will not help in generating the correct asymmetry. Similar pattern can be observed for light neutrino mass $m_\nu^1 = 10^{-8}$ eV in the right panel plot of figure \ref{fig:MBHvsM1}. Moreover, in the left panel, for $\lambda_5=0.0003$ eV, the washout effects become dominant for $M_1\gtrsim~1800$ TeV, and fails to give the observed asymmetry.

\section{Dark Matter in the presence of PBH}\label{sec:DM}
PBH evaporation can play an important role in generating DM relic density, for both thermal as well as non-thermal DM. For earlier works studying such effects in different contexts, see \cite{Morrison:2018xla, Gondolo:2020uqv, Bernal:2020bjf, Green:1999yh, Khlopov:2004tn, Dai:2009hx, Allahverdi:2017sks, Lennon:2017tqq, Hooper:2019gtx, Chaudhuri:2020wjo, Masina:2020xhk, Baldes:2020nuv, Bernal:2020ili, Kitabayashi:2021hox, Masina:2021zpu}. In our scenario, the lightest scalar of the inert doublet is the dark matter candidate. It can be in thermal equilibrium in the early universe by virtue of its electroweak gauge interactions. As usual, when the interaction rate of dark matter becomes less than the Hubble, $\Gamma_{\rm DM}<H$, the DM goes out of equilibrium and eventually freezes out to give a relic abundance. 

Now, in the presence of PBH, this typical WIMP phenomena can be modified. There can be different scenarios \cite{Bernal:2020bjf,Gondolo:2020uqv} depending on the interplay between the freeze-out temperature of WIMP DM $T_{\rm fo}$ and the  PBH evaporation temperature $T_{\rm ev}$. Two other important energy scales are given by $T_{\rm eq}$ and $T_c$ (see figure \ref{fig:after}). Here, $T_{\rm eq}$ corresponds to the temperature, such that for $T>T_{\rm eq}$, PBH cannot dominate the energy density of the universe. This is shown by the region marked as I in the top left panel of figures \ref{fig:b4}, \ref{fig:dur}, \ref{fig:after}. If freeze-out happens during this epoch, we would have the usual WIMP case in a radiation dominated universe, with a subsequent dilution of the relic due to PBH evaporation. Now, for $T_c<T<T_{\rm eq}$ (marked as II in the top left panel of figures \ref{fig:b4}, \ref{fig:dur}, \ref{fig:after}), PBH can dominate the energy density of the universe, if $\beta>\beta_c$. However, in this region SM radiation is still found to behave as free radiation, i.e., $\rho_{\rm Rad}\propto a^{-4}$ or equivalently $T\propto a^{-1}$. If freeze-out takes place during this epoch, one has to solve the WIMP dynamics assuming a matter-dominated background. Next, in the region III, PBH affect both the Hubble and the SM radiation evolution, which now scales as $\rho_{\rm Rad}\propto a^{-3/2}$ and the temperature of the plasma goes like $T\propto a^{-3/8}$ \cite{Arias:2019uol}. If DM freezes out during this epoch, one has to solve the WIMP dynamics in a matter dominated background as in the earlier case, but using the new temperature-scale factor relation mentioned above. Finally, region IV corresponds to the case when DM freezes out after PBH evaporation, $T_{\rm fo}<T_{\rm ev}$. In this region, the universe is radiation dominated again, since the PBH have faded away completely, and the WIMP dynamics take place in a radiation dominated universe, with no entropy dilution present like the earlier cases. Also, since DM can still be in equilibrium after PBH evaporation, the DM produced from PBH evaporation enters into the thermal bath and gives no extra contribution to the relic.

Now, in order to investigate the effect of PBH on the WIMP dynamics, we first  study the cases of PBH evaporation happening after DM freeze-out, i.e., $T_{\rm fo}>T_{\rm ev}$, which corresponds to the first three cases discussed above. Here, the contribution to the DM relic can be divided into two parts \cite{Gondolo:2020uqv}: one produced by Hawking radiation, $\Omega_{\rm DM}^{\rm BH} h^2$ and the other from the WIMP freeze-out $\Omega_{\rm DM}^{\rm fo} h^2$, with a subsequent entropy dilution because of the PBH evaporation. Thus, $\Omega_{\rm DM}^{\rm total} h^2 = \Omega_{\rm DM}^{\rm BH} h^2 +\Omega_{\rm DM}^{\rm fo} h^2 $. Now, since PBH have to evaporate before the BBN scale, one should have $T_{\rm fo}\gtrsim 0.1$ MeV. Assuming a typical freeze-out around $z_{\rm fo}=M_{\rm DM}/T_{\rm fo} \sim 30$, we can get an approximate lower bound on the DM mass for the case of $T_{\rm fo}>T_{\rm ev}$ as
\begin{equation}
M_{\rm DM}\gtrsim 3~{\rm MeV}.
\end{equation} 
This is trivially satisfied for scalar doublet DM as constraints from direct search restrict its mass to be much heavier \cite{Aaboud:2019rtt,Lundstrom:2008ai}. Moreover, to be consistent with leptogenesis, note that the mass of DM should be less than the right handed neutrino, i.e. $M_{\rm DM}<M_1$, so that the right handed neutrino can decay to produce lepton asymmetry and scalar DM remains stable.

\begin{figure}
\includegraphics[scale=0.29]{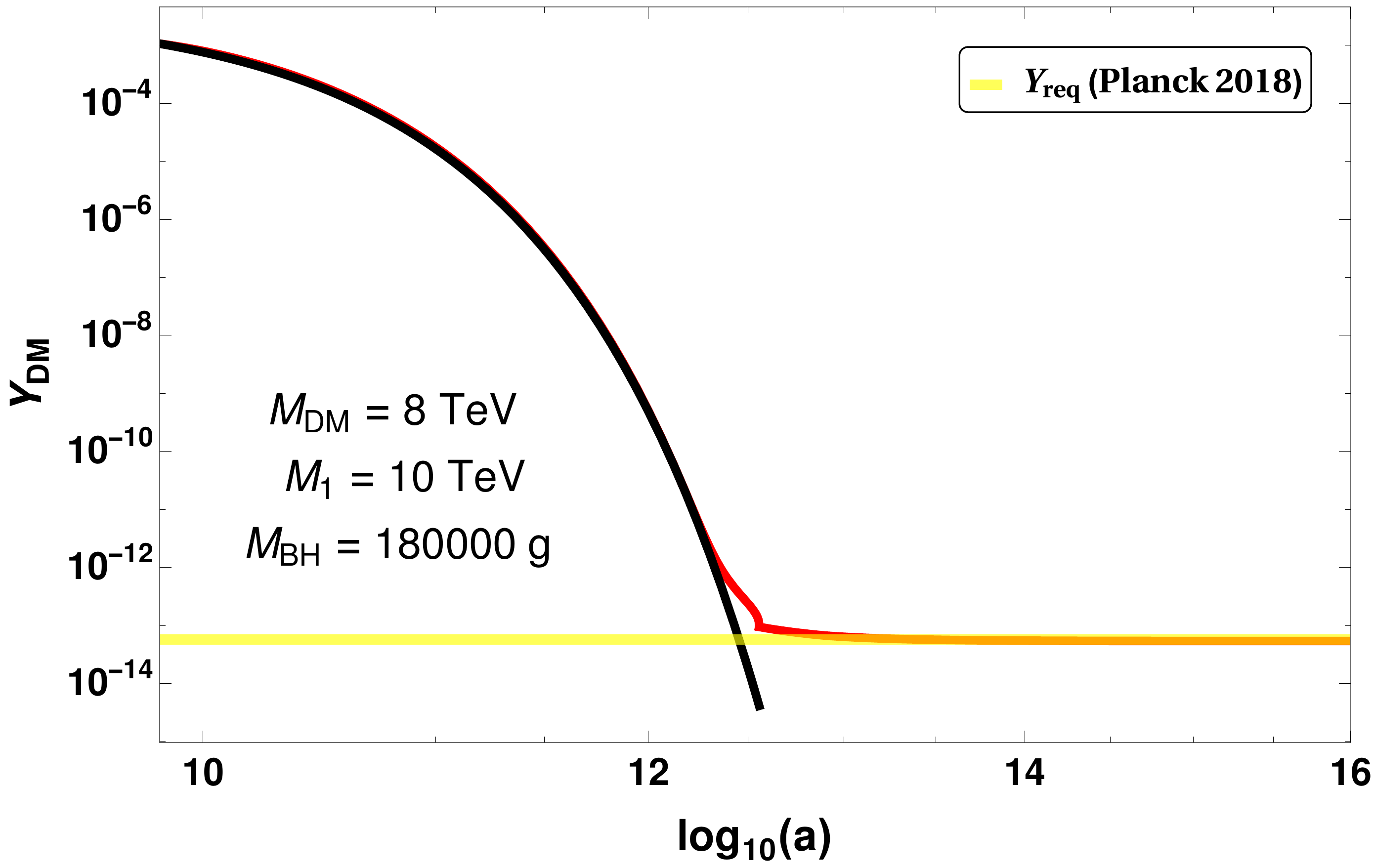}
\caption{Evolution of $Y_{\rm DM}$ with scale factor for the benchmark values shown and considering $\beta=10^{-3}$. The black line corresponds to the evolution of the equilibrium number density and the yellow line represents the value of $Y_{\rm DM}$ required to give the observed relic abundance of DM. Here, $\lambda_5 = 0.00008$ and mass of the charged scalar, $M_{\eta^+} = 8010$ TeV.}
\label{fig:WIMP}
\end{figure}

We first analyse the thermal DM production. We solve the Boltzmann equation for dark matter around the freeze-out time, considering DM mass to be $8$ TeV. The relevant equation for its comoving number density $Y_{\rm DM}$ in terms $z=M_{\rm DM}/T$ can be written as 
\begin{equation}
k \frac{d~Y_{\rm DM}}{dz}=-\frac{<\sigma v>s}{Hz}\left(Y_{\rm DM}^2-Y_{\rm eq}^2\right),
\end{equation} 
where $k=1~{\rm or}~3/8$, depending on whether we are in region $I/II$ or region $III$ respectively. In figure \ref{fig:WIMP}, we show the evolution of $Y_{\rm DM}$ with the scale factor $a$. It can be seen that for the chosen benchmark values, DM production is thermally overabundant. However, entropy dilution from PBH evaporation around ${\rm log}(a)\sim 12.6$ can eventually lead to the correct relic abundance. Thus, small values of $<\sigma v>$, which otherwise give a large relic, can now be allowed. However, as will see in a while, PBH not only lead to entropy dilution but also overproduction of DM specially for large $\beta$.

\begin{figure}
\includegraphics[scale=0.29]{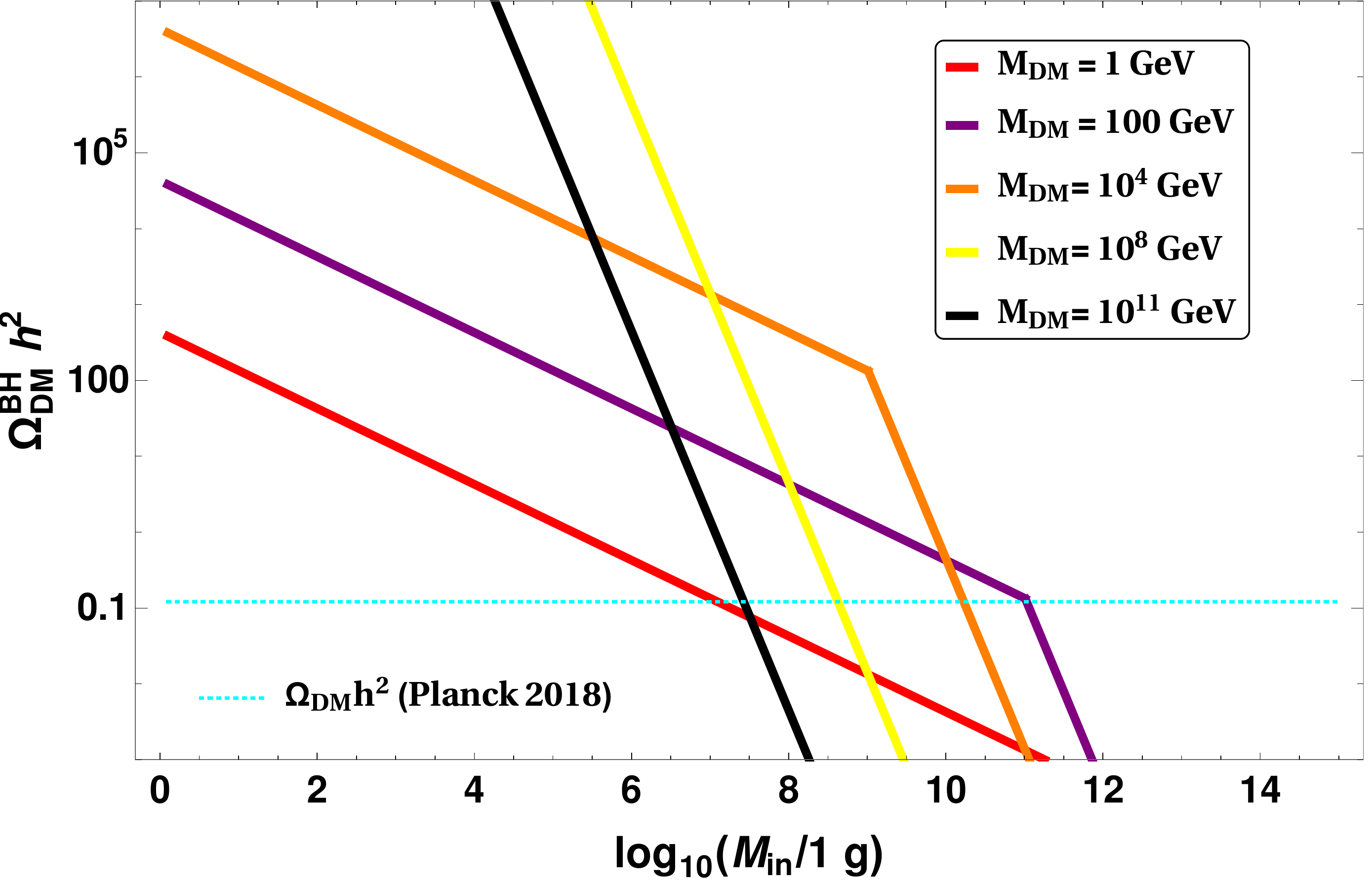}~~
\includegraphics[scale=.29]{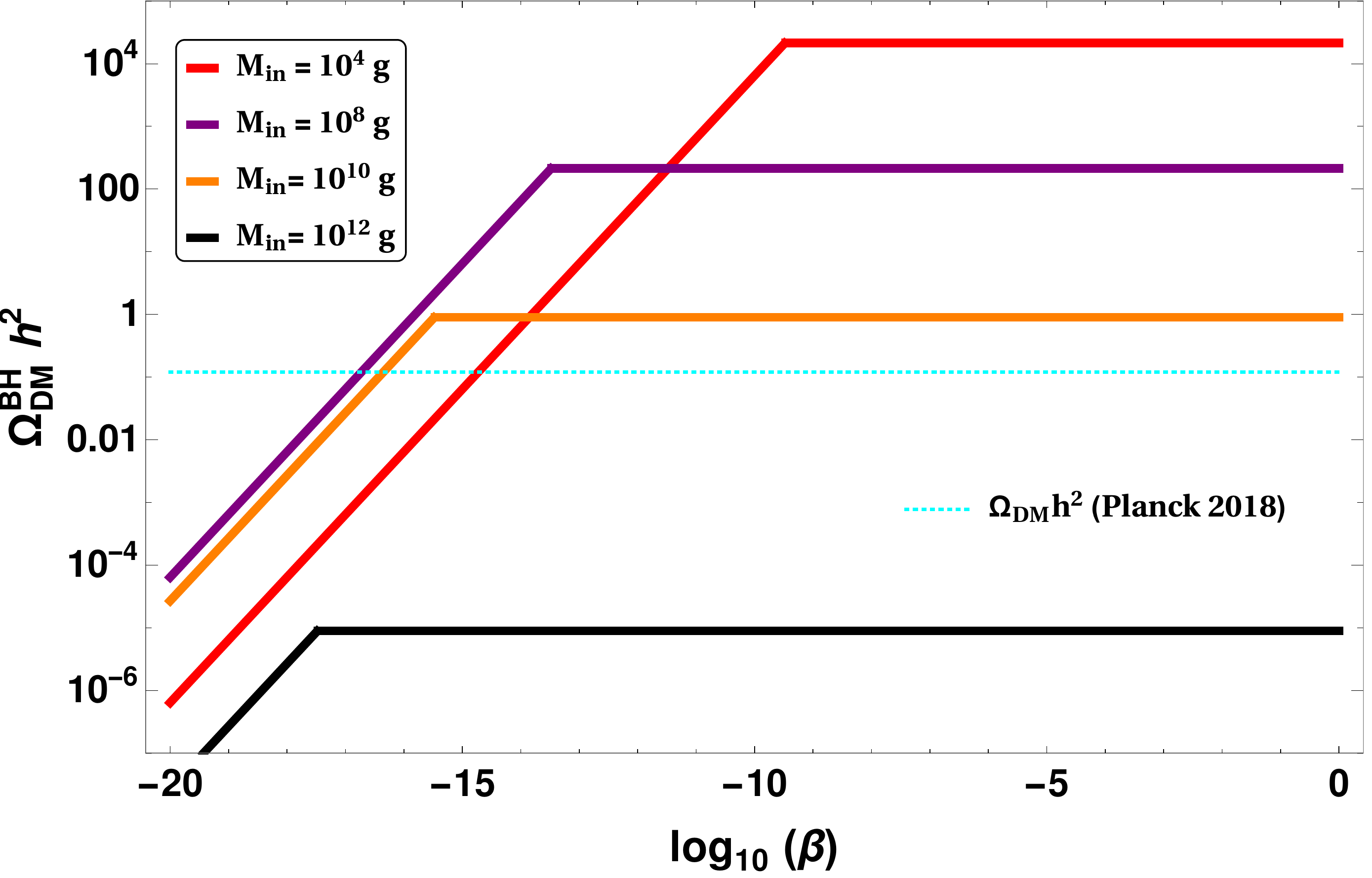}
\caption{Variation of Dark Matter relic abundance $\Omega_{\rm DM}^{\rm BH} h^2$ coming from  PBH evaporation with initial PBH mass $M_{\rm in}$ (left) and with initial PBH abundance $\beta$ (right). The dotted line represents the observed relic abundance. In the left panel, $\beta = 10^{-4}$, such that we have a PBH dominated universe. The DM mass is taken to be $5$ TeV in the right panel.}
\label{fig:DMnonth}
\end{figure}

For the non-thermal contribution to the DM relic coming from PBH evaporation, the present DM relic can be estimated as \cite{Gondolo:2020uqv,Bernal:2020bjf} 
\begin{equation}\label{eqn:YDM_nt}
Y_{\rm DM}=\frac{n_{\rm DM}(T_{0})}{s(T_{0})}\simeq\frac{3}{4}N_{\rm DM}\times
\begin{cases}
        \beta\frac{T_{\rm in}}{M_{\rm in}}\,\quad &\text{for radiation domination, }\beta \leq \beta_c,\\[8pt]
       \frac{\bar{T}_{\rm ev}}{M_{\rm in}}\,\quad &\text{for matter domination, }\beta \geq \beta_c,\,
     \end{cases} 
 \end{equation}
 where $n_{\rm DM}$ and $s$ corresponds to the number density and entropy density respectively, $T_0$ is the present temperature and $\bar{T}_{\rm ev}=\frac{2}{\sqrt{3}}T_{\rm ev}$ \cite{Bernal:2020bjf}.
 The number of DM particles emitted, $N_{\rm DM}$ is given by 
\begin{equation}
N_{\rm DM}=\frac{15\zeta(3)}{\pi^4}\frac{1}{g_*(T_{\rm BH})}\times
\begin{cases}
        (\frac{M_{\rm in}}{M_{\rm Pl}})^2\,\quad {\rm for}~~M_{\rm DM} \leq T_{\rm BH}^{\rm in},\\[8pt]
       (\frac{M_{\rm Pl}}{M_{\rm DM}})^2\,\quad {\rm for}~~M_{\rm DM} \geq T_{\rm BH}^{\rm in},\,
     \end{cases}
\end{equation}  
In the left panel of figure \ref{fig:DMnonth}, we show the DM relic abundance generated from PBH evaporation as a function of initial PBH mass $M_{\rm in}$ for different DM mass. The right panel of the figure shows the same as a function of $\beta$ for different values of $M_{\rm in}$. It can be seen that for DM masses between $4~{\rm GeV}\lesssim M_{\rm DM}\lesssim 10^{9}$ GeV, DM is overproduced, unless we are in a radiation dominated universe (small $\beta$). This was also illustrated in \cite{Bernal:2020kse,Bernal:2020bjf}. Note that in this regime, higher PBH masses which can give rise to the correct relic fall outside the BBN bound (equation \eqref{eqn:BBNbound}).

\section{$N_{2}$ leptogenesis and fermion dark matter}\label{sec:N2lepto}
\begin{figure}
\includegraphics[scale=0.29]{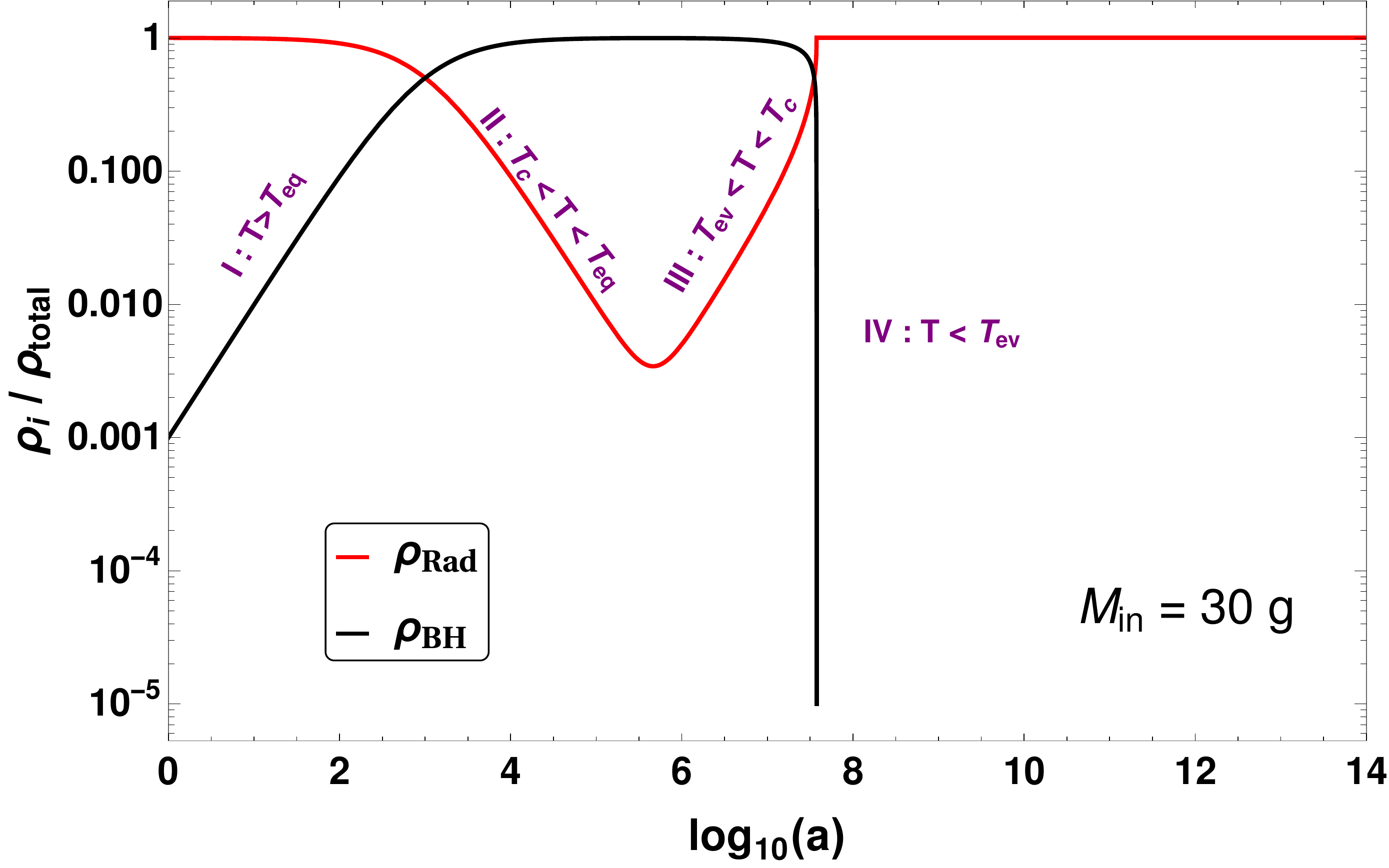}~~
\includegraphics[scale=.29]{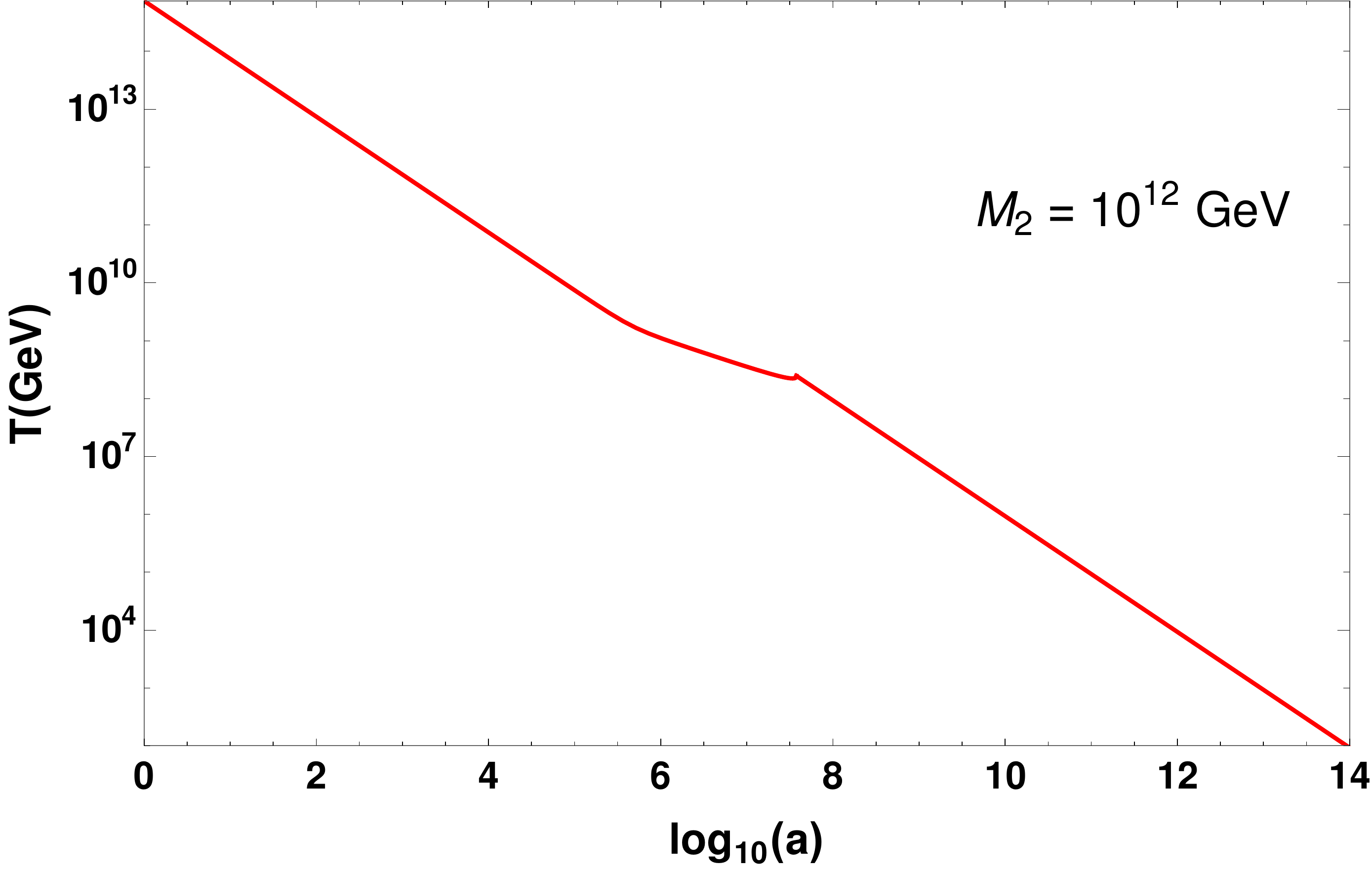}\\
\includegraphics[scale=.29]{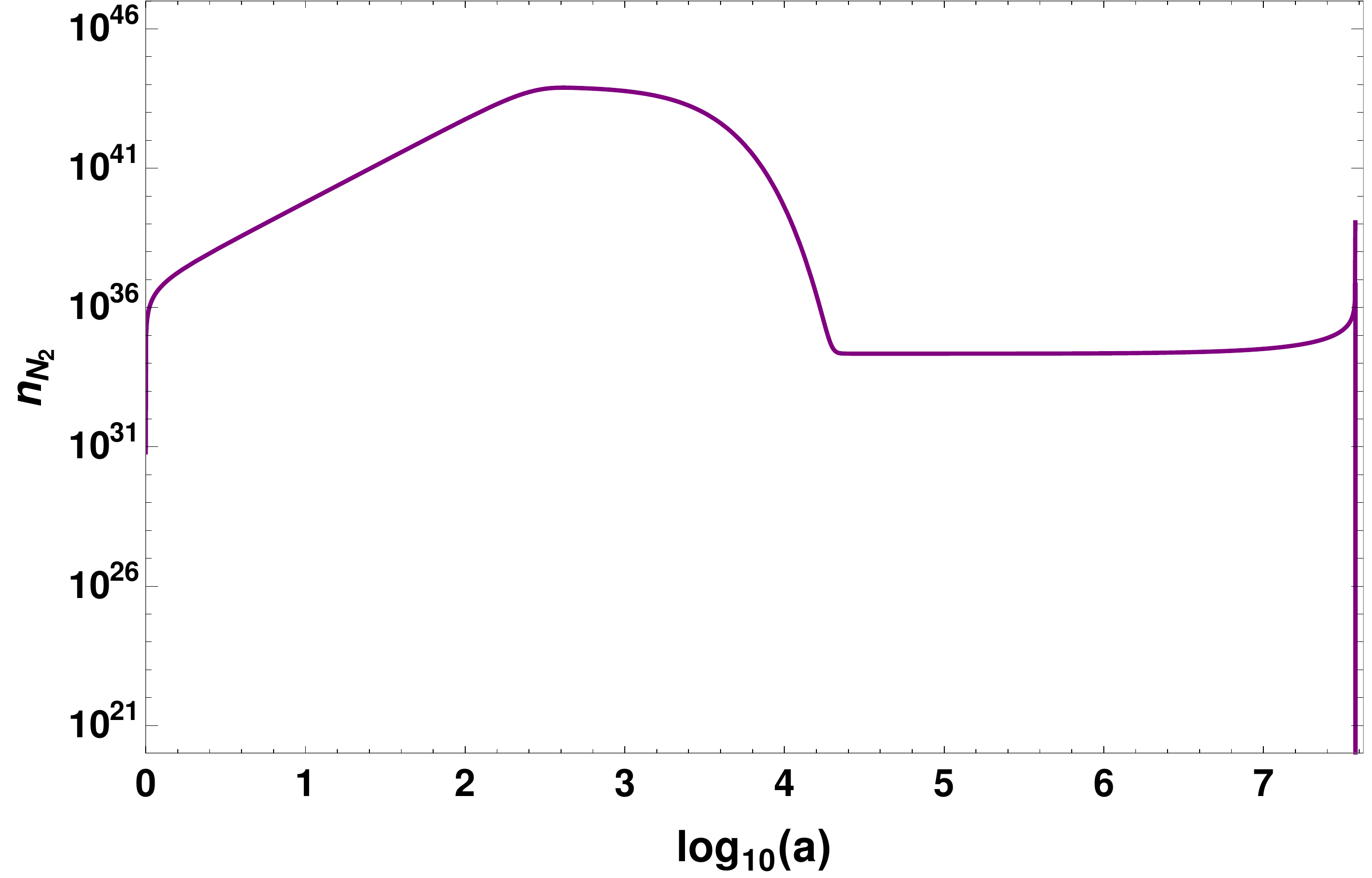}~~
\includegraphics[scale=.29]{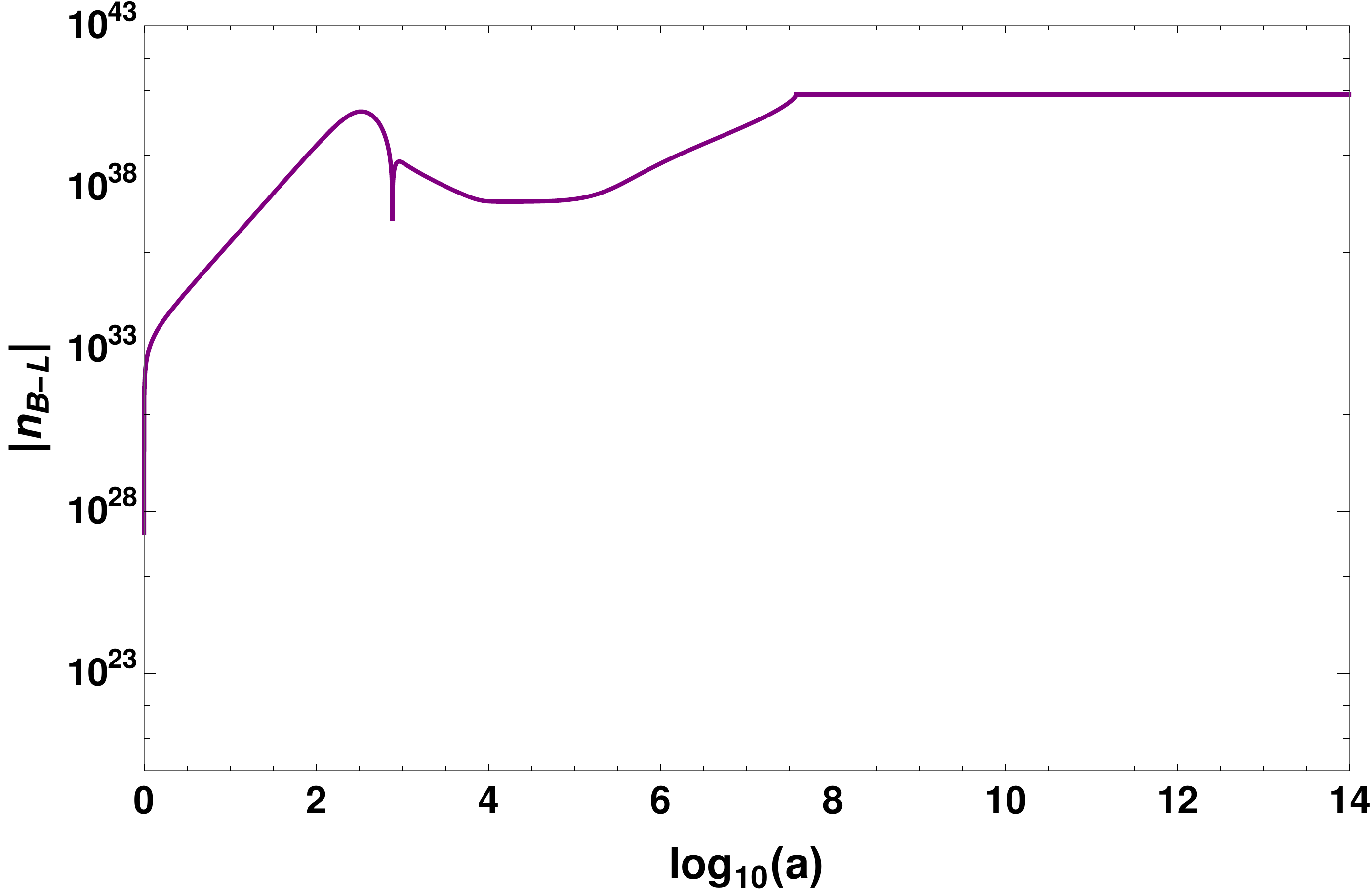}
\caption{Upper panel : Evolution of the energy densities (left) and temperature of the thermal plasma (right), taking $M_{\rm in}=30~g$, $M_2=10^{12}$ GeV, $\lambda_5=0.5$, $m_\nu^1 = 10^{-18}$ eV. Lower plot : Evolution of the comoving number densities of $N_1$ (left) and $B-L$ (right) for the same parameters.}
\label{fig:afterN2}
\end{figure}

In this section, we explore the possibility of having leptogensis from $N_{2}$ decays while considering the lightest RHN $N_{1}$ as a dark matter candidate. Although $N_{3}$ decay can also generate lepton asymmetry in principle, we consider the asymmetry generated from $N_3$ decay or any pre-existing asymmetry to be washed out due to strong washout effects mediated either by $N_2$ or $N_3$ themselves. Such a scenario of $N_2$ leptogenesis in scotogenic model was studied in details recently in \cite{Mahanta:2019gfe}, and we follow the same prescription here, considering a mass hierarchy: $M_3=10^3~M_2$. While the evolution of different quantities remain similar as before, the scale of leptogenesis gets pushed up for $N_2$ leptogenesis. Therefore, the typical PBH masses for "before", "during" and "after" scenarios described earlier can be much smaller. Additionally, since $N_1$ is much lighter compared to the scale of leptogenesis, apart from the $\Delta L=2$ washout introduced by $l \eta \longrightarrow \bar{l}\eta^{*}$ scattering we can have $\Delta L=1$ washout processes : $l W^\pm (Z)\longrightarrow N_{1} \eta$ and $l \eta \longrightarrow W^\pm (Z)N_1$ as well, which can affect leptogenesis. In our numerical analysis we take all these processes and solve the same Boltzmann equations as discussed in section \ref{sec:lepto}. Therefore, in the $N_2$ leptogenesis scenario we are in a strong washout region which pushes the leptogenesis scale above $\mathcal{O} \geq 10^{10}$ GeV. The produced asymmetry gets further diluted if PBH evaporation takes place after the generation of asymmetry. For illustrative purposes, we show the evolution of different physical quantities for $N_2$ leptogenesis in figure \ref{fig:afterN2} considering a scenario where PBH of 30 g initial mass evaporates after the scale of leptogenesis. Since the overall pattern is similar to $N_1$ leptogenesis discussed before except the change in scale of leptogenesis and PBH mass, we skip the detailed discussion of the other two possibilities in $N_2$ leptogenesis.

\begin{figure}
\includegraphics[scale=0.35]{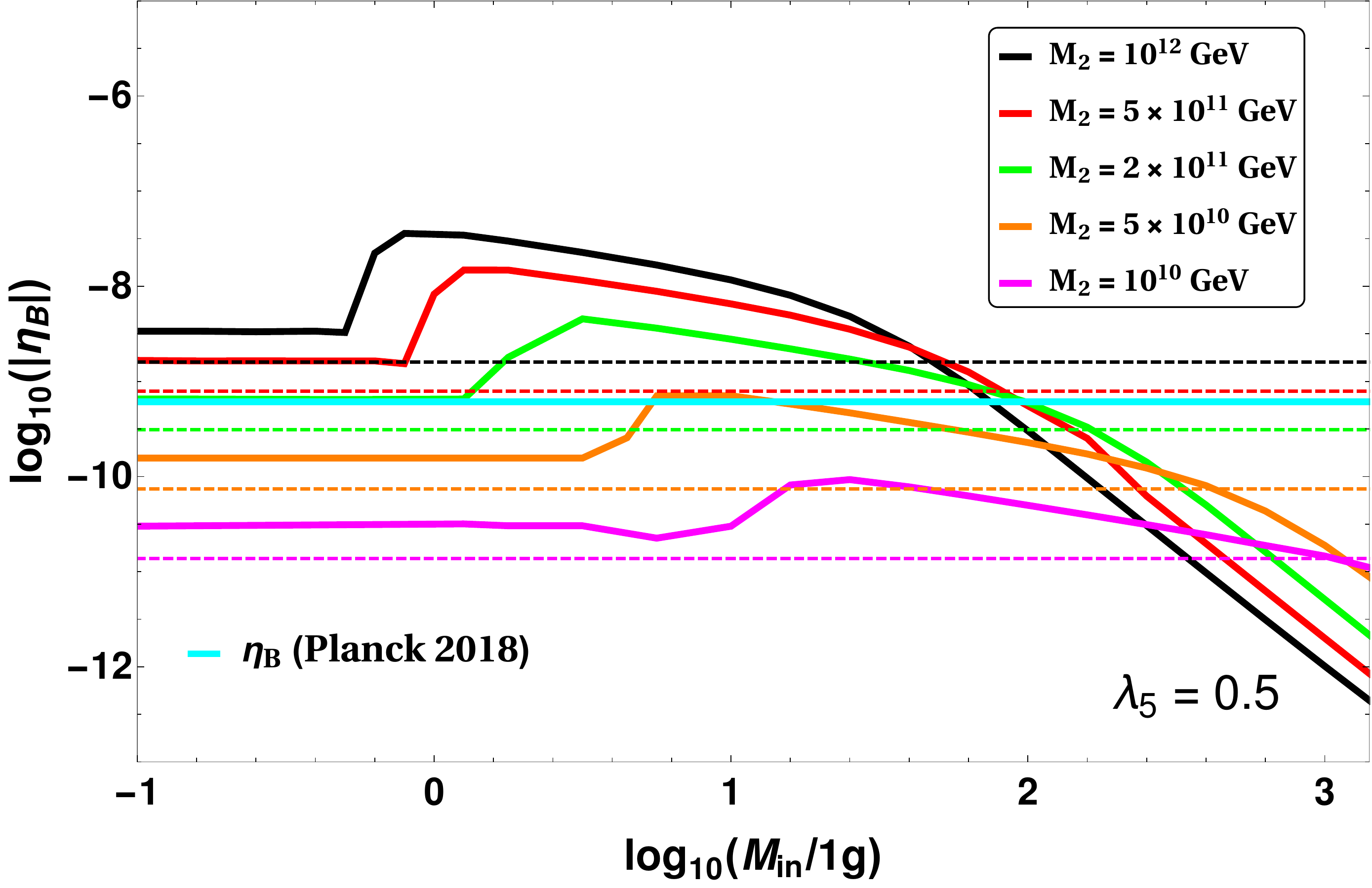}
\caption{Final baryon asymmetry as a function of the PBH mass, for different leptogenesis scales $M_2$, taking $\lambda_5=0.5$ and lightest neutrino mass $m_\nu^1=10^{-18}$ eV. The Cyan line indicates the observed asymmetry and the dashed lines represent the asymmetry produced in the absence of PBH. }
\label{fig:mvseta3}
\end{figure}

In figure \ref{fig:mvseta3}, we show the variation of the final baryon asymmetry with initial PBH mass $M_{\rm in}$ for different leptogenesis scales, namely $M_2$, keeping $\beta$ fixed at $10^{-3}$. Here, in contrast to the behavior seen in our earlier case (refer to figure \ref{fig:meta}) of leptogenesis from $N_1$ decay, there is an increase in the asymmetry compared to the purely thermal contribution (dashed lines) for a certain region of initial PBH mass. This is because, here, in addition to the entropy injection by PBH, the RHNs emitted by PBH also contribute significantly to the asymmetry. This effect is more dominant when leptogenesis is over and the strong washout effects had also decreased by the time PBH starts evaporating. For instance, with $M_2=10^{12}$ GeV, there is an increase in the asymmetry between $0.5~{\rm g}\lesssim M_{\rm in} \lesssim 1~{\rm g}$. At higher values of $M_{\rm in}$, the effect of dilution by PBH starts dominating over the enhancement by RHNs from PBH, which explains the decreasing behavior of the asymmetry. As we keep on decreasing the leptogenesis scale, this enhancement effect is shifted to the right, since now PBH with a higher initial mass evaporate after the leptogenesis scale as well as after the dominance of strong washout is over. For instance, with $M_2=10^{10}$ GeV, the contribution from RHNs emitted by PBHs is more dominant after $M_{\rm in}\gtrsim 10~{\rm g}$, which leads to an increase in the asymmetry.

While both thermal and non-thermal $N_1$ DM with masses around the TeV corner were studied in \cite{Mahanta:2019gfe}, here $N_1$ is constrained to be in the light mass regime, since $M_1\gtrsim4~{\rm GeV}$ is overproduced in a PBH dominated universe, as discussed in section \ref{sec:DM}. Now, depending on the lightest active neutrino mass which determines the Yukawas, $N_{1}$ can either be a thermal or a non-thermal dark matter candidate. Realising such light thermal dark matter is often challenging due to the Lee-Weinberg bound \cite{Lee:1977ua}. Unless there exists additional new particles in such low mass regime to assist in DM annihilation or coannihilation, such light thermal DM is typically overproduced due to small cross section. Hence, we consider the case of non-thermal dark matter, by taking the $N_1$ Yukawas small. Such DM, with negligible initial abundance gets frozen in at later epochs from the particles present in the thermal bath, either via decay or scattering. A recent review of such DM, also known as the feebly interacting (or freeze-in) massive particle (FIMP) paradigm, can be found in \cite{Bernal:2017kxu}.

The DM candidate $N_{1}$ can be as light as a few keV (typical lower bound for fermion DM \cite{Drewes:2016upu}) and as shown in earlier work \cite{Mahanta:2019gfe}, only normal ordering (NO) of light neutrinos is consistent with FIMP type $N_1$.  On the other hand, inverted ordering (IO) cannot give rise to the required Yukawa coupling of $N_1$ DM. Thus, we stick to NO and consider tiny Yukawa coupling of $N_1$ via Casas-Ibarra (CI) parametrisation \cite{Casas:2001sr}.

\begin{figure}
\includegraphics[scale=0.35]{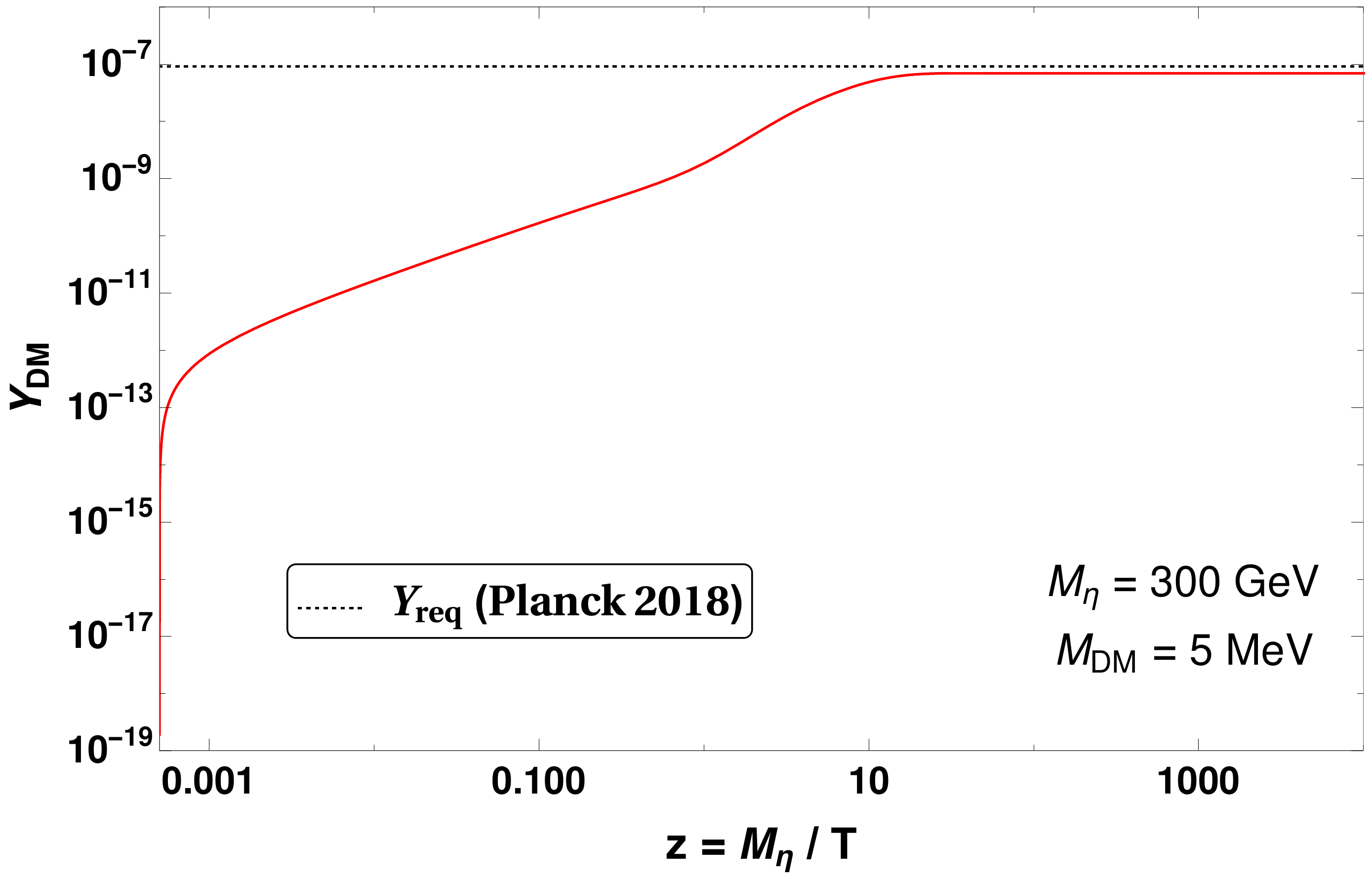}
\caption{Evolution of comoving number density of DM considering $M_{\rm DM}=5$ MeV, $M_\eta=300$ GeV, $\lambda_5=0.5$ and $m_\nu^1 = 10^{-18}$ eV.  }
\label{fig:freezein}
\end{figure}

Now, for small $N_1$ Yukawas, $N_1$ cannot reach thermal equilibrium and hence we are in the ballpark of FIMP dark matter as mentioned above. Hence, $N_1$ needs to be produced from the decays or scatterings of particles in the thermal bath. The dominant scattering processes are $l W^\pm (Z)\longrightarrow N_{1} \eta$ and $l \eta \longrightarrow W^\pm (Z)N_1$ whereas the dominant decay mode is $\eta \rightarrow N_1 l$. While the production from decay continues till $\eta$ gets Boltzmann suppressed, the contribution from scattering to DM yield gets saturated when the equilibrium abundance of $W^\pm / Z$ is Boltzmann-suppressed. For the chosen mass range of PBH so as to affect leptogenesis, light FIMP DM production from the bath particles happens after the evaporation of PBH. Thus, the total contribution to the DM relic will be the sum of DM density produced from the PBH (which now cannot enter into thermal equilibrium unlike the WIMP case discussed in section \ref{sec:DM}) and the freeze-in production which would happen after PBH evaporation in a radiation dominated universe.  We solve the following Boltzmann equations for $\eta$ and $N_1$ to find the relic abundance from freeze-in.

\begin{align}
\dfrac{dY_{\eta}}{dz}&=-\dfrac{4\pi^{2}}{45}\dfrac{M_{\rm Pl}m_{\eta}}{1.66}\dfrac{\sqrt{g_{*}(z)}}{z^{2}}\bigg [\displaystyle\sum_{p \equiv \rm SM \;particles} \langle \sigma  v \rangle_{\eta\eta\rightarrow pp}\bigg(Y_{\eta}^{2}-(Y_{\eta}^{\rm eq})^{2}  \bigg)\bigg]\\&-\dfrac{M_{\rm Pl}}{1.66}\dfrac{z}{m_{\eta}^{2}}\dfrac{\sqrt{g_{*}(z)}}{g_{s}(z)}\Gamma_{\eta\rightarrow N_{1} l}Y_{\eta} \nonumber\\
\dfrac{dY_{N_{1}}}{dz}&=\dfrac{M_{\rm Pl}}{1.66}\dfrac{z}{m_{\eta}^{2}}\dfrac{\sqrt{g_{*}(z)}}{g_{s}(z)}\Gamma_{\eta\rightarrow N_{1} l}Y_{\eta} \nonumber\\&+\frac{4 \pi^2}{45} 
\frac{M_{\rm Pl} M_{\eta}}{1.66}\frac{\sqrt{g_{\star}(z)}}{z^2} \times\Bigg(\sum_{x = W^\pm, Z, \eta}
\langle {\sigma {\rm v}}_{lx\rightarrow N_1 x}\rangle
\,\,{({{Y}_x^{eq}}{{Y}_l^{eq}} -Y_{N_1}Y_{x})}
\Bigg)\label{eqn:FIMP}
\end{align}
where $z = M_\eta/T$ and $g_\star(z)$ is given by:
\begin{eqnarray*}
\sqrt{g_\star(z)} = \frac{g_{\rm s}(z)}{\sqrt{g_{\rho}(z)}}
\left(1 - \frac{1}{3}\frac{d\,{\rm ln}\,g_{\rm s}(z)}{d\,{\rm ln}z}\right) \, .
\end{eqnarray*}  
$g_{\rho}(z)$ denotes the effective degrees of freedom
related to the energy density of the universe at $z$. The evolution plot of the comoving number density of DM is shown in figure \ref{fig:freezein}. 

\section{Summary and Conclusion}
\label{sec:conclude}
\begin{figure}
\includegraphics[scale=0.29]{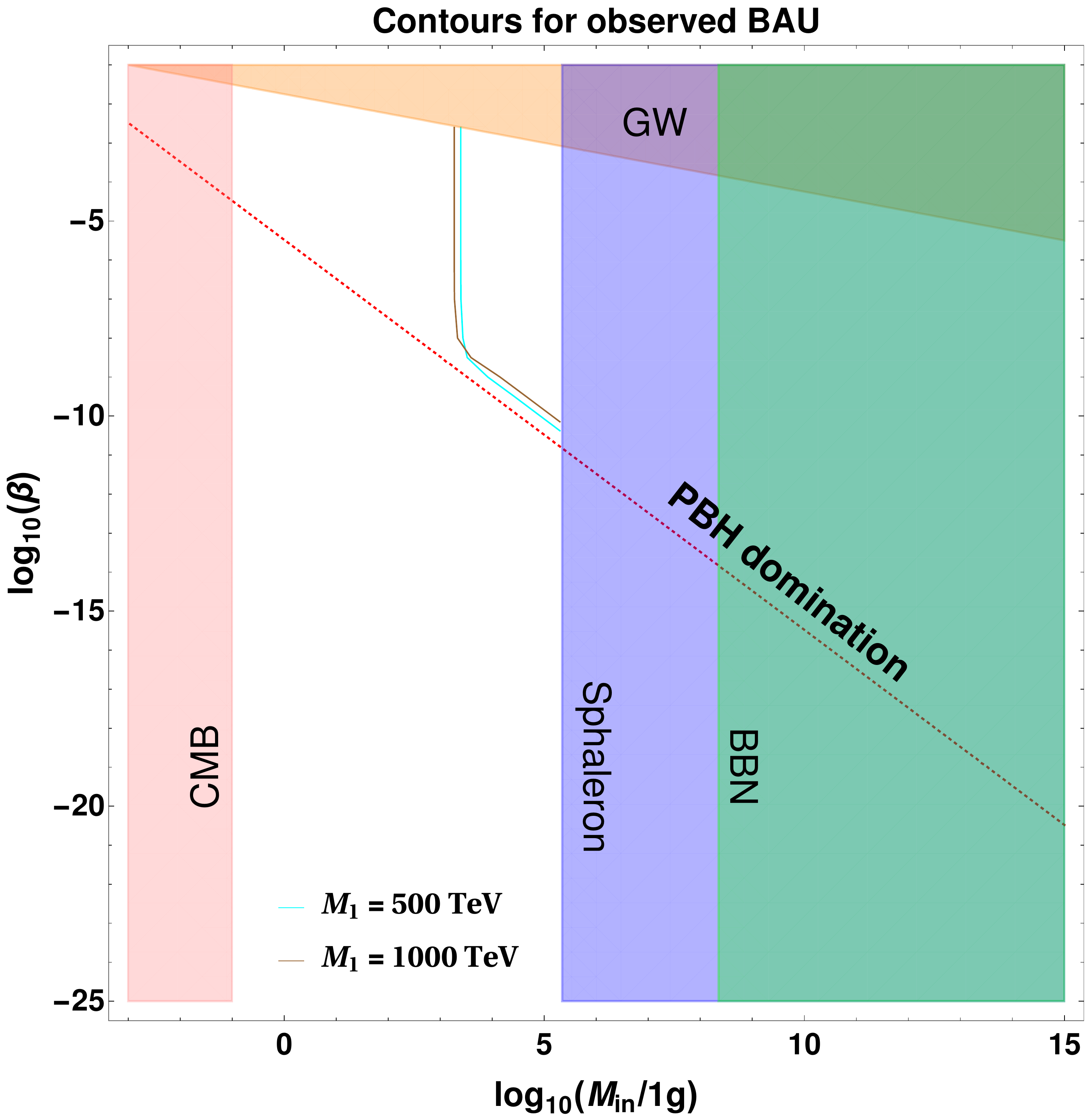}~~
\includegraphics[scale=0.29]{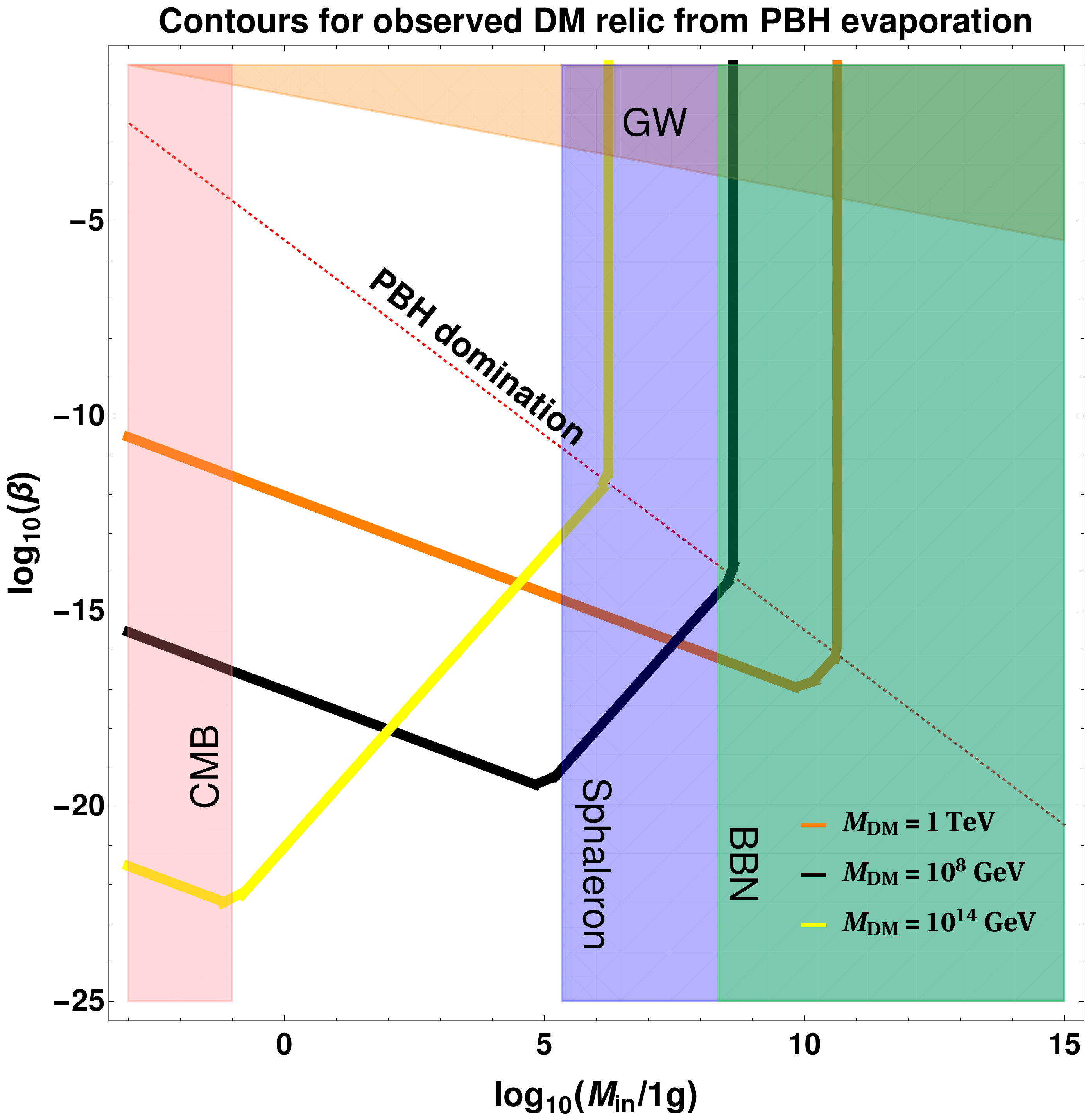}
\caption{Parameter space in the $M_{\rm in}-\beta$ plane giving the observed baryon asymmetry of the universe (left panel) from leptogenesis due to $N_1$ decay and observed DM relic from PBH evaporation $\Omega_{\rm DM}^{\rm BH} h^2$ (right). Here, $\lambda_5 = 0.0004$ and $m_\nu^1 = 10^{-11}$ eV. The shaded regions are excluded from different observable as described in the text. }
\label{fig:parspace}
\end{figure}

In the previous sections, we have highlighted the key features of the minimal scotogenic model in terms of leptogenesis and dark matter in the presence of primordial black holes. We now perform a numerical scan over the two key parameters of PBH, the initial PBH mass $M_{\rm in}$ (equation \eqref{eqn:Mini}) and the initial PBH fraction $\beta$ (equation \eqref{eqn:beta}). The relevant parameter space along with different bounds is shown in figure \ref{fig:parspace} for low scale leptogenesis scenario.

\begin{figure}
\includegraphics[scale=0.29]{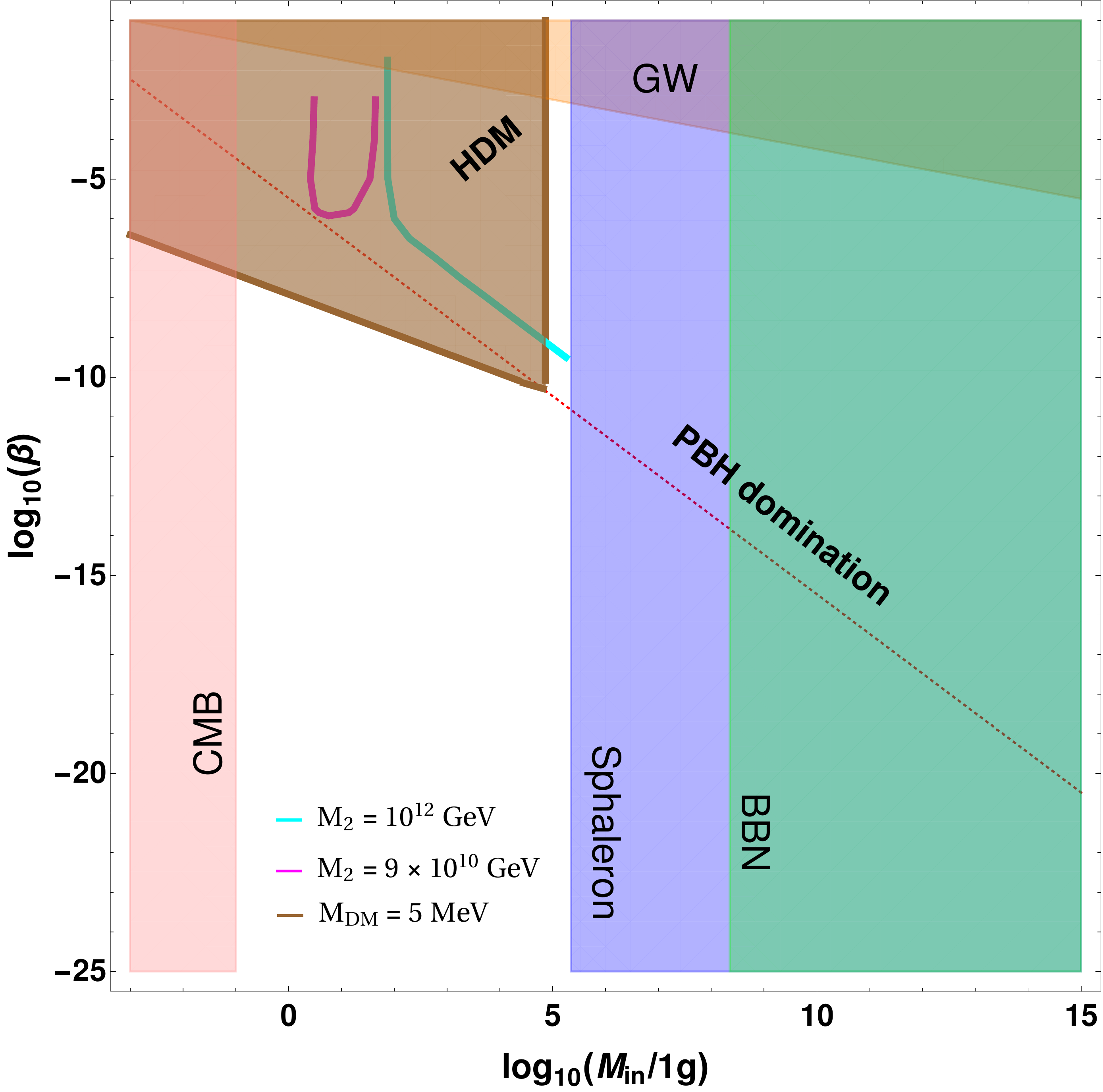}~~
\includegraphics[scale=0.29]{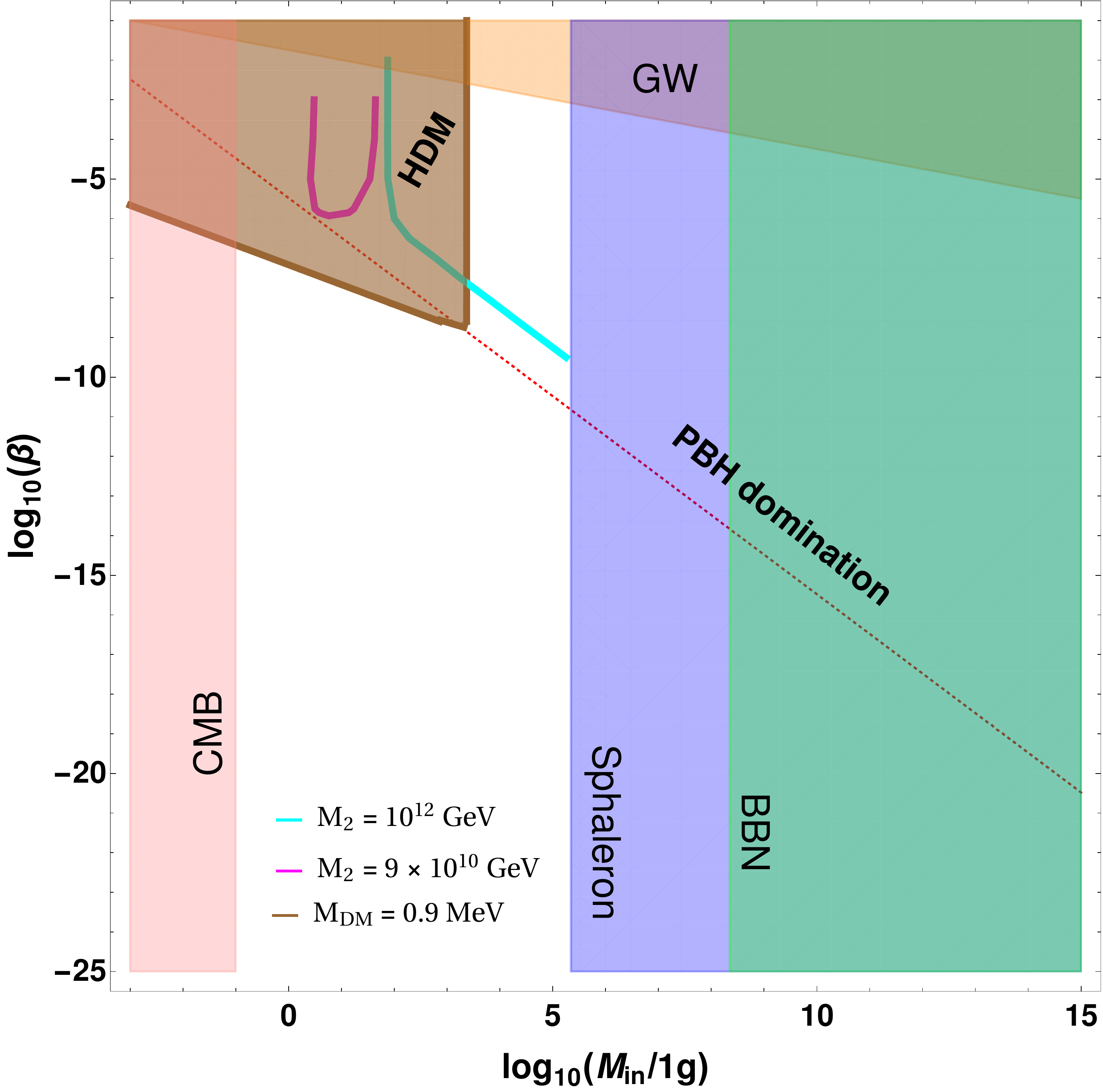}
\caption{Parameter space in the $M_{\rm in}-\beta$ plane giving the observed baryon asymmetry of the universe BAU (Cyan) and observed DM relic. The shaded regions are excluded from different observable as described in the text. Here, $M_{\rm DM}=5$ MeV (left panel) and $0.9$ MeV (right panel), $M_\eta=300$ GeV, $\lambda_5=0.5$ and $m_\nu^1 = 10^{-18}$ eV.}
\label{fig:N2lepto}
\end{figure}

The pink shaded region is ruled out from the upper bound on the inflation scale (equation \eqref{eqn:infbnd}). The green and blue shaded regions are ruled out due to the criteria of PBH evaporation before BBN (equation \eqref{eqn:BBNbound}) and the sphaleron scale (equation \eqref{eqn:sphbound}) respectively. The red dotted line marks the boundary between PBH domination and radiation domination (equation \eqref{eqn:betac}). In the orange shaded region, gravitational waves (GW) are overproduced leading to a backreaction problem, as discussed in \cite{Papanikolaou:2020qtd}. To avoid this, we require \cite{Papanikolaou:2020qtd} 
\begin{equation}
\beta<10^{-4}\left(\frac{10^9~g}{M_{\rm in}}\right)^{1/4}.
\end{equation} 

The contours giving rise to the correct baryon asymmetry for two different leptogenesis scales are shown in the left panel plot of figure \ref{fig:parspace}. The key model parameter $\lambda_5$ is fixed at $0.0004$ and the lightest neutrino mass is taken to be $10^{-11}$ eV. For such choice of parameters, thermal leptogenesis leads to overproduction of baryon asymmetry. PBH with a sufficiently high mass can give rise to the correct asymmetry through entropy dilution. To understand the behavior with $\beta$, let us first concentrate on the leptogenesis scale of $500$ TeV. As evident from the  cyan coloured contour, the asymmetry is almost independent of $\beta$ (the vertical region). However, as we keep on decreasing $\beta$, the effect of PBH starts to diminish, which increases the asymmetry. Hence, to produce the correct asymmetry a higher value of PBH mass is required. This explains the departure of the cyan coloured contour from the vertical pattern. Similar behavior can be seen for the other leptogenesis scale as well.     

Now, as discussed in section \ref{sec:DM}, DM relic is overproduced because of Hawking radiation from PBH for large $\beta$. In the right panel plot of figure \ref{fig:parspace}, we show the contours giving the  correct DM relic abundance with contribution only from PBH evaporation (assuming that thermal abundance is negligible either due to large DM annihilation or entropy dilution from PBH evaporation). The lower values of $\beta$, say less than around $\mathcal{O}(-10)$, can give the correct non-thermal DM relic abundance. However, for such tiny values of $\beta$, the leptogenesis setup will merge with the thermal leptogenesis, with no significant effect due to the presence of PBH. Note that non-thermally generated relic for DM with lower mass can be consistent with larger values of $\beta$ as well. It turns out that, if DM mass is below the GeV regime, non-thermal DM relic can be consistent with observed abundance even with large values of $\beta$. However, DM being part of a scalar doublet, such tiny masses are ruled out by direct search constraints. If the lightest $Z_2$ odd particle were one of the neutral singlet fermions, then this possibility could have been realistic. Thus, for scalar doublet DM, the only realistic possibility is to have its mass around a TeV or lower so that it freezes-out below the sphaleron decoupling temperature. Since PBH of chosen mass evaporates before sphaleron decoupling, the DM produced from such evaporation enter the thermal bath without affecting final DM relic.

Next, in figure \ref{fig:N2lepto}, we show the parameter space in the $M_{\rm in}-\beta$ plane, which gives the correct observed asymmetry along with the observed DM relic for $N_2$ leptogenesis scenario. Here since the leptogenesis scale is pushed high, PBH with a smaller mass can also start affecting the asymmetry. Hence, the contour for the correct observed asymmetry shifts towards left compared to that of figure \ref{fig:parspace} (left panel). The behavior of the observed asymmetry contour for the leptogenesis scale of $10^{12}~{\rm GeV}$ (shown in cyan color) is the same as that of $N_1$ leptogenesis discussed before, since it is thermally overproduced (refer to figure \ref{fig:mvseta3}) and entropy dilution by PBH can give rise to the correct observed asymmetry. However, for a lower leptogenesis scale of $9\times 10^{10}$ GeV (shown in magenta color), the asymmetry is thermally underproduced and contribution from RHNs emitted by PBH can increase the asymmetry as discussed in section \ref{sec:N2lepto}. Now, as can be seen from figure \ref{fig:mvseta3}, such asymmetry contours will intersect the observed asymmetry contour at two points. Thus, two values of initial PBH mass $M_{\rm in}$ can give rise to the correct asymmetry for a fixed value of $\beta$. Moreover, if we keep on decreasing $\beta$ further, the enhancement effect induced by PBH keeps on decreasing and at some point, we will not be able to get the observed asymmetry. This explains the U-shaped pattern obtained in figure \ref{fig:N2lepto} for $N_2$ leptogenesis scale of $9\times 10^{10}$ GeV.

Now, our scenario corresponds to a case of mixed dark matter, where along with the cold or warm DM component from freeze-in, we have a potential hot component produced by PBH evaporation. Data from observations related to the CMB and baryon acoustic oscillation (BAO) allow us to put an upper bound on the fraction of this hot component with respect to the total DM, depending on the value of DM mass \cite{Diamanti:2017xfo}. Here, we apply a conservative $10~\%$ upper bound on such hot dark matter (HDM) component \cite{Bernal:2020bjf}. Considering DM mass as $5$ MeV, the region which gives more than $10~\%$ of the total DM from PBH evaporation is shown by the brown region (labelled as HDM) in the left panel plot of figure \ref{fig:N2lepto}. The remaining contribution is given by the freeze-in component, which is independent of the PBH parameters. Thus, the white regions represent the region allowed by the observed DM relic. For example, the brown contour gives $\Omega_{\rm DM}^{\rm BH}h^2=0.1\times0.12$. The remaining $90\%$ can be obtained from freeze-in by suitably choosing the other parameters of the model. For instance, figure \ref{fig:freezein} shows the evolution plot for the parameters which gives freeze-in contribution $\Omega_{\rm DM}^{\rm FIMP}h^2=0.9\times0.12$. While for DM mass of 5 MeV, the brown shaded region disfavours most parts of the cyan colored line, for lower DM mass say 0.9 MeV, the brown shaded region shrinks allowing more of the cyan colored line showing leptogenesis favoured parameter space, as shown in the right panel plot of figure \ref{fig:N2lepto}. Decreasing DM mass further can also allow the other leptogenesis scale of $9\times 10^{10}$ GeV.

\begin{figure}
\includegraphics[scale=0.28]{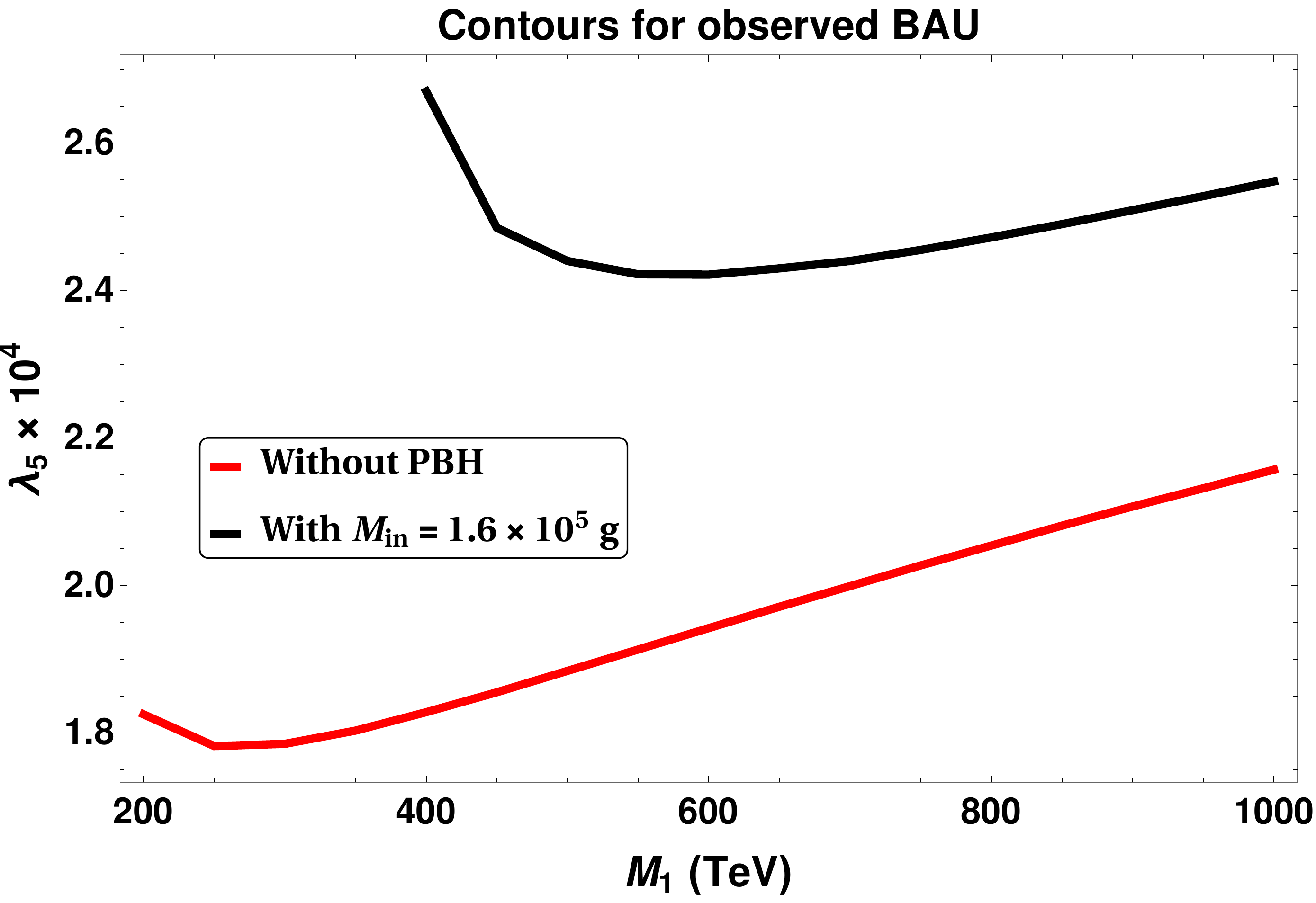}
\includegraphics[scale=0.28]{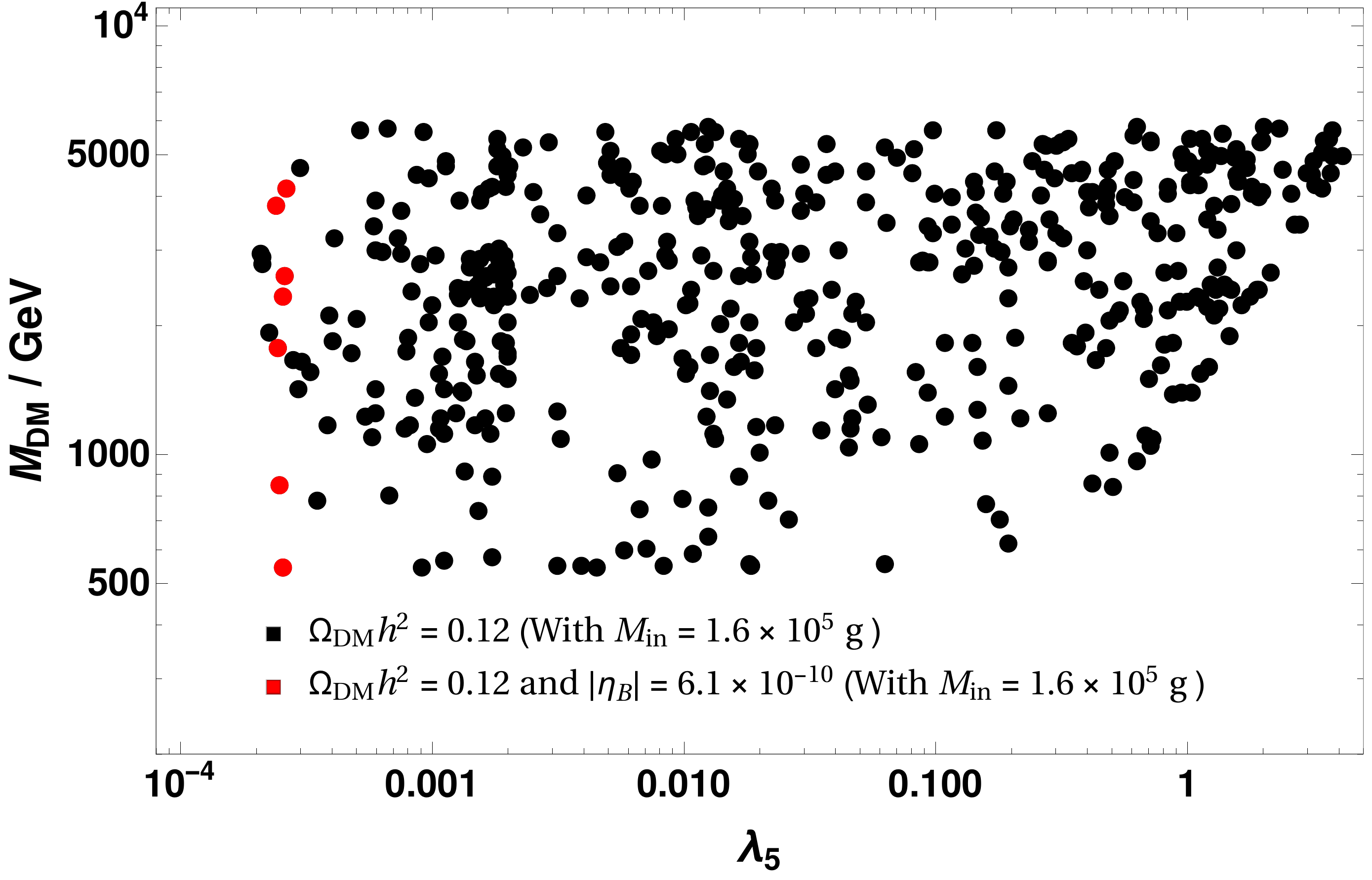}
\caption{Left panel: Parameter space in the $\lambda_5-M_{1}$ plane giving the observed BAU with and without the presence of PBH for $N_1$ leptogenesis with $\beta=4 \times 10^{-11}$, $m_\nu^1 = 10^{-11}$ eV. Right panel: Scalar DM parameter space consistent with the requirement that its freeze-out occurs after PBH evaporation.}
\label{fig:wtwttPBH}
\end{figure}

Hence, we conclude that in our setup, considering leptogenesis from $N_1$ decay with scalar doublet dark matter, it is not possible to find a region of PBH parameters which can simultaneously affect DM genesis as well as leptogenesis. For PBH to affect leptogenesis, one requires large PBH fraction $\beta$. Typically, for the region of parameter space leading to overproduction of baryon or lepton asymmetry in the scotogenic model, presence of PBH with sufficient fraction $\beta$ can lead to entropy dilution bringing the prediction to agreement with observations. However, such large $\beta$ values lead to overproduction of DM from PBH evaporation, for heavy DM mass regime where DM freeze-out occurs before PBH evaporation. Therefore, in order to have non-trivial effects of PBH on leptogenesis, we need to have DM in low mass regime so that its freeze-out occurs after PBH evaporation and relic is generated purely by thermal freeze-out. While successful DM genesis and leptogenesis in the presence of PBH can occur independently in this setup with above-mentioned choices of parameters, scenarios where such cogenesis occurs in a dependent manner can have interesting constraints on PBH properties. To explore this direction, we have also considered the scenario where leptogenesis occurs due to $N_2$ decay while the lightest right handed neutrino $N_1$ is the DM candidate. While the scale of leptogenesis is pushed to high scale regime in this case, the gauge singlet nature of $N_1$ allows us to consider it in the light mass window (few GeV or below) where it is not overproduced from PBH even with $\beta$ values leading to PBH domination. However, such light DM originating from PBH can lead to a hot DM component, tightly constrained from astrophysical bounds. While considering a conservative upper bound on such hot DM fraction of $10\%$, we found the PBH model parameters consistent with successful $N_2$ leptogenesis and non-thermal $N_1$ DM with mixed contribution from PBH as well as SM bath. 

\begin{figure}
\includegraphics[scale=0.29]{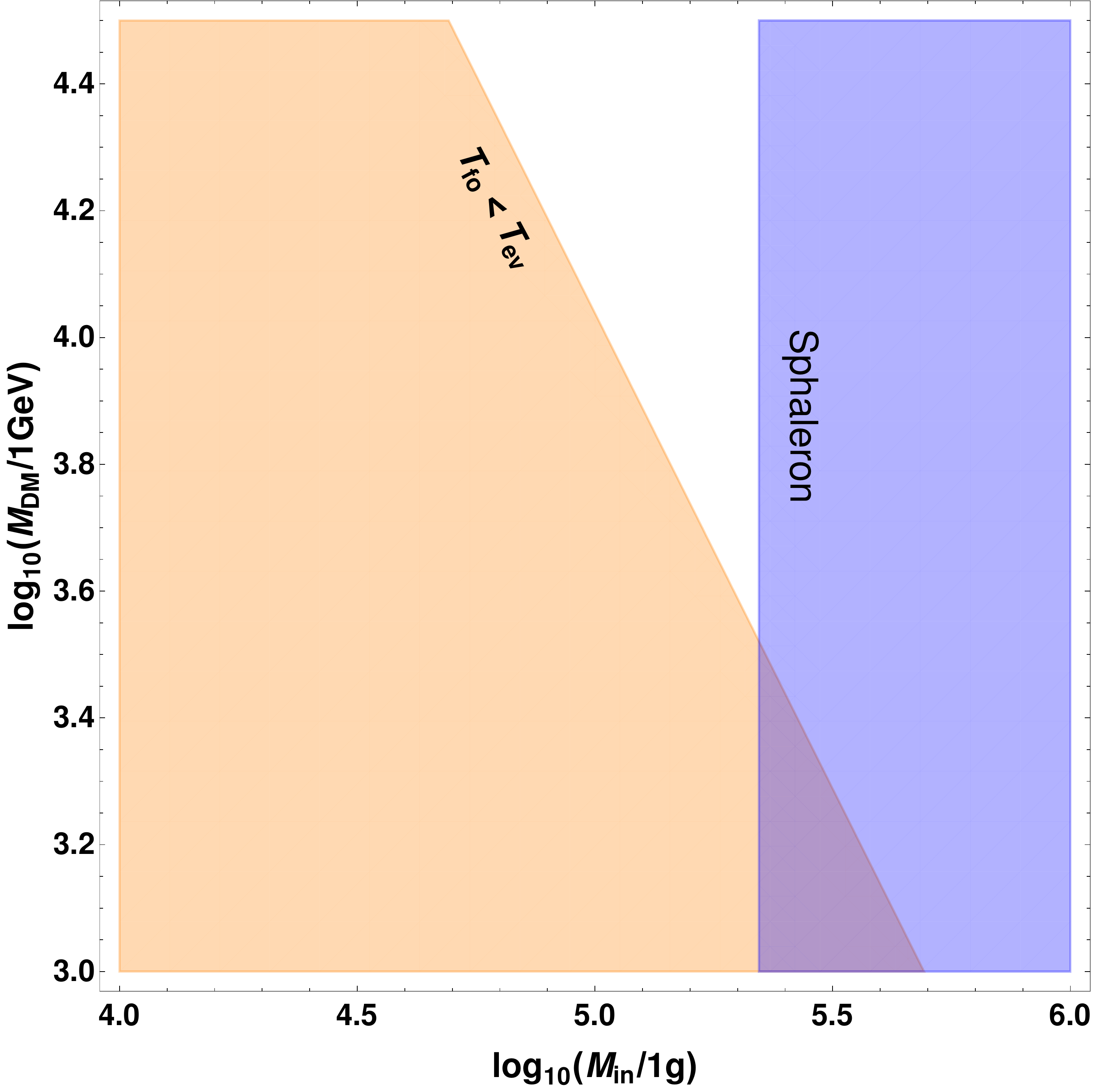}~~
\caption{Region showing $T_{\rm fo}<T_{\rm ev}$ (orange shaded) in the $M_{\rm in}-M_{\rm DM}$ plane, taking $z_{\rm fo}=M_{\rm DM}/T \approx 25$, typical for thermal freeze-out.}
\label{fig:MinMDM}
\end{figure}

Finally, for illustrating more specific estimates of the model parameters in the presence of PBH, we show the parameter space for $N_1$ leptogenesis and scalar DM in figure \ref{fig:wtwttPBH}. From the left panel plot it is clear that the asymmetry contour gets modified in the presence of PBH, with the behavior determined by the interplay of entropy dilution, enhancement from PBH evaporation and washout effects discussed in section \ref{sec:lepto}. The PBH initial mass and initial fraction are chosen to be $1.6\times10^{5}$ g and $\beta=4\times10^{-11}$ respectively, such that they evaporate long after the scale of leptogenesis and at the same time do not dilute the asymmetry much, because of a small value of $\beta$ (see left panel of figure \ref{fig:parspace}). Such a high value of PBH mass can also constrain the scalar DM parameter space, from the condition that they evaporate before the DM freeze-out, to prevent the overproduction of DM relic discussed above. This is illustrated in the right panel of figure \ref{fig:wtwttPBH}, where, by running a scan over the relevant parameters, we show the parameter space in the $M_{\rm DM}-\lambda_5$ plane giving the correct DM relic in the presence of PBH, having the same mass of $1.6\times10^{5}~g$. Clearly, DM with high masses are ruled out because they freeze-out at a higher temperature before PBH evaporation, leading to overproduction. On the other hand lower DM mass below around 500 GeV are disallowed due to large annihilation rates into electroweak gauge bosons, keeping thermal relic density suppressed. The slope in large $\lambda_5$ region is observed as DM annihilation mediated by components of the inert doublet is enhanced in this regime for larger mass splitting within inert doublet components (where splitting is proportional to $\lambda_5$), requiring larger DM mass to keep the annihilation rates within required rate for generating correct thermal relic. The points shown in red color in the right panel plot are also consistent with correct baryon asymmetry (compare with left panel). Note that for a lower PBH mass, the condition of $T_{\rm fo}<T_{\rm ev}$ will be trivially satisfied. Hence, in general, PBH and DM with higher masses are more constrained. This is further demonstrated in figure \ref{fig:MinMDM}, where in the $M_{\rm in}-M_{\rm DM}$ plane, we show the region of $T_{\rm fo}<T_{\rm ev}$ (orange shaded), which excluding the sphaleron bound, prohibits an extra small range of high PBH mass (white region) due to DM overproduction.

\begin{figure}
\includegraphics[scale=0.28]{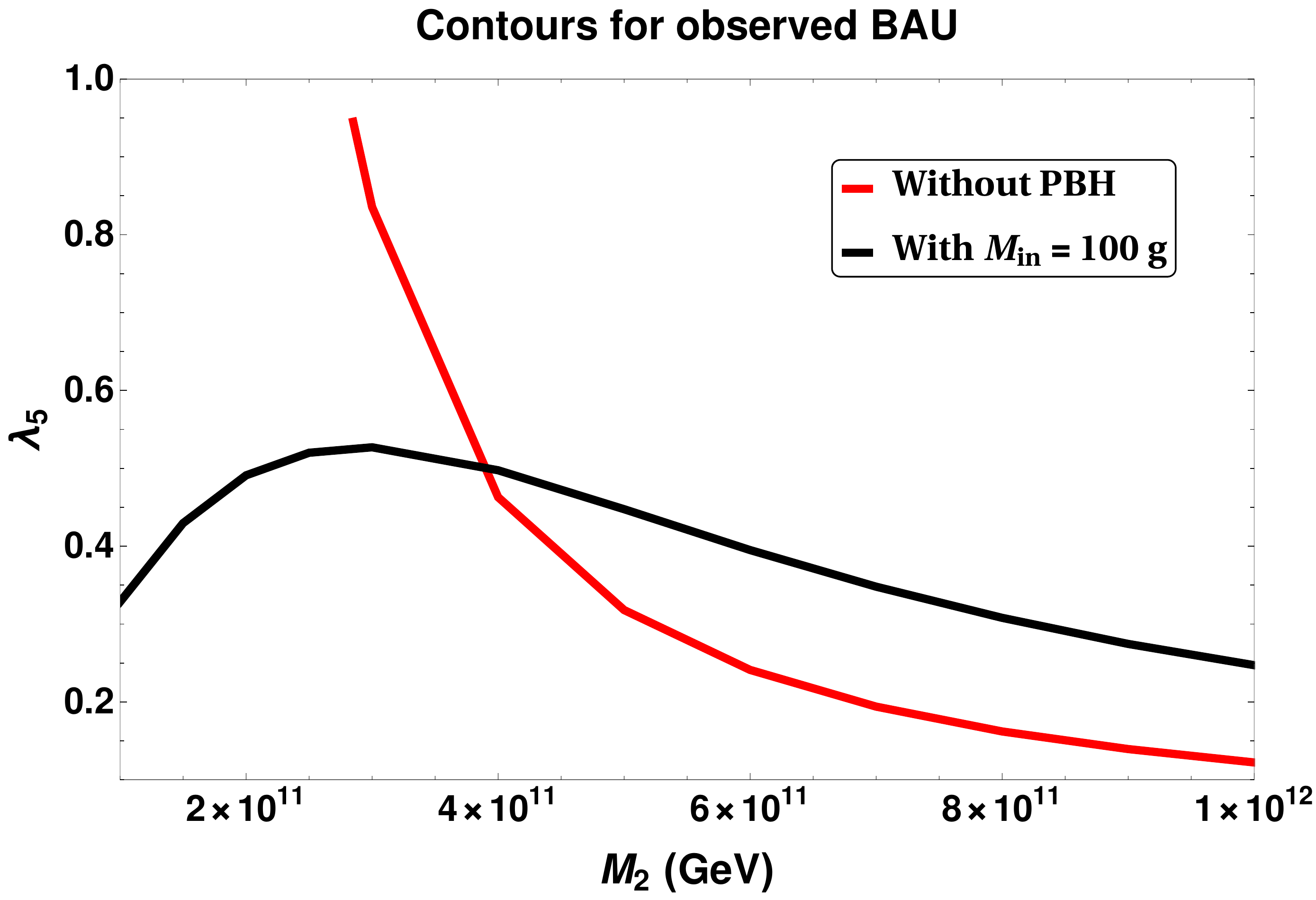}
\includegraphics[scale=0.28]{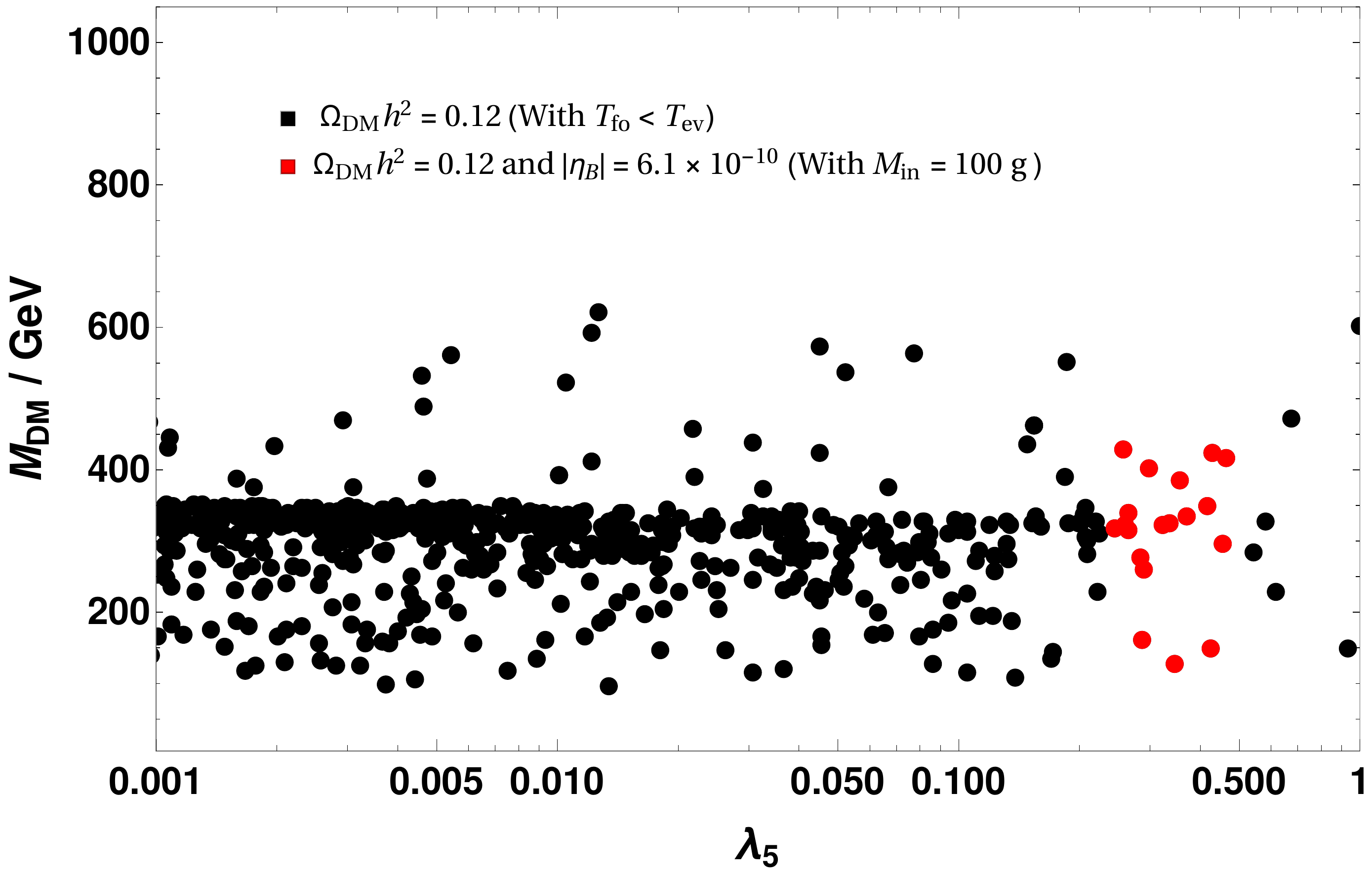}~~
\caption{Left panel: Parameter space in the $\lambda_5-M_{2}$ plane giving the observed BAU with and without the presence of PBH for $N_2$ leptogenesis with $\beta=10^{-3}$, $m_\nu^1 = 10^{-18}$ eV. Right panel: Parameter space in the $\lambda_5-M_{\rm DM}$ plane giving the observed fermion DM relic of the universe with the condition $T_{\rm fo}<T_{\rm ev}$.}
\label{fig:FDM}
\end{figure}

Similarly, we show the parameter space for $N_2$ leptogenesis and fermion DM in the presence of PBH in figure \ref{fig:FDM}. While non-thermal fermion DM has been discussed already, the $N_2$ leptogenesis favoured parameter space can be more tightly correlated with thermal fermion DM scenario \cite{Ahriche:2017iar}, while also showing the crucial differences arising due to the presence of PBH. While the contours for successful leptogenesis are different with and without PBH, the fermion DM mass range gets shifted to smaller values from the requirement $T_{\rm fo}<T_{\rm ev}$. Since fermion DM Yukawa couplings remain small due to large $\lambda_5$ required for high scale $N_2$ leptogenesis, its relic is governed primarily due to coannihilation with inert doublet scalars, requiring them to be light and hence within discovery reach of ongoing experiments like the large hadron collider (LHC).

To summarize, we have performed a detailed analysis of leptogenesis and dark matter production in the presence of PBH considering the well-known scotogenic model as a working example which naturally allows a low scale seesaw and leptogenesis possibility. The important parameters of this particle physics set up are the leptogenesis scale $M_{1,2}$, DM mass $M_\eta$ (or $M_1$), $\lambda_5$ and the lightest neutrino mass $m_\nu^1$, which also determine the Yukawa couplings. Our main motivation was to investigate how the presence of PBH can affect the otherwise allowed (disallowed) parameter space of this model in a radiation dominated universe, specially focusing on low scale leptogenesis which has not been discussed in earlier works. We indeed obtained some dependency on the PBH parameters: the initial PBH mass $M_{\rm in}$ and the initial PBH fraction $\beta$. These dependencies have been illustrated in figures \ref{fig:meta}, \ref{fig:MBHvsM1} and \ref{fig:mvseta3}, and further summarized in figures \ref{fig:parspace} and \ref{fig:N2lepto} along with the different bounds discussed above. Clearly, different values of parameters namely $M_{1,2}$, $M_\eta$, $\lambda_5$ and $m_\nu^1$ than that in the purely radiation dominated case is required now in the presence of PBH to generate the same baryon asymmetry. This is primarily due to the production of RHNs and DM by PBH and entropy injection by PBH. Although such a similar scenario was also considered in some earlier works, for example in \cite{Perez-Gonzalez:2020vnz} in the framework of high scale type I seesaw leptogenesis, here we explore the possibility of low scale seesaw in radiative neutrino mass model with the added bonus of a stable dark matter candidate. Since typical PBH initial mass and fraction affecting leptogenesis also lead to overproduction of DM, we are either forced to consider light thermal DM which freezes out after PBH evaporation or light non-thermal DM. We have constrained the PBH parameter space with both the observed baryon asymmetry and DM relic of the universe, specially in our second case of fermionic DM, where the interplay between the three entities (leptogenesis, DM and PBH) is more involved as explained earlier. We also found the contribution to leptogenesis from RHN produced by PBH to dominate compared to the entropy dilution effect in the second case, for a region of PBH masses  (see figure \ref{fig:mvseta3}). This can relax the lower bound on the leptogenesis scale by an order or so compared to the thermal $N_2$ leptogenesis case in a purely radiation dominated universe. The scalar DM scenario with $N_1$ leptogenesis is also predictive in the sense that large $\beta$ values for PBH affecting leptogenesis pushes DM mass into low mass regime. In such cases, DM can freeze-out after PBH evaporation without getting overproduced. Since scalar DM is part of an $SU(2)$ doublet, such light mass range can have interesting implications for discovery at collider or direct search experiments \cite{Belanger:2015kga, Belyaev:2016lok, Belyaev:2018ext, Arhrib:2013ela}. The comparison of leptogenesis and DM favoured parameter space for specific choices of PBH mass and initial fraction are shown in figure \ref{fig:wtwttPBH} and figure \ref{fig:FDM} for scalar DM ($N_1$ leptogenesis) and fermion DM ($N_2$ leptogenesis) respectively.

Apart from predicting different model parameters consistent with DM, BAU compared to the standard scenario (which can be looked up at particle physics based experiments), the PBHs can lead to observational consequences themselves. The different regions of PBH masses relevant for different leptogenesis scales and scenarios discussed in this work can give rise to a stochastic background of gravitational waves (GW) produced from PBH evaporation \cite{Anantua:2008am}\footnote{Other ways of generating GW from PBH are discussed in \cite{Saito:2008jc, Hooper:2020evu, Papanikolaou:2020qtd}.}. Such GW are in the high frequency regime (above THz), beyond the reach of presently functioning GW experiments. There have been several experimental proposals to detect such high frequency GW in future \cite{Arvanitaki:2012cn, Holometer:2015tus, Ito:2019wcb, Ejlli:2019bqj, Aggarwal:2020olq}. If evidence of a PBH-dominated universe with the above mass range is found, it would put tight constraints on purely thermal leptogenesis and dark matter models. In particular, future observations of PBH with mass around $10^3-10^4$ g can tightly constrain intermediate scale type I seesaw leptogenesis as found by \cite{Perez-Gonzalez:2020vnz} while scotogenic leptogenesis can still survive at low scale (or intermediate scale for $N_2$ case) as discussed earlier. While we have considered an ideal case with all PBHs having the same mass, similar to earlier works, a more detailed analysis with a realistic mass function of PBH while also including PBH formation mechanism is left for our future works.

\acknowledgments
DB acknowledges the support from Early Career Research Award from DST-SERB, Government of India (reference number: ECR/2017/001873). SJD would like to thank Yuber F. Perez-Gonzalez for an important clarification through e-mail, during the course of this work.

\appendix
\begin{figure}
\includegraphics[scale=.5]{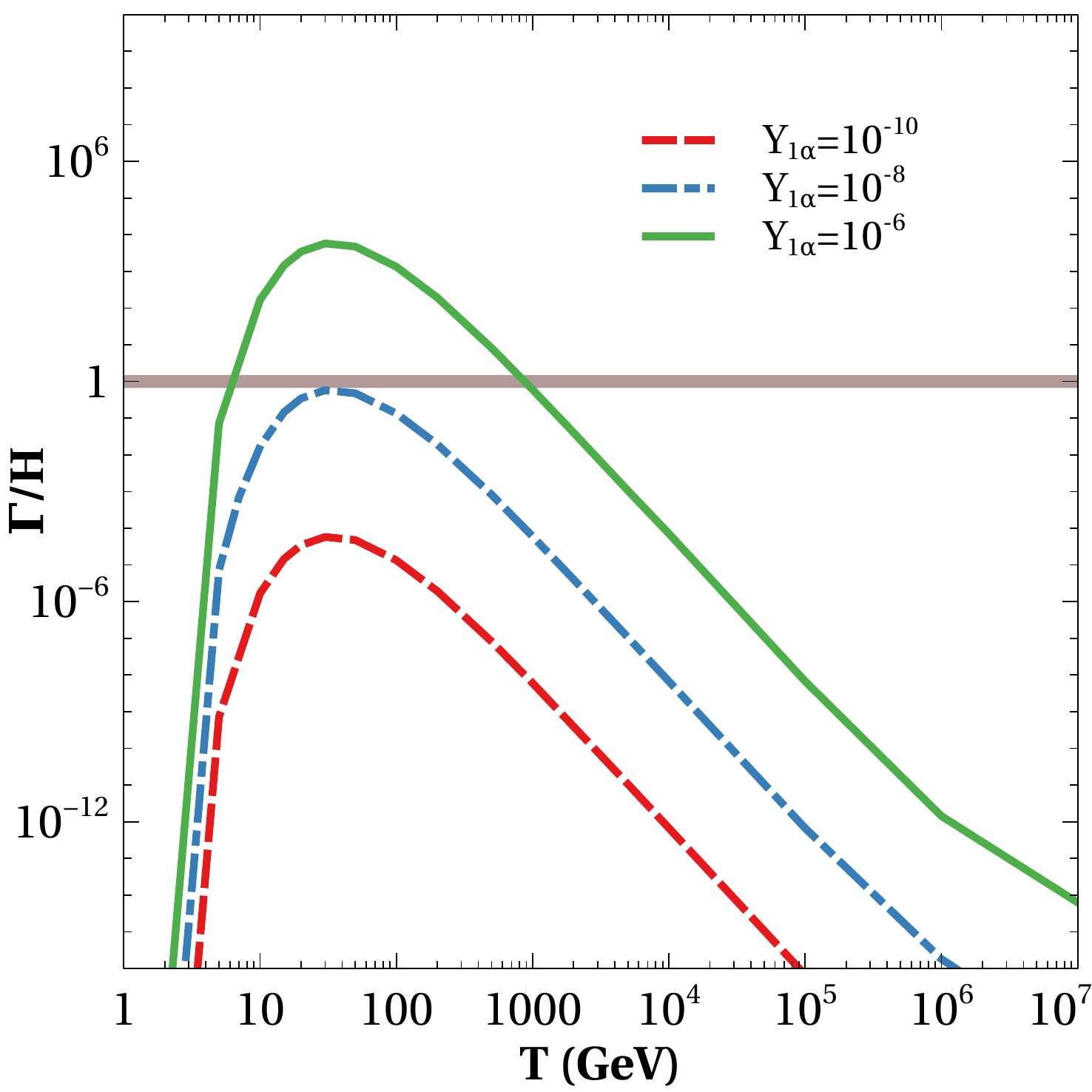}~~
\includegraphics[scale=.5]{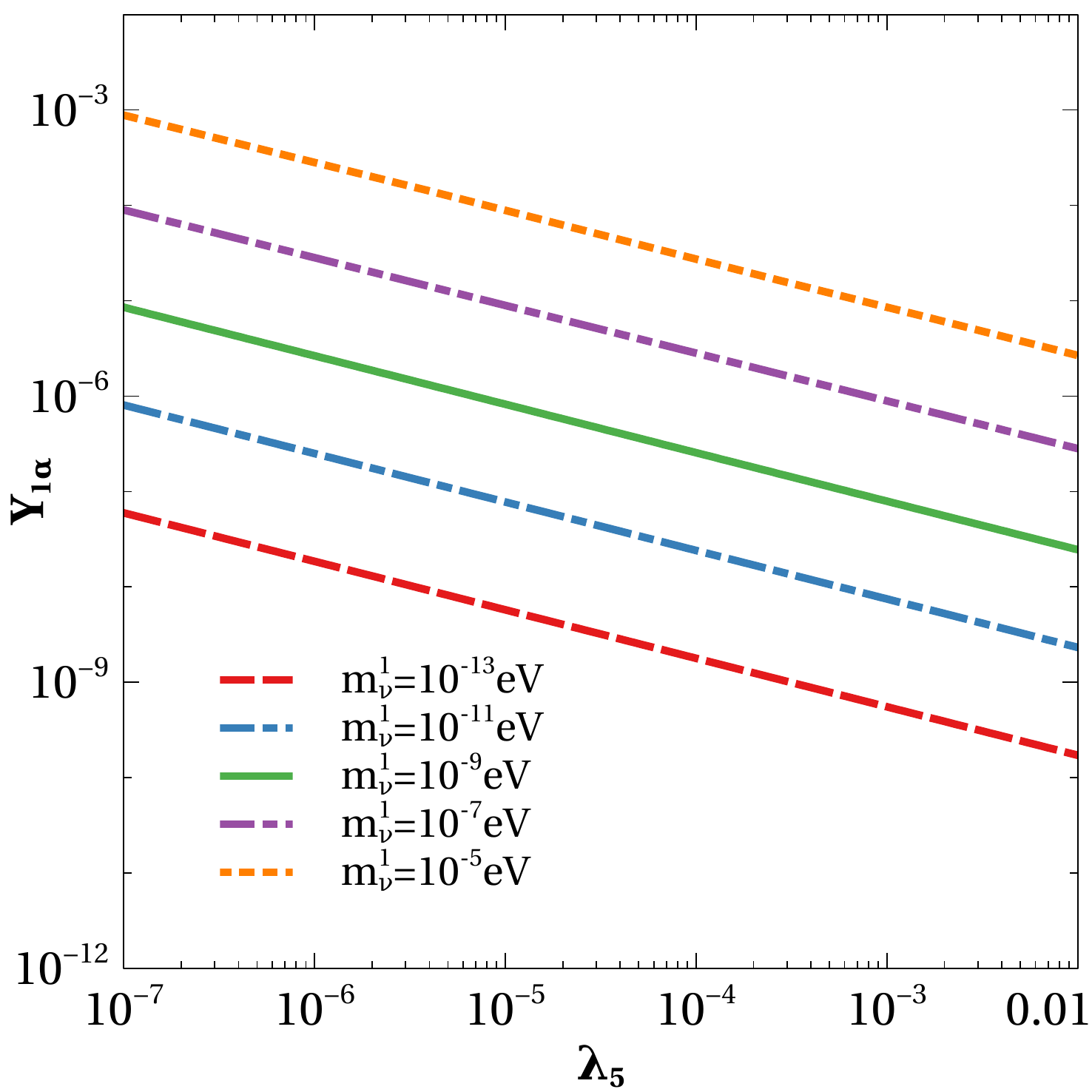}
\caption{Left panel: Interaction rates of right handed neutrino $N_1$ with SM bath taking $M_1=10$ TeV. Right panel: Dependence of Yukawa coupling on $\lambda_5$ and lightest neutrino mass $m_\nu^1$ via Casas-Ibarra parametrisation.}
\label{fig:rates}
\end{figure}
\section{On the initial abundance of $N_{1}$}
\label{appen:1}
The initial abundance of $N_{1}$ plays a crucial role in determining the asymmetry generation. The initial abundance $n_{N_{1}^{\rm ini}}=n_{N_{1}^{\rm eq}}$ is not guaranteed always. For that one has to look at the rate of the production channels of $N_{1}$ at the initial temperature and compare it with the Hubble expansion rate at that particular temperature. In the minimal scotogenic model we have these production channels of $N_{1}$: $l W^{\pm} (Z) \longrightarrow N_{1} \eta$ and $l l \longrightarrow N_{1} N_{1}$. Out of these, the process $l l \longrightarrow N_{1}N_{1}$ is Yukawa suppressed compared to the other two processes. We, therefore, consider the other two processes to find the rate of production of $N_{1}$. We use the package CalcHEP \cite{Belyaev:2012qa} to find the cross-sections for these two relevant processes. In figure \ref{fig:rates} (left panel), we show the interaction rates of $N_1$ for different Yukawa couplings which justify the use of negligible initial number density of $N_1$ for small couplings as it never enters equilibrium. The right panel plot of figure \ref{fig:rates} shows the dependence of Yukawa coupling on $\lambda_5$ and lightest neutrino mass $m_\nu^1$ (assuming normal ordering) through the 
Casas-Ibarra (CI) parametrisation \cite{Casas:2001sr} extended to radiative seesaw model \cite{Toma:2013zsa}. From the temperature evolution plot of interaction rates, we can infer that at a very high temperature the rates lie below the Hubble expansion rate for almost all the values of Yukawa couplings. For thermal low scale leptogenesis scenario where one usually solves the Boltzmann equations from a temperature of $T^{\rm initial}\sim 10 M_{1}-100 M_{1}$, it is not safe to take $n_{N_{1}^{\rm ini}}=n_{N_{1}^{\rm eq}}$ unless some gauge interaction is assumed at high energies which can bring $N_{1}$ into equilibrium. In our case we are solving the Boltzmann equations from the PBH formation temperature ($T^{\rm initial}=T_{\rm BH}^{\rm in}$)which is even higher compared to $T^{\rm initial} \sim 10 M_{1}-100 M_{1}$ and our Yukawas are also small ($Y_{1\alpha}\sim \mathcal{O}(10^{-8})$). Therefore we take a vanishing initial abundance of $N_{1}$ in the whole analysis.

\section{Davidson-Ibarra bound in scotogenic model}
\label{appen:2}
The Davidson-Ibarra bound \cite{Davidson:2002qv} in the usual type I seesaw model can be obtained by using the simplified form of the CP asymmetry parameter (for hierarchical right handed neutrinos) as
\begin{equation}
 \epsilon_{1} \lesssim \dfrac{3}{8 \pi} \dfrac{M_{1}}{v^{2}}(m_{3}-m_{1}).
 \label{DIseesaw}
\end{equation}
Thus, the CP asymmetry becomes zero when the light neutrinos become mass-degenerate. A hierarchical right handed neutrino spectrum also suggests a hierarchical light neutrino spectrum. Therefore, for a hierarchical right handed neutrino spectrum one can safely assume 
\begin{equation}
 \mid \epsilon_{1} \mid \lesssim \dfrac{3}{8 \pi} \dfrac{M_{1}}{v^{2}}m_{3}
 \label{DIseesaw2}
\end{equation}
From equation \eqref{DIseesaw2}, clearly one does not have enough freedom to increase the CP asymmetry except increasing the mass of the right handed neutrino $N_{1}$. This bound is very strong and not weakened significantly even if one considers moderately hierarchical right handed neutrino spectrum \cite{Davidson:2002qv}. For the correct order of baryon asymmetry to be generated, given the sphaleron conversion factor, one needs $\mid \epsilon_{1} \mid > 10^{-6}$ which further implies $M_{1}>10^{9}$ GeV.

On the other hand, the Davidson-Ibarra bound for the scotogenic model can be found by using the simplified expression for corresponding CP asymmetry parameter (in hierarchical right handed neutrino limit) \cite{Hugle:2018qbw} 
\begin{equation}
 \mid \epsilon_{1} \mid \lesssim \dfrac{3\pi}{4 \lambda_{5}v^{2}} \xi_{3} M_{1}(m_{3}-m_{1}),
 \label{DIscoto}
\end{equation}
where 
\begin{equation}
 \xi_{i}=\left( \dfrac{1}{8} \dfrac{M_{i}^{2}}{m_{H_{0}}^{2}-m_{A_{0}}^{2}}\left[ L(m_{H_{0}}^{2})-L(m_{A_{0}}^{2}) \right] \right)^{-1}.
\end{equation}
The values of $\xi_{i}$ are of the order of unity for most of the parameter space of interest. From equation \eqref{DIscoto}, one can see that by choosing small values of $\lambda_{5}$ the required CP asymmetry can be generated even with small $M_{1} \sim \mathcal{O}$(10 TeV). The reason for the appearance of the coupling $\lambda_{5}$ in the Davidson-Ibarra bound is because of the radiative mass generation of the neutrinos through the quartic coupling of Higgs and inert doublet $\eta$. By choosing a small value of $\lambda_{5}$ one can have large values of the Yukawa couplings which satisfy the light neutrino data and also it makes the CP asymmetry parameter large. However, in the usual type I seesaw model smaller Yukawa is needed to satisfy the light neutrino data and therefore the asymmetry remains under-abundant unless $M_{1}>10^{9}$ GeV.


\providecommand{\href}[2]{#2}\begingroup\raggedright\endgroup

\end{document}